\documentclass{cernyrep}
\usepackage{graphicx}
\usepackage{subfig}
\usepackage{multirow}
\usepackage{array}
\usepackage{amssymb}
\usepackage[T1]{fontenc}
\usepackage[bookmarks, colorlinks=true, linktoc=page, pdftex, linkcolor=black, citecolor=black, urlcolor=blue]{hyperref}
\usepackage{siunitx}
\usepackage{csquotes}
\usepackage{xargs}                      

\usepackage{tikz}
\usepackage{tikz-3dplot}
\usetikzlibrary{shapes,arrows}
\usetikzlibrary{positioning}
\tikzstyle{block} = [rectangle, draw, fill=blue!20, 
    text width=5em, text centered, rounded corners, minimum height=4em]
\tikzstyle{line} = [draw, -latex']

\usepackage{fancyhdr}
\fancyhfoffset{4 mm}
\fancypagestyle{ARTTITLE}{%
\fancyhf{} 
\lhead{\small{Proceedings of the 2018 CERN--Accelerator--School course on\\ 
\it{Numerical Methods for Analysis, Design and Modelling of Particle Accelerators}, Thessaloniki, (Greece)}} 
\lfoot{Available online at \url{https://cas.web.cern.ch/previous-schools}}
\rfoot{\thepage\hspace*{3mm}}
 
}

\begin{document}
\title{Numerical analysis techniques for non-linear dynamics}
\author{Yannis Papaphilippou
                 }
\institute{CERN, Geneva, Switzerland}

\begin{abstract}
The content of this contribution is based on the course on numerical analysis techniques for non-linear dynamics.
After introducing basic concepts as the visual analysis of trajectories in phase space and the importance
of the nature of fixed points in their topology and dynamics, the motion close to a resonance is presented, 
with simple non-linear map examples. The onset of chaotic motion and the modern methods used for their detection
are detailed with a focus on frequency map analysis and concrete examples for a variety of rings and non-linear effects.
\end{abstract}


\maketitle

\thispagestyle{ARTTITLE}

\section{Introduction}

The purpose of this lecture is to present techniques for the analysis of  non-linear motion, with direct applications to particle
accelerators. Starting from the basic principles of phase-space dynamics and in particular the stability of fixed points
(or periodic orbits), the concept of Poincar\'e map is introduced, where the motion close to a resonance and the onset
of chaotic motion can be analysed based on simple numerical examples. The second part is dedicated to numerical
methods for the detection of chaos, with an emphasis to Frequency Map Analysis (FMA) and its application to
various concrete beam dynamics problems.

\section{Phase space dynamics}

\begin{figure}[ht]
\begin{center}
\begin{tikzpicture}
 \node (img1)  {\includegraphics[width=6cm]{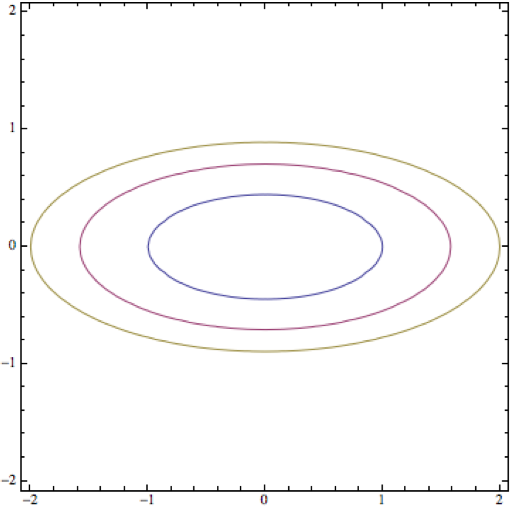}};
 \node[below=of img1, node distance=0cm, xshift=0.1cm,yshift=1cm,font=\color{black}] {$u$};
 \node[left=of img1, node distance=0cm, rotate=0, anchor=center,xshift=0.5cm,yshift=0.1cm,font=\color{black}] {$p_u$};
\node[right=of img1,yshift=0.1cm] (img2)  {\includegraphics[width=6cm,height=6.05cm]{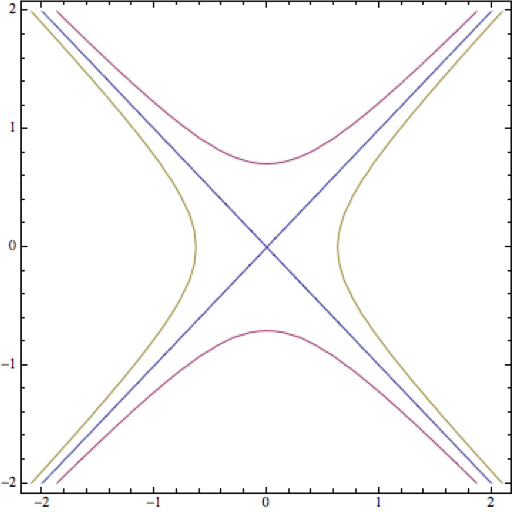}};
 \node[below=of img2, node distance=0cm, xshift=0.1cm,yshift=1cm,font=\color{black}] {$u$};
 \node[left=of img2, node distance=0cm, rotate=0, anchor=center,xshift=0.5cm,yshift=0.1cm,font=\color{black}] {$p_u$};
 \end{tikzpicture}
\caption{Phase-space plot $(u,p_u)$ of a simple harmonic oscillator corresponding to the Hamiltonian of eq.~\eqref{eq:harmosc} (left) and one eq.~\eqref{eq:harmoscII} (right).}
\label{fig:harmosc}
\end{center}
\end{figure}
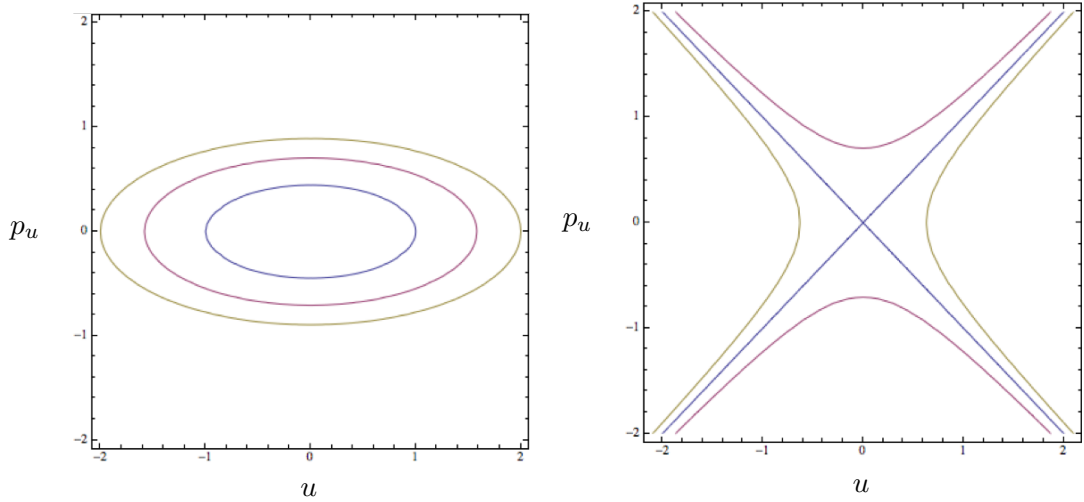

Following the evolution of trajectories in phase space $u,p_u$, with $u$
the ``position" and $p_u$ the conjugate momentum, provides a 
valuable description for understanding the evolution of a dynamical
system. In particular, the existence of an integral of motion, i.e.
a constant quantity with time (e.g. the value of the "Hamiltonian" 
or the "energy" for a conservative system)  imposes 
geometrical constraints on phase space flow. 
For the simple example of an harmonic oscillator, with eigen-frequency $\omega_0$
\begin{equation}
H=\frac{1}{2}\left( p_u^2 + \omega_0^2 u^2\right)
\label{eq:harmosc}
\end{equation}
the phase space curves are ellipses parameterized by the Hamiltonian (energy) 
around the equilibrium point (see Fig.~\ref{fig:harmosc} left).

\begin{figure}[ht]
\begin{center}
\begin{tikzpicture}
 \node (img1)  {\includegraphics[width=4cm]{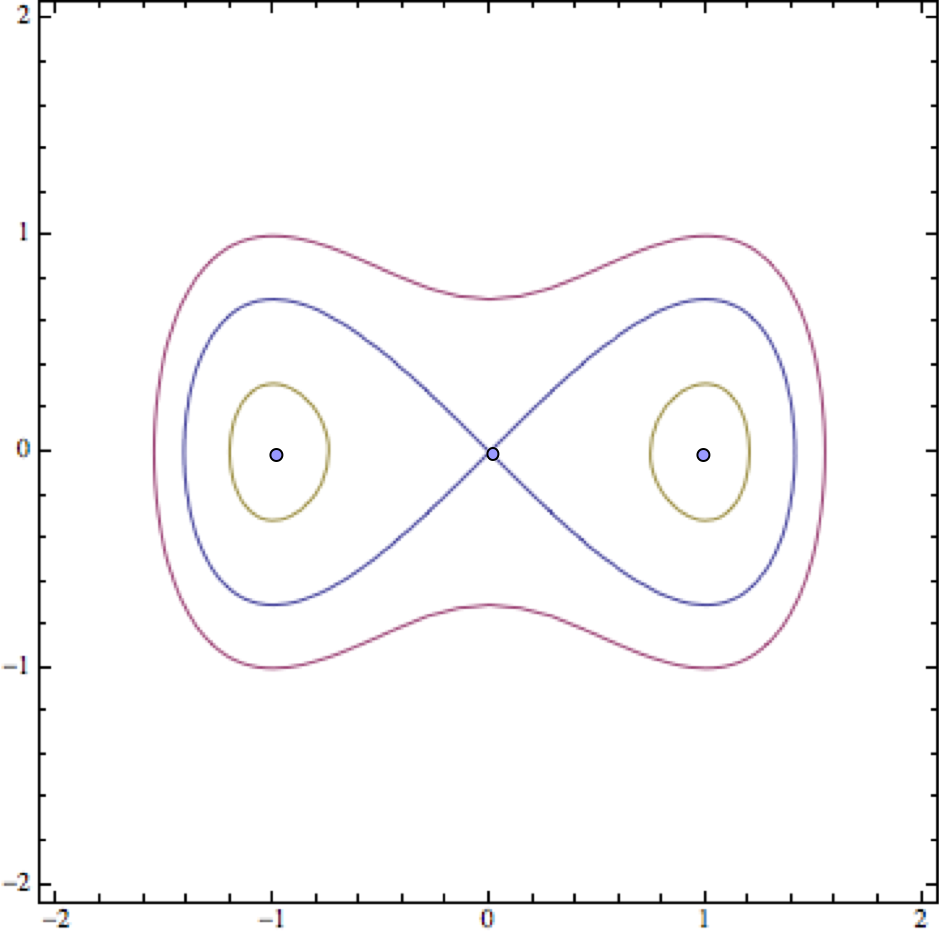}};
 \node[below=of img1, node distance=0cm, xshift=0.1cm,yshift=1cm,font=\color{black}] {$u$};
 \node[left=of img1, node distance=0cm, rotate=0, anchor=center,xshift=0.8cm,yshift=0.1cm,font=\color{black}] {$p_u$};\node[right=of img1,yshift=0cm] (img2)  {\includegraphics[width=4cm]{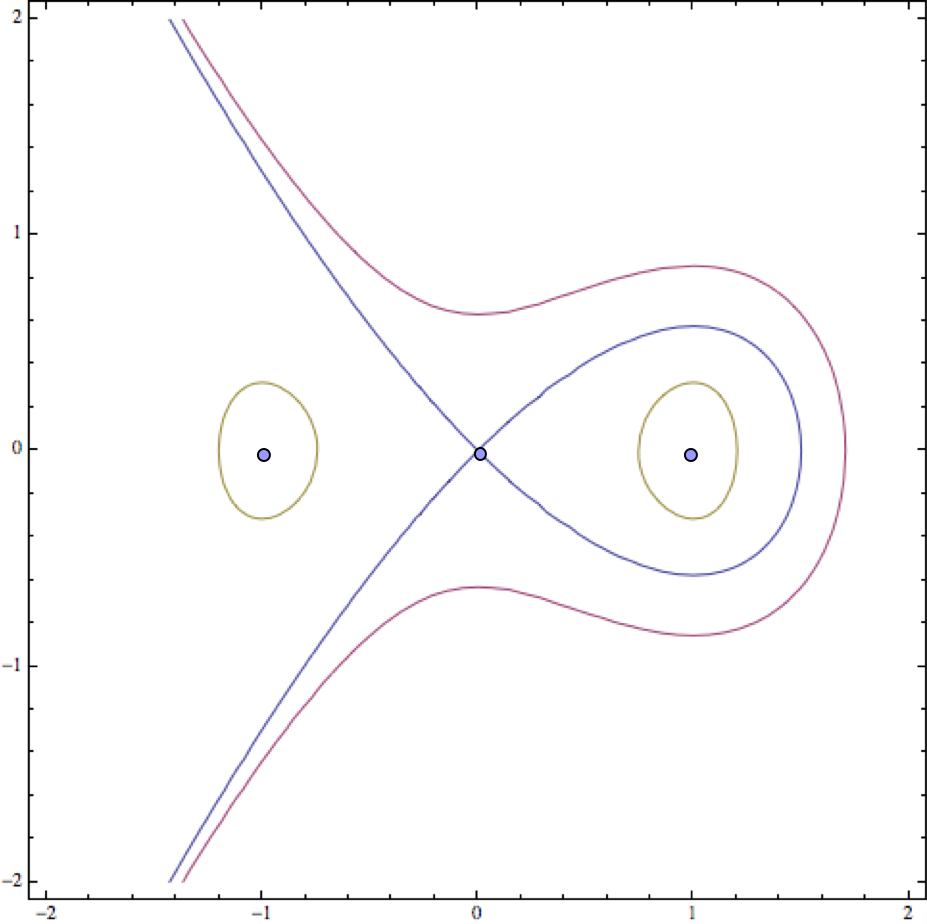}};
  \node[below=of img2, node distance=0cm, xshift=0.1cm,yshift=1cm,font=\color{black}] {$u$};
 \node[left=of img2, node distance=0cm, rotate=0, anchor=center,xshift=0.8cm,yshift=0.1cm,font=\color{black}] {$p_u$};
\node[right=of img2,yshift=0cm] (img3)  {\includegraphics[width=4cm]{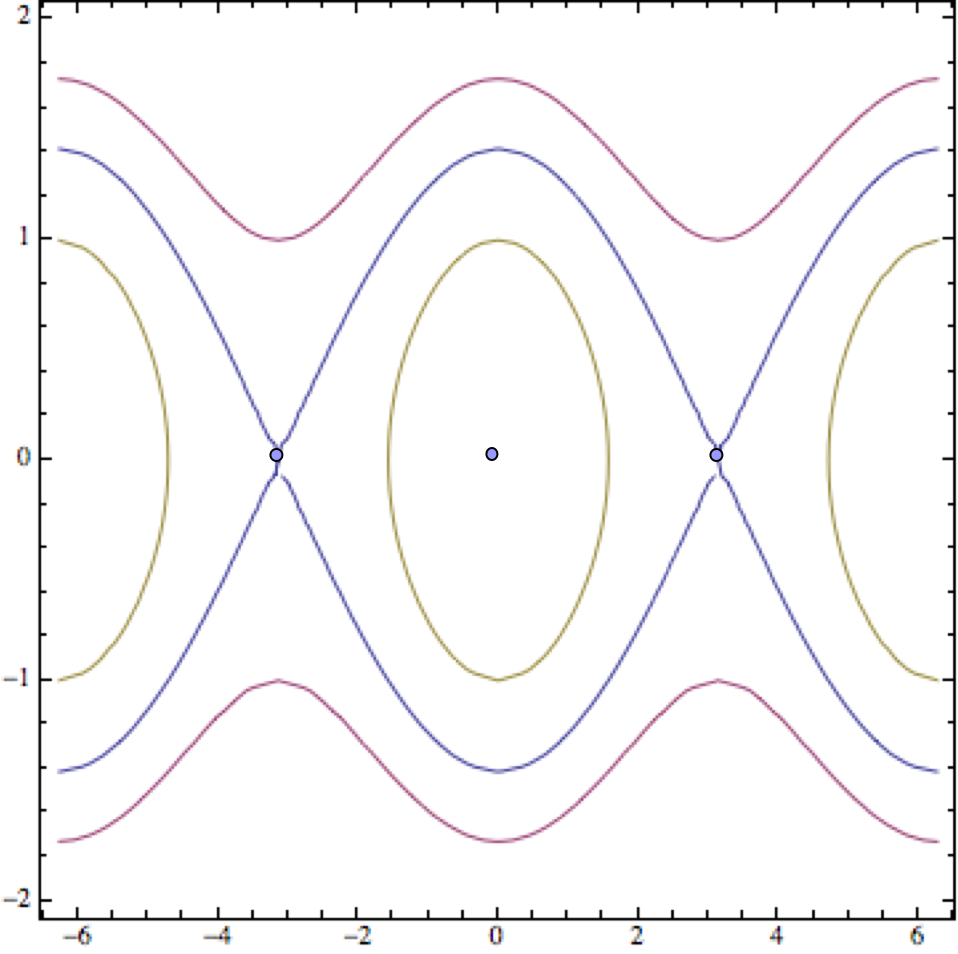}};
 \node[below=of img3, node distance=0cm, xshift=0.1cm,yshift=1cm,font=\color{black}] {$u$};
 \node[left=of img3, node distance=0cm, rotate=0, anchor=center,xshift=0.8cm,yshift=0.1cm,font=\color{black}] {$p_u$};
 \end{tikzpicture}
\caption{Phase-space plot $(u,p_u)$ of three non-linear oscillators with a quartic potential (left), a cubic potential (center) and the pendulum (right).}
\label{fig:nonosc}
\end{center}
\end{figure}
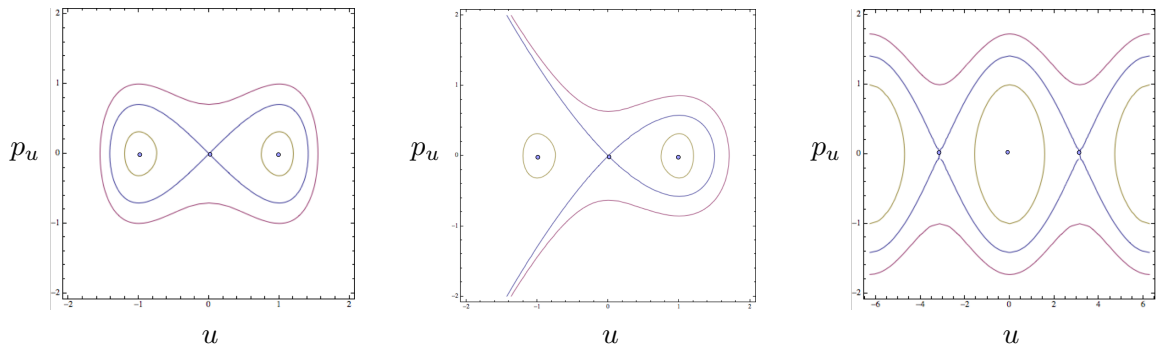

By simply changing the sign of the potential in the harmonic oscillator, i.e. for a Hamiltonian
written as:
\begin{equation}
H=\frac{1}{2}\left( p_u^2 - \omega_0^2 u^2\right)
\label{eq:harmoscII}
\end{equation}
the phase trajectories become hyperbolas, which are symmetric around the equilibrium point,
where two straight lines cross, moving towards and away from it.
It is straightforward to generalise this picture for conservative non-linear oscillators having the Hamiltonian
\begin{equation}
H=\frac{1}{2} p_u^2 - V(u)
\label{eq:hamiltnonosc}
\end{equation}                             								              									              with the potential $V(u)$ being a general polynomial function of positions.
The equilibrium points are associated with extrema of the potential. As
examples, three non-linear oscillators phase space plots are presented in Fig.~\ref{fig:nonosc}: 
one with a quartic potential with Hamiltonian $H = \frac{1}{2}p_u^2 - \frac{1}{2}u^2 +\frac{1}{4}u^4$ 
with three equilibrium points corresponding to two minima and one maximum (left);  
a cubic potential (center) with Hamiltonian $H = \frac{1}{2}p_u^2 - \frac{1}{2}u^2 +\frac{1}{3}u^3$ 
(one minimum and one maximum); and the well-known pendulum (right) $H=\frac{1}{2}p_\phi^2 - \frac{g}{L}\cos\phi$, 
presenting periodic minima and maxima.

\subsection{Fixed point analysis}

Consider a general second order differential equation system 
\begin{equation}
 \begin{aligned}
\frac{du}{dt}  &= f_1(u,p_u) \;\;,\\
\frac{dp_u}{dt}&= f_2(u,p_u) \;\;.
\label{eq:difeq}
\end{aligned}
\end{equation}
If the system is described by the Hamiltonian, \eqref{eq:hamiltnonosc}, the above equations correspond to the Hamilton's equations:
\begin{equation}
 \begin{aligned}
 \frac{du}{dt}  &= \frac{\partial H(u,p_u)}{\partial p_u} = p_u \;\;,\\
\frac{dp_u}{dt}&= -\frac{\partial H(u,p_u)}{\partial u} = -\frac{\partial V(u)}{\partial u}  \;\;.
\end{aligned}
\label{eq:difeqham}
\end{equation} 
The equilibrium or “fixed” points at $(u,p_u)=(u_0,p_{u0})$ are defined by $ f_1(u_0,p_{u0}) = f_2(u_0,p_{u0})=0$ 
and are determinant for the topology of the close-by trajectories. In the case of the Hamiltonian \eqref{eq:hamiltnonosc}, they are located at $(u,p_u)=(u_0,0)$, where the position is obtained by $\frac{\partial V(u)}{\partial u}|_{u=u_0}=0$.
The linearized equations of motion at their vicinity are
\begin{equation}
\frac{d}{dt}\begin{bmatrix} \delta u \\ \delta p_u\end{bmatrix} =  {\cal M}_J \begin{bmatrix} \delta u \\ \delta p_u\end{bmatrix} = \begin{bmatrix} \dfrac{\partial f_1(u_0,p_{u0})}{\partial{u}} & \dfrac{\partial f_1(u_0,p_{u0})}{\partial{p_u}}\\ \dfrac{\partial f_2(u_0,p_{u0})}{\partial{u}} & \dfrac{\partial f_2(u_0,p_{u0})}{\partial{p_u}} \end{bmatrix} \begin{bmatrix} \delta u \\ \delta p_u\end{bmatrix}\;\;,
\label{eq:jacob}
\end{equation}
where ${\cal M}_J$ is the Jacobian matrix of the differential equation system. The fixed point nature is revealed 
by the nature of the eigenvalues of the Jacobian matrix ${\cal M}_J $, given by the solutions of the characteristic polynomial  
$\det |{\cal M}_J -\lambda {\mathbf I}| = 0$.

For a conservative system with one degree of freedom, the second order characteristic polynomial for any fixed point has two possible solutions corresponding to:
\begin{itemize}
\item Two {\it complex} eigenvalues with opposite sign, representing {\it elliptic} fixed points. The phase space flow is described by ellipses, with particles evolving clockwise or anti-clockwise (see left side of Fig.~\ref{fig:harmosc}).
\item Two {\it real} eigenvalues with opposite sign, representing {\it hyperbolic} (or saddle) fixed points. The phase space flow is described by two lines (or manifolds), which are incoming (stable) and outgoing (unstable) (see right side of Fig.~\ref{fig:harmosc}).
\end{itemize}

The pendulum, with length $L$, gravitational constant $g$ conjugate position and momenta $(\phi,p_\phi)$, represented by the Hamiltonian $H=\frac{1}{2}p_\phi^2 - \frac{g}{L}\cos\phi$ can serve as an example of the described fixed point analysis. The “fixed” points can be found by using eqs.~\eqref{eq:difeqham} and they are located at $(\phi_n,p_\phi) = (\pm n \pi,0)$, for $n=0,1,2\dots$. The Jacobian matrix is
$
\begin{bmatrix}
0 & 1 \\ -\frac{g}{L}\cos{\phi_n} & 0 
\end{bmatrix}\;\;,
$
with its eigenvalues $
\lambda_{1,2}=\pm i \sqrt{\frac{g}{L}\cos{\phi_n}}  $.
Two cases can be distinguished: 
\begin{itemize}
\item For even multiples of the position (phase) $\phi_n= 2 n \pi$, the eigenvalues are purely imaginary $\lambda_{1,2}=\pm i \sqrt{\frac{g}{L}} $ corresponding to elliptic fixed points. 
\item For odd multiples of the phase $\phi_n= (2 n+1) \pi$, the eigenvalues are real $\lambda_{1,2}=\pm  \sqrt{\frac{g}{L}} $	corresponding to hyperbolic fixed points.
\end{itemize}
The separatrix represented by the blue curve in Fig.~\ref{fig:nonosc} (right) is defined by the stable and unstable manifolds through the hyperbolic points, separating bounded librations (green curves) and unbounded rotations (red curves).

\subsection{Phase space for time-dependent systems}

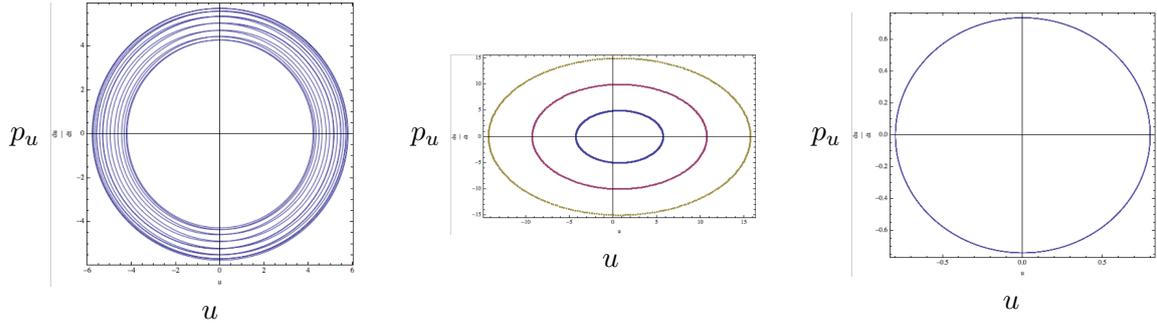
\begin{figure}[ht]
\begin{center}
\begin{tikzpicture}
 \node (img1)  {\includegraphics[width=4cm]{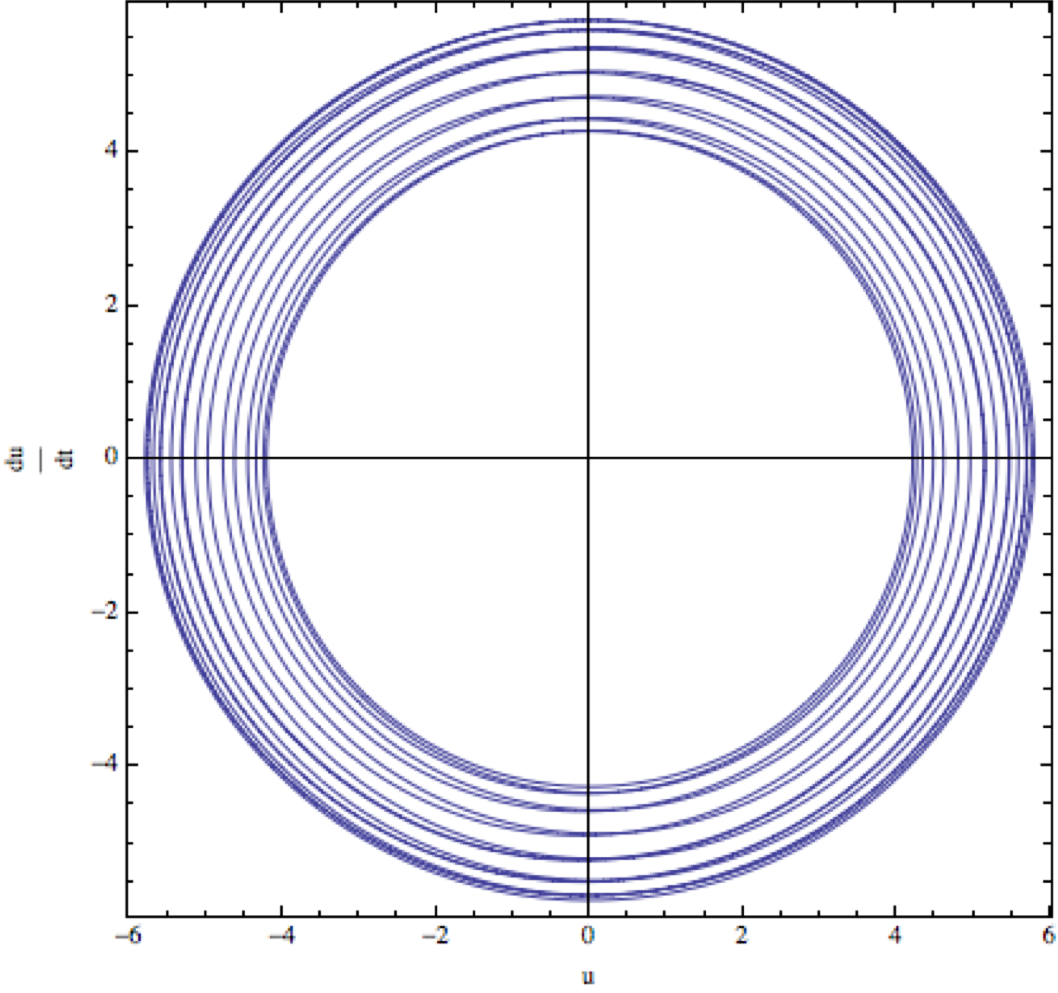}};
 \node[below=of img1, node distance=0cm, xshift=0.1cm,yshift=1cm,font=\color{black}] {$u$};
 \node[left=of img1, node distance=0cm, rotate=0, anchor=center,xshift=0.8cm,yshift=0.1cm,font=\color{black}] {$p_u$};\node[right=of img1,yshift=0cm] (img2)  {\includegraphics[width=4cm]{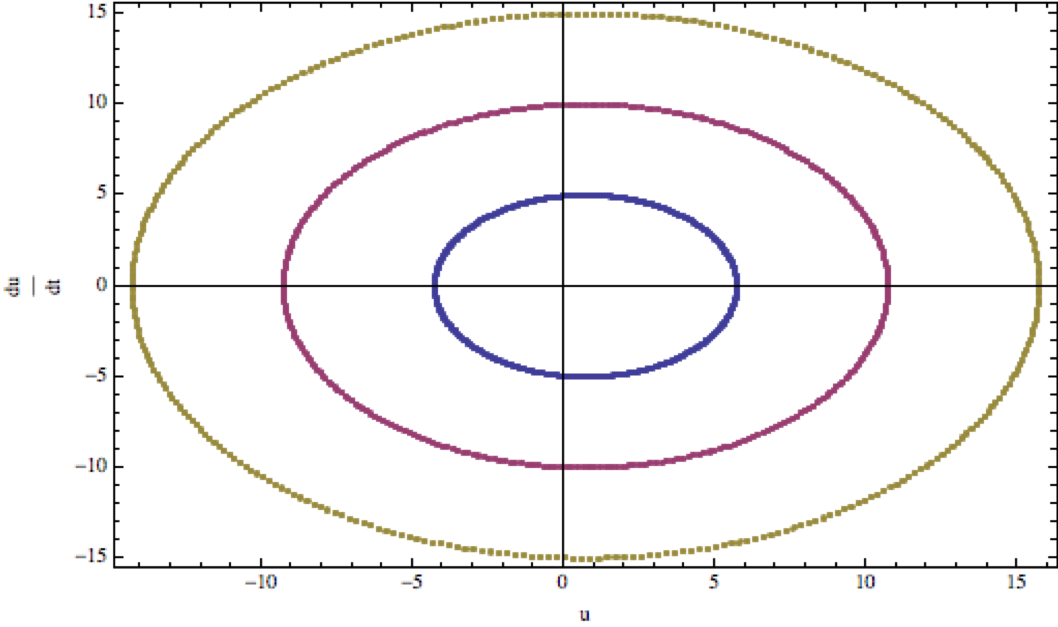}};
  \node[below=of img2, node distance=0cm, xshift=0.1cm,yshift=1cm,font=\color{black}] {$u$};
 \node[left=of img2, node distance=0cm, rotate=0, anchor=center,xshift=0.8cm,yshift=0.1cm,font=\color{black}] {$p_u$};
\node[right=of img2,yshift=0cm] (img3)  {\includegraphics[width=4cm]{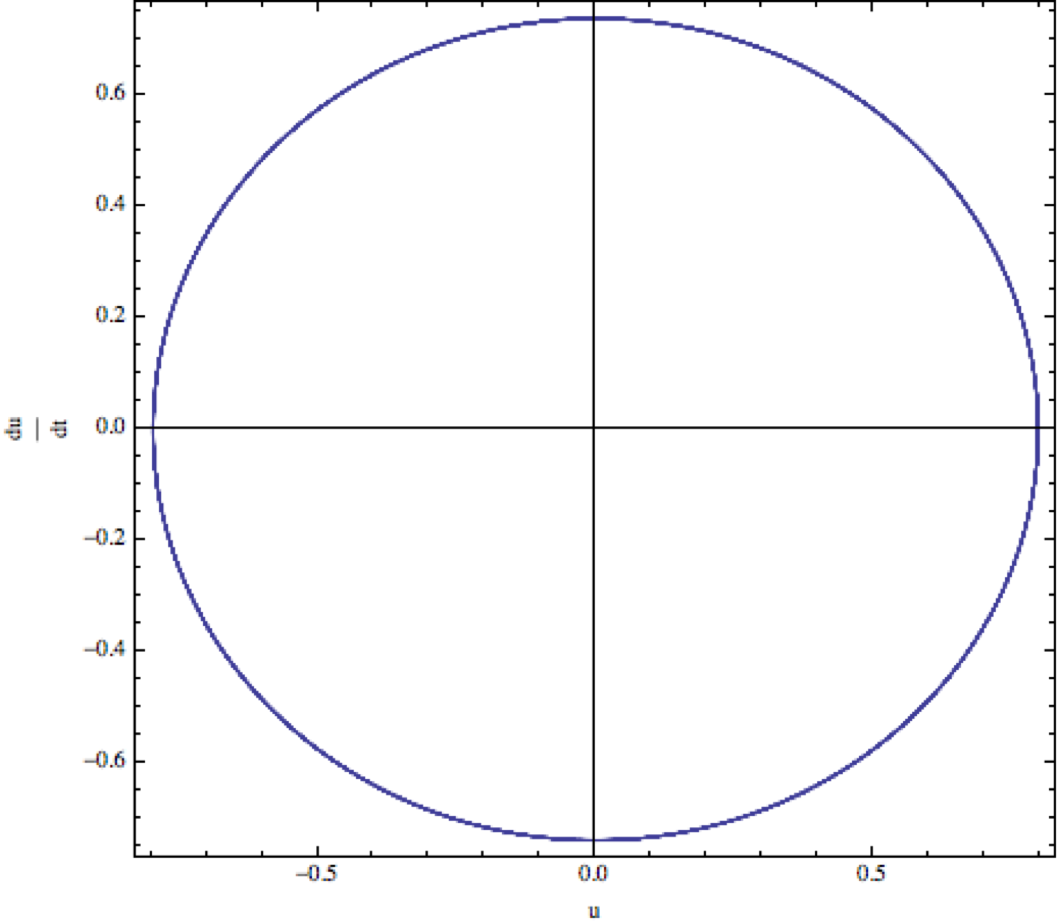}};
 \node[below=of img3, node distance=0cm, xshift=0.1cm,yshift=1cm,font=\color{black}] {$u$};
 \node[left=of img3, node distance=0cm, rotate=0, anchor=center,xshift=0.8cm,yshift=0.1cm,font=\color{black}] {$p_u$};
 \end{tikzpicture}
\caption{Phase-space plot $(u,p_u)$ for one trajectory of the time-dependent system \eqref{eq:hamperiod} (left), Poincar\'e map for integer multiples of the period $T_0$ (center) and the periodic orbit passing through the origin  $(u,p_u)=(u_0,0)$ (right).}
\label{fig:periodosc}
\end{center}
\end{figure}

Consider a simple harmonic oscillator where the frequency $\omega_0(t)$ is time-dependent and is
represented by the Hamiltonian
\begin{equation}
H= \frac{1}{2} \left(  p_u^2 +\omega_0^2(t) u^2 \right)  \;\;.
\label{eq:hamperiod}
\end{equation} 
Plotting the evolution of a trajectory in phase space in Fig.~\ref{fig:periodosc} (left), 
provides curves that intersect each other, due to the fact that now the phase space is 3-dimensional, 
where time $t$ is the extra dimension. When the time dependence is periodic, as in the case 
of accelerator rings, the elimination of time can be achieved by rescaling it to become  $\tau = \omega_0 t$. 
Considering every integer interval of the new “time” variable (i.e. every period $T_0=\frac{2\pi}{\omega_0}$), 
the phase space looks like the one of the harmonic oscillator (see center part of Fig.~\ref{fig:periodosc}).
This is the simplest version of a {\it Poincar\'e map} or {\it surface of section}, 
which is useful for studying geometrically the phase space of multi-dimensional systems. 
The fixed point at the location $(u,p_u)=(u_0,0)$ of the map is indeed a periodic orbit, as 
represented in Fig.~\ref{fig:periodosc} (right).

\section{Poincar\'e map}

 The first recurrence or {\it Poincar\'e map}~\cite{Poincare} or surface of section is defined by the successive intersections of trajectories of a dynamical system, with a fixed surface transversal to the phase space flow, i.e. for which trajectories intersect this surface and do not run parallel to it. For an autonomous Hamiltonian system $H(\mathbf{q},\mathbf{p})$  with no explicit time dependence, it can be chosen to be any surface in phase space, e.g. $q_i=0$. In the case of a non-autonomous Hamiltonian system $H(\mathbf{q},\mathbf{p}, t)$, with explicit periodic time dependence, this surface can be defined by multiples of the period $t=k T_0$, with $k$ and integer. 

In a system with $n$ degrees of freedom (or	$n+1$ including the time), the phase space has $2n$ (or $2n+2$) dimensions. 
 By fixing the value of the Hamiltonian to $H=H_0$, a Poincar\'e map reduces the phase space dimensions to $2n-2$ (or $2n$).
This shows that the Poincar\'e map is particularly useful to be constructed and explored visually for a system with $n=2$ degrees of freedom, 
or 1 degree of freedom plus time, as the motion on the map is described by 2-dimensional curves.
 For continuous system, there are numerical techniques~\cite{Henon} to compute the surface exactly.

\begin{figure}[ht]
\begin{center}
{\includegraphics[width=5cm]{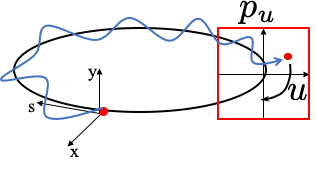}}
{\includegraphics[width=5cm]{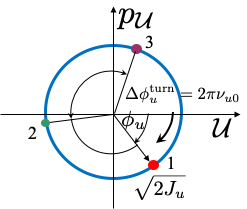}}
{\includegraphics[width=5cm]{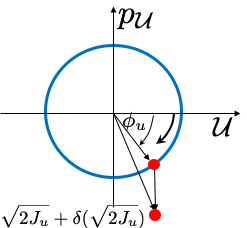}}
\caption{Schematic plot of a Poincar\'e map $(u,p_u)$ for an accelerator ring (left), phase space trajectory in normalised coordinates $({\cal U},p_{\cal U})$ for a linear system (center) and introduction of a non-linear "kick" (right).}
\label{fig:poinmap}
\end{center}
\end{figure}

For an accelerator ring, the construction of the {\it Poincar\'e map} is very natural as the periodicity is imposed by geometry itself. 
One could imagine a virtual Beam Position Monitor which is able to record positions and momenta at a location of the ring. 
This will give the usual ellipses for linear motion, which can be transformed to circles (see Fig.~\ref{fig:poinmap}a) by the well-known Floquet transformation 
from the original variables $(u,u')\approx(u,p_u)$~\footnote{The approximation $u'\approx p_u$ assumes a term $\frac{p_u^2}{2}$ in the 
relativistic Hamiltonian which is only true for transverse magnetic fields, on-momentum motion and expansion of the square root involving momenta
to lower orders, i.e. the "paraxial" approximation.} to normalised coordinates $({\cal U},{\cal U}')\approx({\cal U},p_{\cal U})$, with 
\begin{equation}
\begin{pmatrix}
{\cal U} \\ p_{\cal U}
\end{pmatrix}
=
\begin{pmatrix}
\frac{1}{\sqrt{\beta}} & 0 \\
\frac{\alpha}{\sqrt{\beta}} & \sqrt{\beta}
\end{pmatrix}
\begin{pmatrix}
u \\ p_u
\end{pmatrix}
= \sqrt{2 J_u}
\begin{pmatrix}
\cos({\phi_u}) \\
-\sin({\phi_u})
\end{pmatrix}
\;\;,
\label{eq:normcoord}
\end{equation}
with $J_u = \frac{1}{2}(\gamma_u u^2 + 2\alpha_u u p_u + \beta_u p_u^2)$ and $\phi_u = \arctan\left( -\beta_u \frac{p_u}{u}-\alpha_u\right)$ the action-angle variables.
The particles are executing simple rotations on a circle of radius $\sqrt{2 J_u}$, which is an integral of motion with a change in the angle variable per turn $\Delta \phi_u^\text{turn} = 2\pi \nu_{u0}$,  with $\nu_{u0}$ the unperturbed frequency of motion or tune. In integrable systems with more degrees of freedom, the phase space trajectories move in products of these circles, i.e. multi-dimensional tori parameterised by the action integrals and the angles are just varying linearly with "time" (or "turns" in the case of an accelerator ring). A resonance condition corresponds to a periodic orbit or fixed points on the Poincar\'e map. If a non-linearity is introduced at a certain location, represented like a change of momentum ${\delta P_{\cal U}}$ (or "kick" in the accelerator jargon), this will translate to a radius  or an action change  $\delta{(\sqrt{2J_u})}$ and thereby the circular phase space curve (or in general the torus) will be distorted.

\subsection{Single Octupole map}

A simple numerical example can be considered, by building the map composed by an uncoupled rotation in phase space with phase advances  $(\mu_x,\mu_y)$ 
\begin{equation}
\begin{pmatrix}
x \\ p_{x} \\ y \\ p_{y}
\end{pmatrix}_{n+1}
=
\begin{pmatrix}
\cos(\mu_x) & \sin(\mu_x) & 0 & 0 \\
-\sin(\mu_x) & \cos(\mu_x) & 0 & 0 \\
& 0 & 0 & \cos(\mu_y) & \sin(\mu_y) \\
& 0 & 0 & -\sin(\mu_y) & \cos(\mu_y) 
\end{pmatrix}
\begin{pmatrix}
x \\ p_{x} \\ y \\ p_{y}
\end{pmatrix}_n
\;\;,
\label{eq:rotation}
\end{equation}
and a single octupole kick with integrated strength ${\overline{k_3}}$: 
\begin{equation}
\begin{pmatrix}
x \\ p_{x} \\ y \\ p_{y}
\end{pmatrix}_{n+2}
=
\begin{pmatrix}
x \\ p_{x} - \overline{k_3} (x^3-3xy^2)\\ y \\ p_{y}- \overline{k_3} (-3x^2y+y^3)
\end{pmatrix}_{n+1}
\;\;.
\label{eq:octupolemap}
\end{equation} 
We can first restrict the motion in the $(x,p_x)$ phase plane, i.e.
we set $y_0=p_{y0}=0$. The map is iterated for a number of “turns” (here 1000)
and a few initial conditions.

\begin{figure}[ht]
\begin{center}
\begin{tikzpicture}
 \node (img1)  {\includegraphics[trim=15 255 5 5,clip,width=5cm]{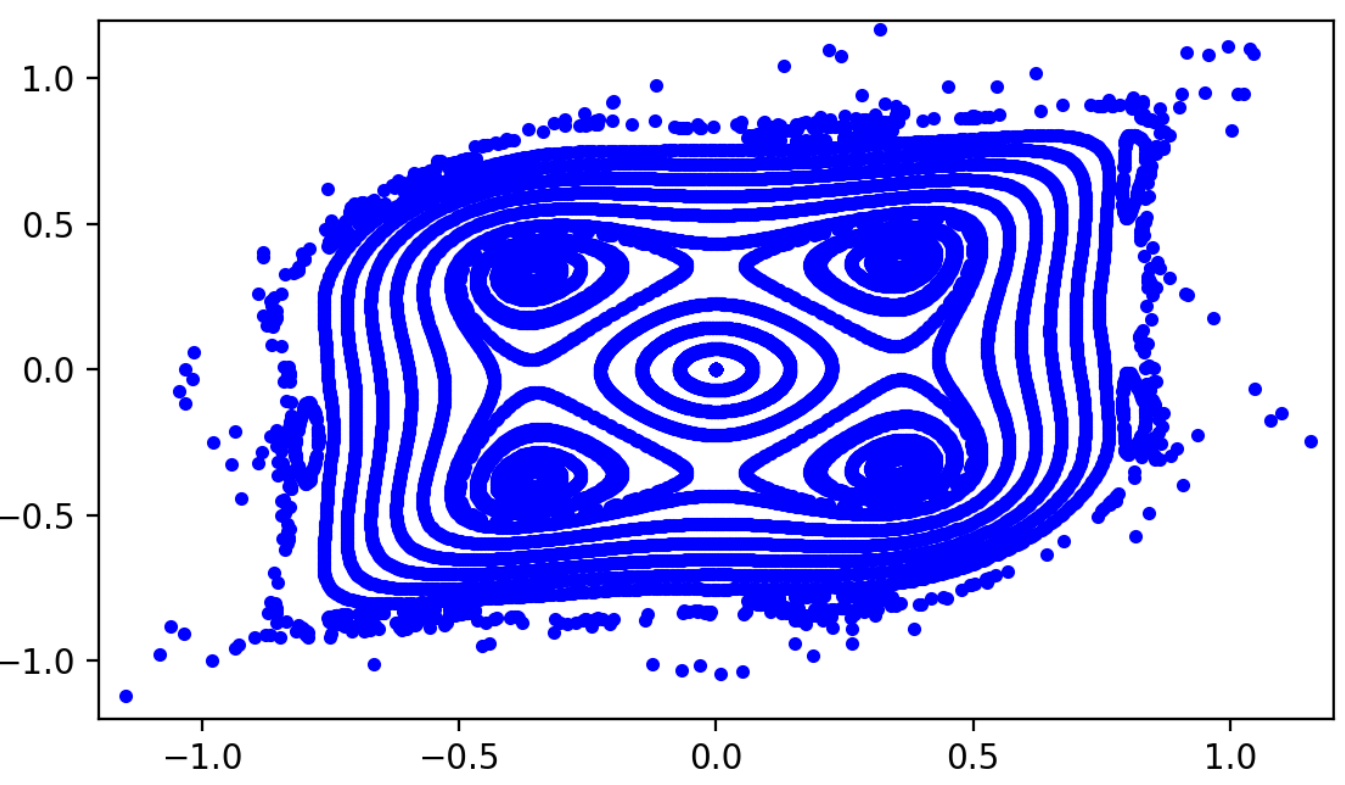}};
 \node[below=of img1, node distance=0cm, xshift=0.1cm,yshift=1cm,font=\color{black}] {$x$};
 \node[left=of img1, node distance=0cm, rotate=0, anchor=center,xshift=0.8cm,yshift=0.1cm,font=\color{black}] {$p_x$};\node[right=of img1,yshift=0cm] (img2)  {\includegraphics[trim=20 15 5 245,clip,width=5cm]{fig5.pdf}};
  \node[below=of img2, node distance=0cm, xshift=0.1cm,yshift=1.1cm,font=\color{black}] {$\phi_x$};
 \node[left=of img2, node distance=0cm, rotate=0, anchor=center,xshift=0.8cm,yshift=0.1cm,font=\color{black}] {$J_x$}; 
 \end{tikzpicture}
\caption{Poincar\'e map $(x,p_x)$  (left), and action-angle space $(J_x,\phi_x)$ for a simple model including an octupole kick and a rotation in phase space, for $y_0=p_{y0}=0$ and a phase advance of $(\mu_x = 0.22)$.}
\label{fig:octupolemap}
\end{center}
\end{figure}
The phase space trajectories (or Poincar\'e map) $(x,p_x)$ are plotted for $(\mu_x = 0.22)$ in Fig.~\ref{fig:octupolemap} (left), whereas the action-angle space is depicted in Fig.~\ref{fig:octupolemap} (right). Close to the closed orbit at $x=p_x=0$, invariant curves appear, where the action variable seems to be almost constant, as it is very close to an integral of motion. As the trajectories are further away from the closed orbit, for larger amplitudes, the circular phase space trajectories are getting more distorted. Resonances are appearing as stable fixed points surrounded by curves forming "islands". The corresponding unstable fixed points are connected by separatrices (or unstable manifolds). At even larger amplitude the trajectories are no longer smooth as the motion becomes strongly chaotic. In the case of electromagnetic fields generated by the multi-pole expansion employed for electro-magnets in accelerators (polynomials in the transverse coordinates), the phase space is not bounded. In this respect, chaotic trajectories may eventually escape to infinity, introducing the concept of Dynamic Aperture (DA), which will be treated in a later section. For some fields like head-on beam-beam or space-charge which depend on the beam distribution line density (i.e. they vanish to infinity) chaotic particles cannot escape to infinity but they form a halo, which can be indeed lost if geometrical aperture restrictions are present.

\section{Motion close to a resonance}

Since the time of Poincar\'e~\cite{Poincare}, it is known that in the vicinity of a resonance $k_1\omega_1+k_2\omega_2=0$, standard canonical perturbation methods 
fail. This is due to the appearance of small denominators in the amplitudes  of the Fourier series expansion representing the 
non-linear part of the Hamiltonian, which prevents its convergence. Secular perturbation theory was indeed introduced to overcome this problem,
by constructing a special ``resonant'' Hamiltonian which represent motion the vicinity of the resonance (see~\cite{Tabor,LicLie} 
for the general mathematical formalism, whereas its application to accelerator rings are described in~\cite{Hagedorn,Schoch,Guignard76,Guignard78,
Ruth:1985ya, Wolski} for Hamiltonians, whereas the elegant approach accelerator maps are treated references~\cite{Forest1,Forest2,chaouspas}).
In brief, a canonical transformation is applied such that the new variables are in a frame remaining on top of the resonance. If one frequency is slow, 
one can average the motion and remain only with a 1 degree of freedom Hamiltonian which looks like the one of the pendulum. 
Thereby, one can find the location and nature of the fixed points measure the width of the resonance.

For simplicity we consider only one plane, for which any polynomial perturbation of the form $x^{k_1}$, the ``resonant'' Hamiltonian is written as
\begin{equation}
\hat{H}_2 = \delta J_2 + \alpha(J_2) + J^{k_1/2} A_{k_1;p} \cos(k_1\psi_2)\;\;,
\label{eq:resonham}
\end{equation}
with the very small distance to the resonance $\delta<<1$ defined as $\nu=\frac{p}{k_1}+\delta$.
The non-linear shift of the tune is described by the term $\alpha(J_2)$. This Hamiltonian is actually very close
to the one of a pendulum, which is completely integrable, and several properties such as the frequency and
resonance width can be computed analytically.
The conditions for the fixed points can be found by setting the equations of motion equal to zero and obtain
\begin{equation}
\sin(k_1\psi_2) = 0\;, \;\; \delta+\frac{\partial\alpha(J_2)}{\partial J_2} +\frac{k_1}{2}J_2^{k_1/2-1} A_{k_1;p}\cos(k_1\psi_2) =0\;\;.
\end{equation}
There are $k_1$ fixed points for which	$\cos(k\psi_{20}) = -1$	and the fixed points are stable (elliptic). They are surrounded by ellipses.
There are also $k_1$ fixed points for which $\cos(k\psi_{20}) = 1$ and the fixed points are unstable (hyperbolic). The trajectories are hyperbolas
passing in their vicinity are hyperbolas.
\begin{figure}[ht]
\begin{center}
{\includegraphics[width=5cm]{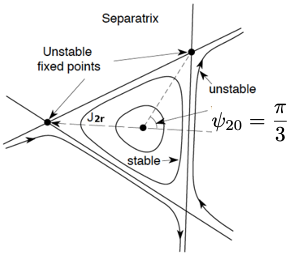}}
\caption{Sketch of the phase space in the vicinity of a third order resonance.}
\label{fig:thirdorder}
\end{center}
\end{figure}

We can apply this to the ``resonant'' Hamiltonian for a sextupole close to a third order resonance for which 
\begin{equation}
\hat{H}_2 = \delta J_2 + J_2^{3/2} A_{3p} \cos(3\psi_2)
;\;.
\end{equation}
Note the absence of the non-linear tune-shift term, as this is a 1st order perturbation theory for a sextupole approximation.
By setting the Hamilton’s equations equal to zero, three fixed points can be found at $\psi_{20} = \frac{\pi}{3}\;,\;\;\frac{3\pi}{3}\;,\;\;\frac{5\pi}{3}\;,\;\; J_{20} = \left(\frac{2\delta}{3A_{3p}}\right)^2$
For	$\frac{\delta}{A_{3p}}>0$ all three points are unstable. Close to the elliptic one at $\psi_{20} = 0$, the motion in phase space 
is described by circles that they get more and more distorted to end up in the “triangular” separatrix uniting the unstable fixed points,
as shown in Fig.~\ref{fig:thirdorder}. The tune separation from the  resonance is $\delta =  \frac{3A_{3p}}{2} J_{20}^{1/2}$.

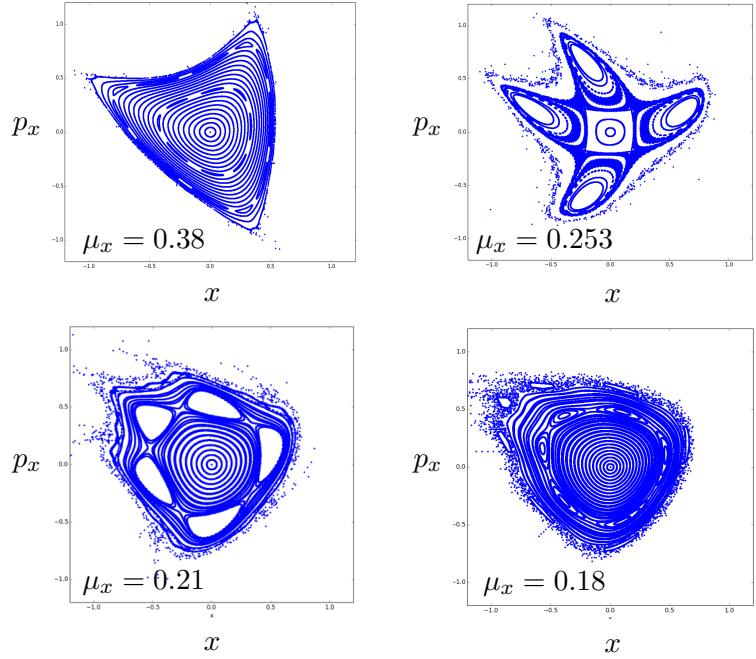
\begin{figure}[ht]
\begin{center}
\begin{tikzpicture}
 \node (img1)  {\includegraphics[trim=10 10 5 5,clip,width=4cm,height=3.55cm]{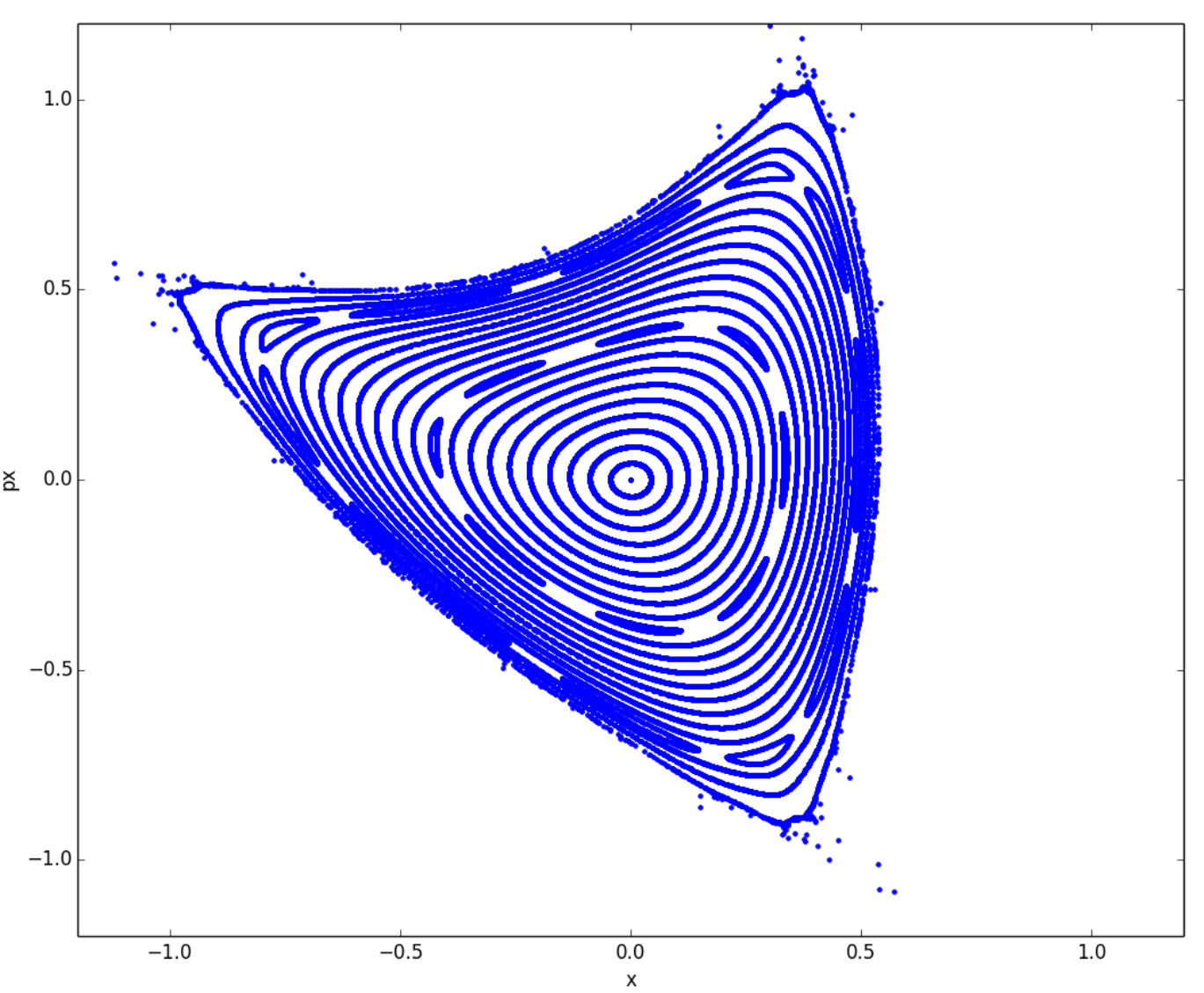}};
 \node[below=of img1, node distance=0cm, xshift=0.1cm,yshift=1cm,font=\color{black}] {$x$};
 \node[left=of img1, node distance=0cm, rotate=0, anchor=center,xshift=0.8cm,yshift=0.1cm,font=\color{black}] {$p_x$};
 \node[below=of img1, node distance=0cm, rotate=0, anchor=center,xshift=-0.8cm,yshift=1.5cm,font=\color{black}] {$\mu_x=0.38$};
 \node[right=of img1,yshift=0cm] (img2) {\includegraphics[width=4cm]{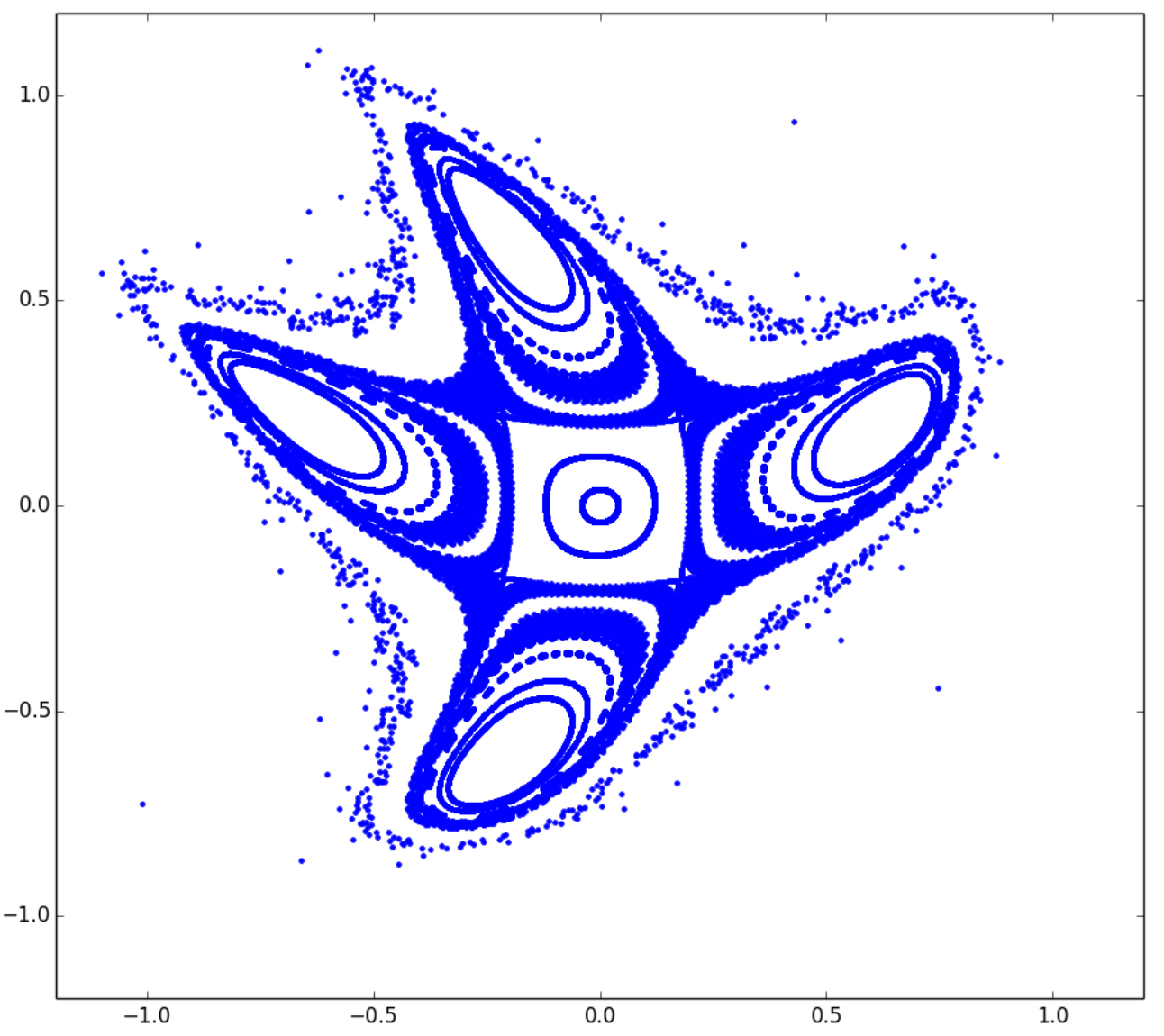}};
  \node[below=of img2, node distance=0cm, xshift=0.1cm,yshift=1cm,font=\color{black}] {$x$};
 \node[left=of img2, node distance=0cm, rotate=0, anchor=center,xshift=0.8cm,yshift=0.1cm,font=\color{black}] {$p_x$};
 \node[below=of img2, node distance=0cm, rotate=0, anchor=center,xshift=-0.8cm,yshift=1.5cm,font=\color{black}] {$\mu_x=0.253$};
 \end{tikzpicture}\\
\begin{tikzpicture}
 \node (img1)  {\includegraphics[width=4cm]{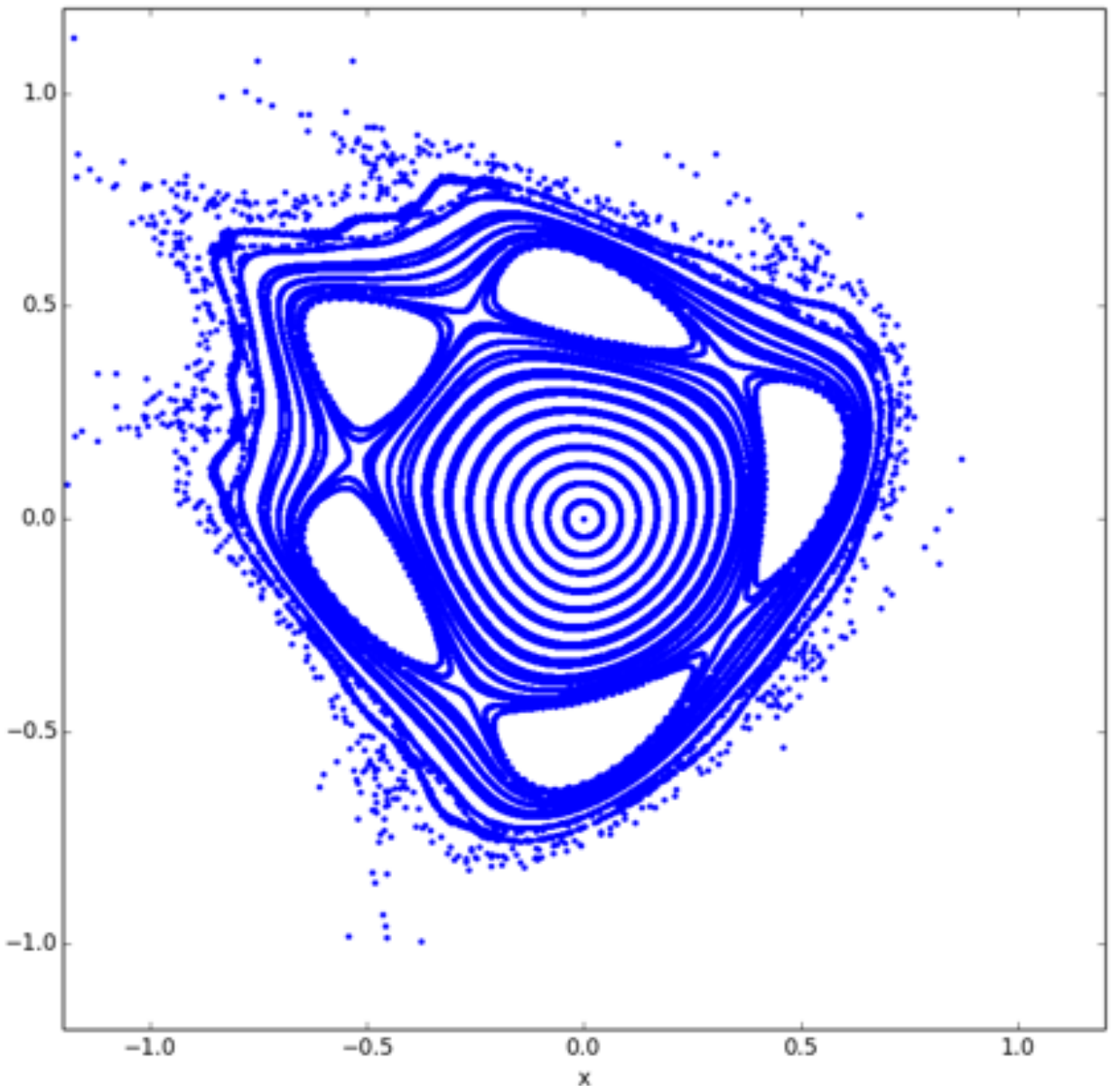}};
 \node[below=of img1, node distance=0cm, xshift=0.1cm,yshift=1cm,font=\color{black}] {$x$};
 \node[left=of img1, node distance=0cm, rotate=0, anchor=center,xshift=0.8cm,yshift=0.1cm,font=\color{black}] {$p_x$};
 \node[below=of img1, node distance=0cm, rotate=0, anchor=center,xshift=-0.8cm,yshift=1.6cm,font=\color{black}] {$\mu_x=0.21$};
 \node[right=of img1,yshift=0cm] (img2) {\includegraphics[width=4cm]{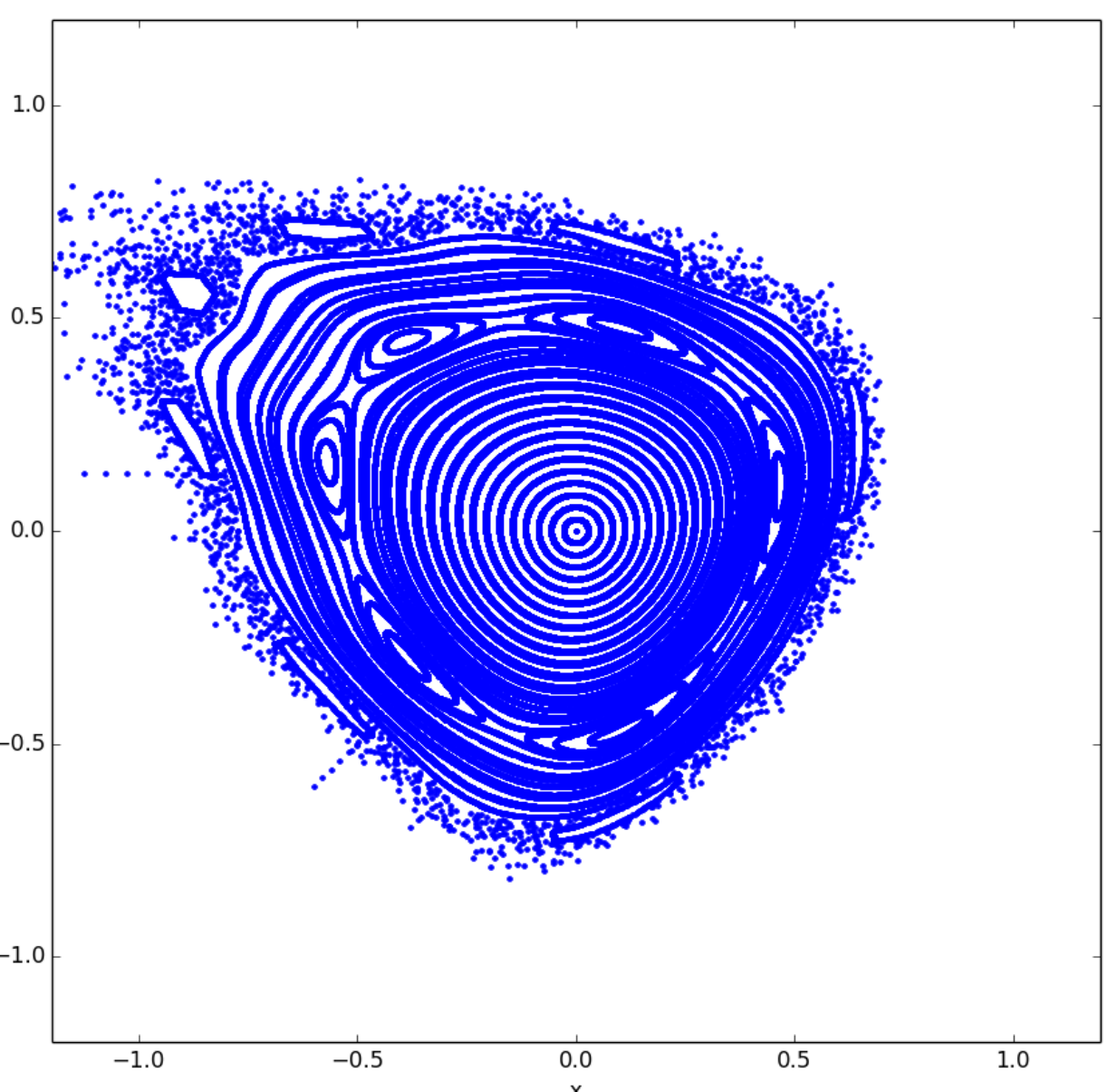}};
  \node[below=of img2, node distance=0cm, xshift=0.1cm,yshift=1cm,font=\color{black}] {$x$};
 \node[left=of img2, node distance=0cm, rotate=0, anchor=center,xshift=0.8cm,yshift=0.1cm,font=\color{black}] {$p_x$};
 \node[below=of img2, node distance=0cm, rotate=0, anchor=center,xshift=-0.8cm,yshift=1.6cm,font=\color{black}] {$\mu_x=0.18$};
 \end{tikzpicture}
\caption{Poincar\'e map $(x,p_x)$ for a simple model including a sextupole kick and a rotation in phase space, for $y_0=p_{y0}=0$ and a phase advance of $(\mu_x = 0.38)$ (top, left), 
$(\mu_x = 0.253)$ (top, right), $(\mu_x = 0.21)$ (bottom, left) and $(\mu_x = 0.18)$ (bottom, right).}
\label{fig:sextrot}
\end{center}
\end{figure}

As a numerical example, we consider a simple map with a single sextupole kick 
\begin{equation}
\begin{pmatrix}
x \\ p_{x} \\ y \\ p_{y}
\end{pmatrix}_{n+2}
=
\begin{pmatrix}
x \\ p_{x} - \overline{k_2} (x^2-y^2)\\ y \\ p_{y}- \overline{k_2} (-2 x y)
\end{pmatrix}_{n+1}
\;\;,
\label{eq:sextmap}
\end{equation} 
with integrated strength  $\overline{k_2}$ and a rotation with phase advances  $(\mu_x,\mu_y)$, see Eq.~\eqref{eq:rotation}.
Restricting the motion again in the $(x,p_x)$-plane i.e. $y_0=p_{y0}=0$ and iterating for a number of “turns” (here 1000), we can produce the Poincar\'e
maps of Fig.~\ref{fig:sextrot} for different phase advances. Indeed, for a phase advance close to the 3rd integer and depending on the tune-shift
with the particles amplitude, the appearance of the 3rd order resonance is evident (top left of Fig~\ref{fig:sextrot}), identified by the triangular
shape of the trajectories at its vicinity. On the other hand,
if the phase advance is away from that resonance, but close to the 4th, 5th, 6th, 7th etc. integer, the corresponding resonances
appear, and several other higher orders (see top, right, and bottom plots of Fig.~\ref{fig:sextrot}). The belief that sextupole resonances 
excite only 3rd order resonances is based on a 1st order perturbation theory approach, whereas at higher orders (i.e. iterating the map more then once,
so that a higher order dependence on the sextupole strength appears), 
all resonances could be potentially excited.

We can apply the same  formalism to the 4th order resonance, through the resonant Hamiltonian of
 Eq.~\eqref{eq:resonham}, which in the case that $k_1=4$ becomes
\begin{equation}
\hat{H}_2 = \delta J_2 + \alpha J_2^2 + J_2^{2} A_{4;p} \cos(4\psi_2)\;\;.
\label{eq:resonhamoct}
\end{equation}
The second term  $\alpha J_2^2$ represents the leading shift of the frequency with the action (e.g.
from an octupole excitation at leading order). The fixed points are found by taking the derivative 
over the two variables and setting them to zero, i.e.
\begin{equation}
\sin(4\psi_2) = 0\;, \;\; \delta+2\alpha J_2 +2J_2 A_{4;p}\cos(4\psi_2) =0\;\;.
\end{equation}
The first trivial solution provides the closed orbit at $(J_{20},\psi_{20}) = (0,0)$, with $\delta=0$. The solutions for the
angles provide eight fixed points for $J_{20}=\frac{\delta }{2 (A_{4;p}- \alpha) }$ or $J_{20} = -\frac{\delta }{2 (A_{4;p}+\alpha)}$. Assuming that $A_{4;p} > \alpha>0$, for half of them at $\psi_{20} = (2k+1)\frac{\pi}{4}$, with $k$ an integer, there is a minimum in the potential, as $\cos(4\psi_{20}) = - 1$,	and they are elliptic. The other half at $\psi_{20} = k\frac{\pi}{2}$ are hyperbolic as $\cos(4\psi_{20}) = + 1$. 

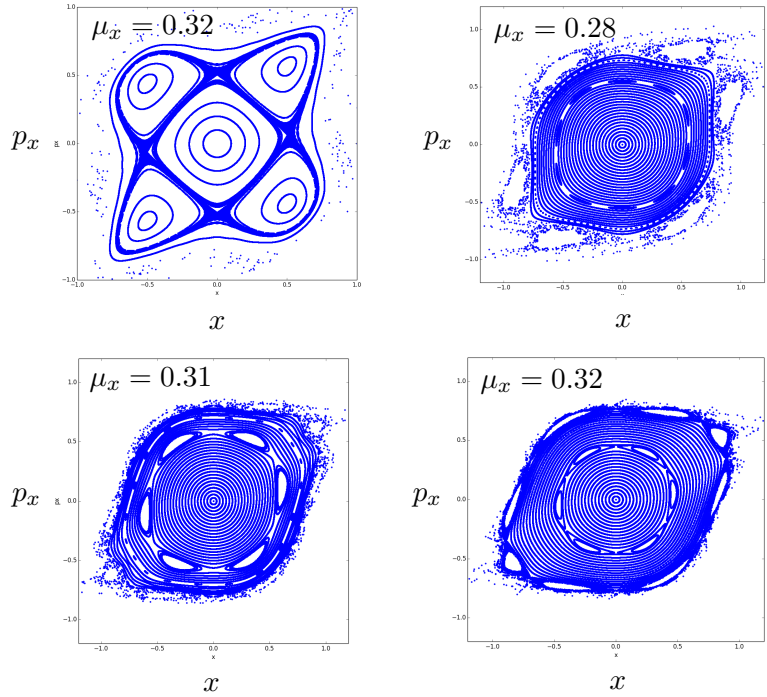
\begin{figure}[ht]
\begin{center}
\begin{tikzpicture}
 \node (img1)  {\includegraphics[width=4.15cm]{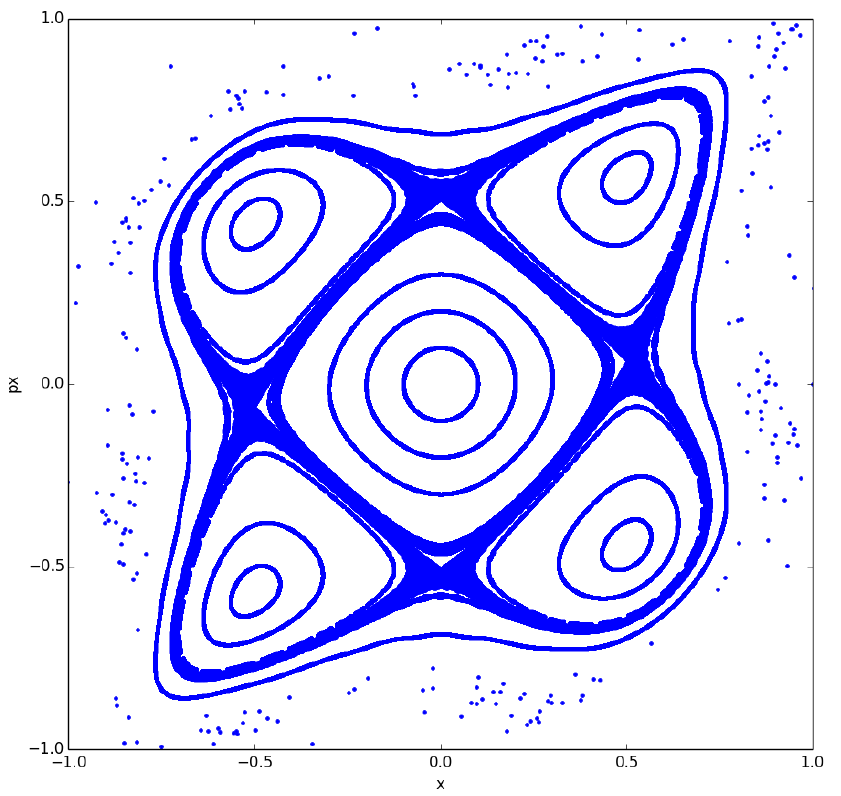}};
 \node[below=of img1, node distance=0cm, xshift=0.1cm,yshift=1cm,font=\color{black}] {$x$};
 \node[left=of img1, node distance=0cm, rotate=0, anchor=center,xshift=0.8cm,yshift=0.1cm,font=\color{black}] {$p_x$};
 \node[above=of img1, node distance=0cm, rotate=0, anchor=center,xshift=-0.75cm,yshift=-1.5cm,font=\color{black}] {$\mu_x=0.32$};
 \node[right=of img1,yshift=0cm] (img2) {\includegraphics[width=4cm]{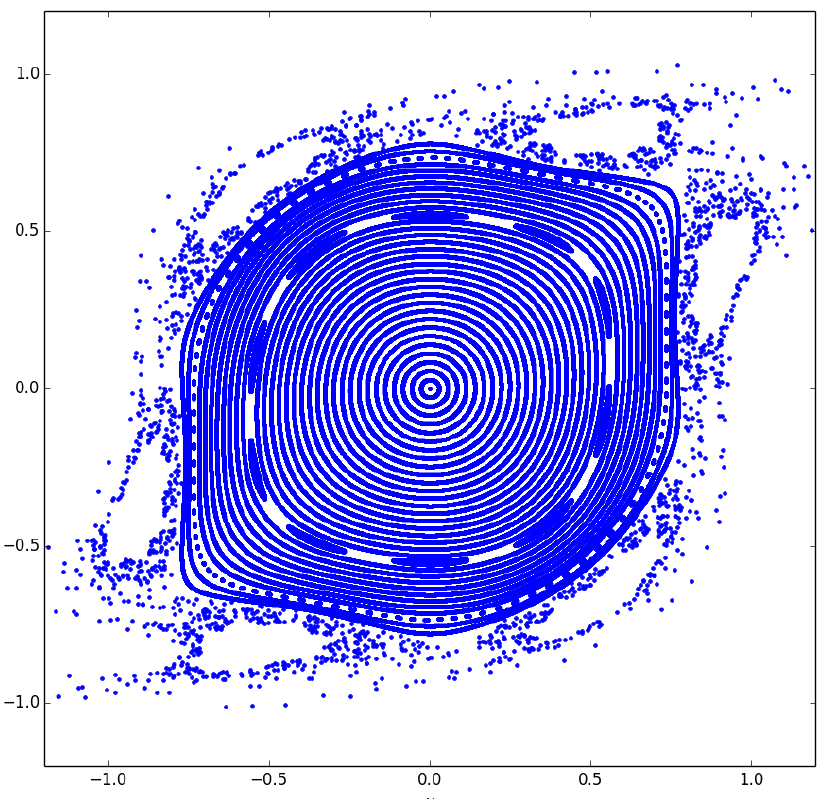}};
  \node[below=of img2, node distance=0cm, xshift=0.1cm,yshift=1cm,font=\color{black}] {$x$};
 \node[left=of img2, node distance=0cm, rotate=0, anchor=center,xshift=0.8cm,yshift=0.1cm,font=\color{black}] {$p_x$};
 \node[above=of img2, node distance=0cm, rotate=0, anchor=center,xshift=-0.8cm,yshift=-1.5cm,font=\color{black}] {$\mu_x=0.28$};
 \end{tikzpicture}\\
\begin{tikzpicture}
 \node (img1)  {\includegraphics[width=4cm]{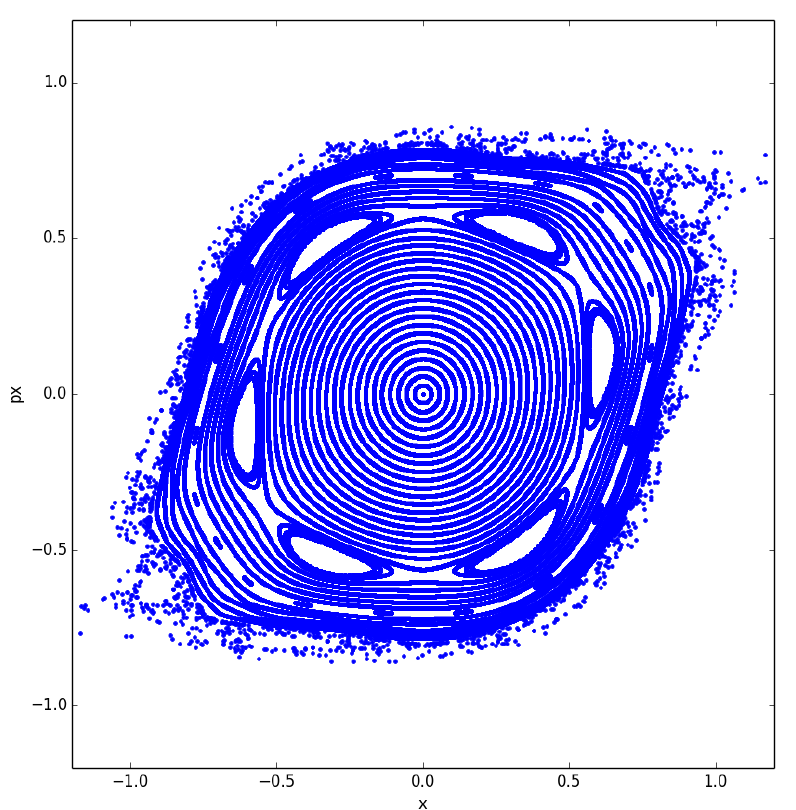}};
 \node[below=of img1, node distance=0cm, xshift=0.1cm,yshift=1cm,font=\color{black}] {$x$};
 \node[left=of img1, node distance=0cm, rotate=0, anchor=center,xshift=0.8cm,yshift=0.1cm,font=\color{black}] {$p_x$};
 \node[above=of img1, node distance=0cm, rotate=0, anchor=center,xshift=-0.7cm,yshift=-1.5cm,font=\color{black}] {$\mu_x=0.31$};
 \node[right=of img1,yshift=0cm] (img2) {\includegraphics[width=4.15cm]{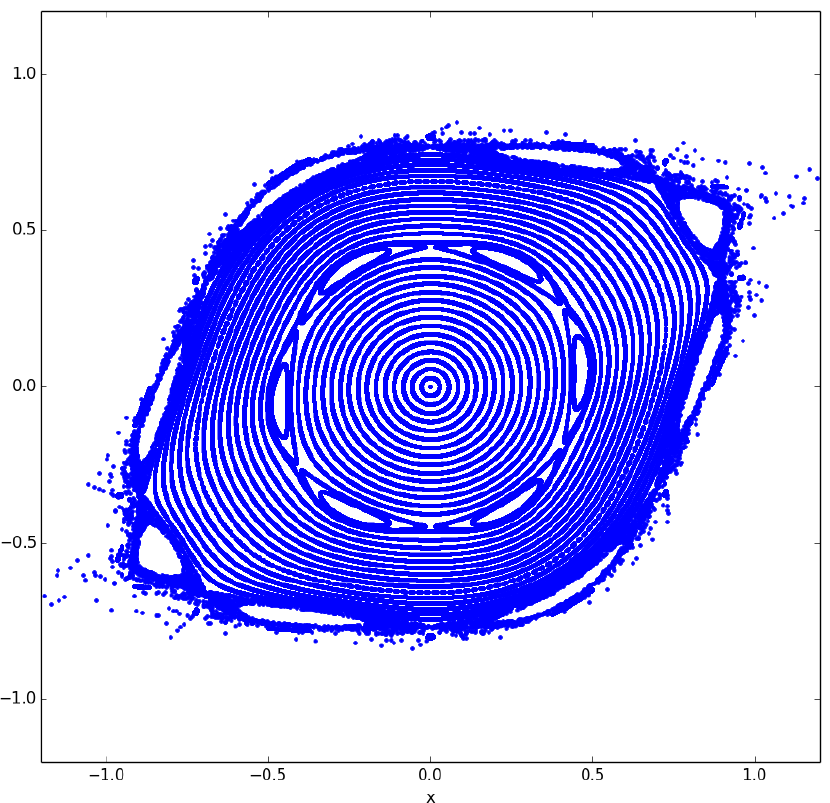}};
  \node[below=of img2, node distance=0cm, xshift=0.1cm,yshift=1cm,font=\color{black}] {$x$};
 \node[left=of img2, node distance=0cm, rotate=0, anchor=center,xshift=0.8cm,yshift=0.1cm,font=\color{black}] {$p_x$};
 \node[above=of img2, node distance=0cm, rotate=0, anchor=center,xshift=-0.9cm,yshift=-1.5cm,font=\color{black}] {$\mu_x=0.32$};
 \end{tikzpicture}
\caption{Poincar\'e map $(x,p_x)$ for a simple model including an octupole kick and a rotation in phase space, for $y_0=p_{y0}=0$ and a phase advance of $\mu_x = 0.32$ (top, left), 
$\mu_x = 0.28$ (top, right), $\mu_x = 0.31$ (bottom, left) and $\mu_x = 0.32$ (bottom, right).}
\label{fig:octrot}
\end{center}
\end{figure}

The topology of a 4th order resonance is shown in~\ref{fig:octrot} (top left). Near the center, and around the closed orbit, 
regular motion is observed. For higher amplitudes, the trajectories are getting more and more deformed towards a rectangular shape, 
as they approach the 4th order. The separatrix passes through four unstable fixed points. Four stable fixed points exist and they are 
surrounded by stable motion, forming the so-called "islands of stability". Motion seems well contained, but this
depends on the number and strength of octupoles. Another interesting observation is that the if $\alpha = 0$, 
i.e. the first order tune-shift with amplitude is eliminated, the solution for the action is $J_2 = 0$ and the eigenvalues of the Jacobian matrix~\eqref{eq:jacob}
are real. This means that there is no minimum in the potential and the central fixed point becomes hyperbolic.

 As for the sextupole, the octupole can excite any resonance, as can be observed in Fig.~\ref{fig:octrot}. Indeed, multi-pole magnets can excite any resonance order, depending on the tunes, the strength of the magnet and particle amplitudes.

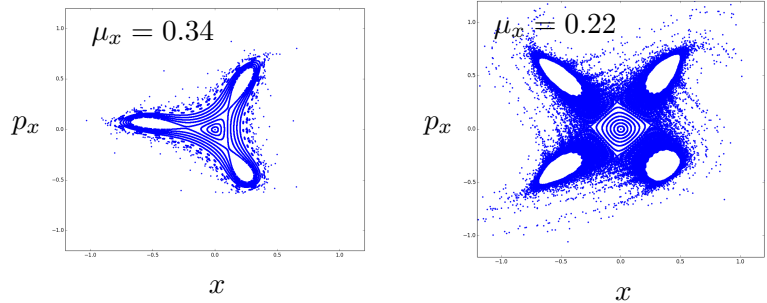
\begin{figure}[ht]
\begin{center}
\begin{tikzpicture}
 \node (img1)  {\includegraphics[width=4.15cm]{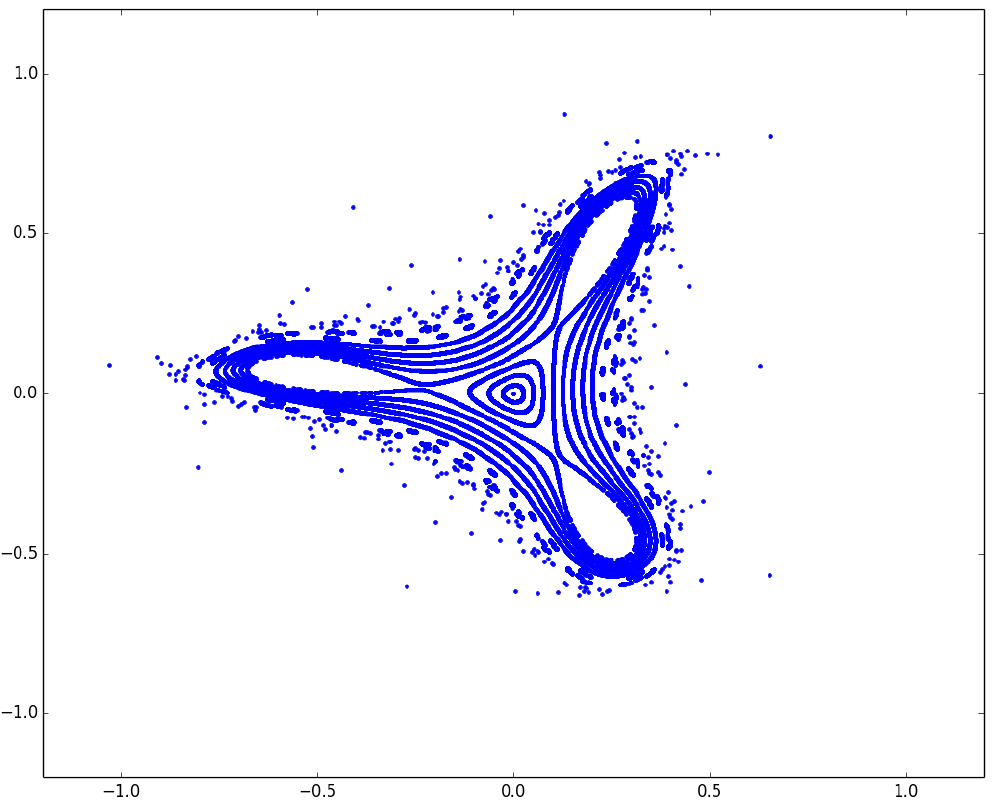}};
 \node[below=of img1, node distance=0cm, xshift=0.1cm,yshift=1cm,font=\color{black}] {$x$};
 \node[left=of img1, node distance=0cm, rotate=0, anchor=center,xshift=0.8cm,yshift=0.1cm,font=\color{black}] {$p_x$};
 \node[above=of img1, node distance=0cm, rotate=0, anchor=center,xshift=-0.75cm,yshift=-1.5cm,font=\color{black}] {$\mu_x=0.34$};
 \node[right=of img1,yshift=0cm] (img2) {\includegraphics[width=4cm]{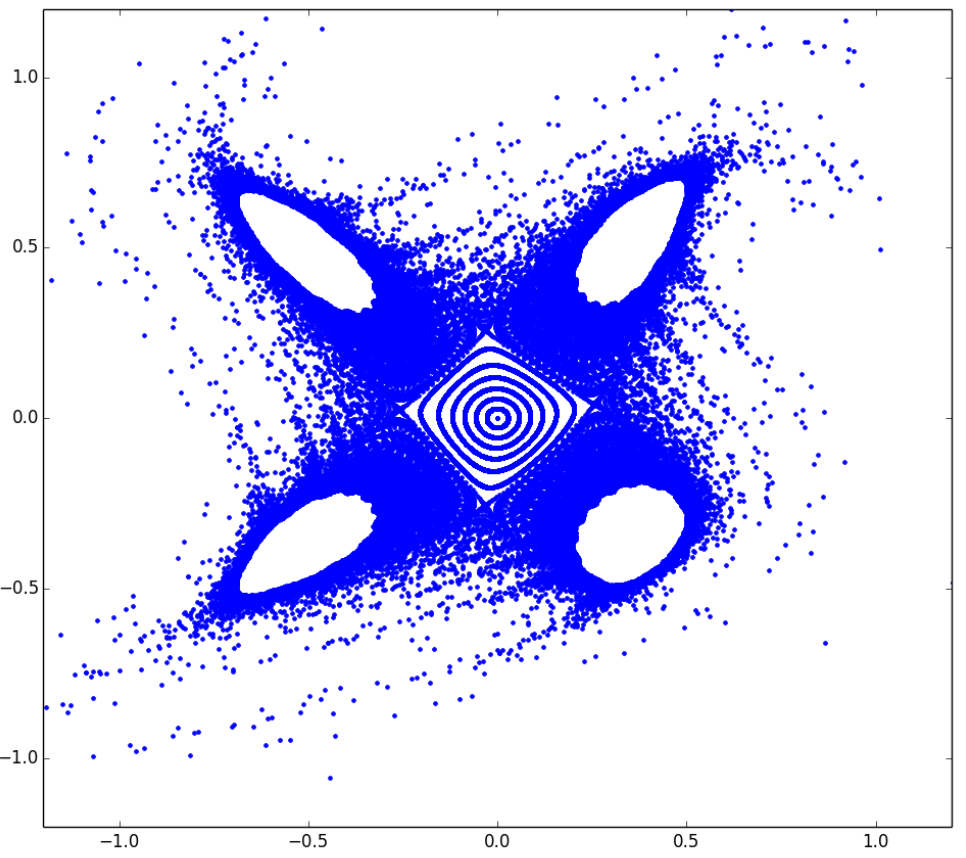}};
  \node[below=of img2, node distance=0cm, xshift=0.1cm,yshift=1cm,font=\color{black}] {$x$};
 \node[left=of img2, node distance=0cm, rotate=0, anchor=center,xshift=0.8cm,yshift=0.1cm,font=\color{black}] {$p_x$};
 \node[above=of img2, node distance=0cm, rotate=0, anchor=center,xshift=-0.8cm,yshift=-1.5cm,font=\color{black}] {$\mu_x=0.22$};
 \end{tikzpicture}
 \caption{Poincar\'e map $(x,p_x)$ for a simple model including a sextupole and an octupole kick with a rotation in phase space, for $y_0=p_{y0}=0$ and a phase advance of $\mu_x = 0.34$ (left) and 
$\mu_x = 0.22$ (right).}
\label{fig:sextoctrot}
\end{center}
\end{figure}
 In Fig.~\ref{fig:sextoctrot}, the Poincar\'e map $(x,p_x)$ for a combined  kick of a sextupole and an octupole, plus a phase space rotation is given, for different phase advances. Indeed, in that case, the chaotic motion region is getting wider when close to the 4th order resonance (see ~\ref{fig:sextoctrot} right). On the other hand, the appearance of a non-zero amplitude detuning term $\alpha$, allows the appearance of 3rd order resonance stable fixed points (see ~\ref{fig:sextoctrot} left).

\section{Onset of chaos}

\begin{figure}[ht]
\begin{center}
\begin{tabular}{cc}
{\includegraphics[width=5cm]{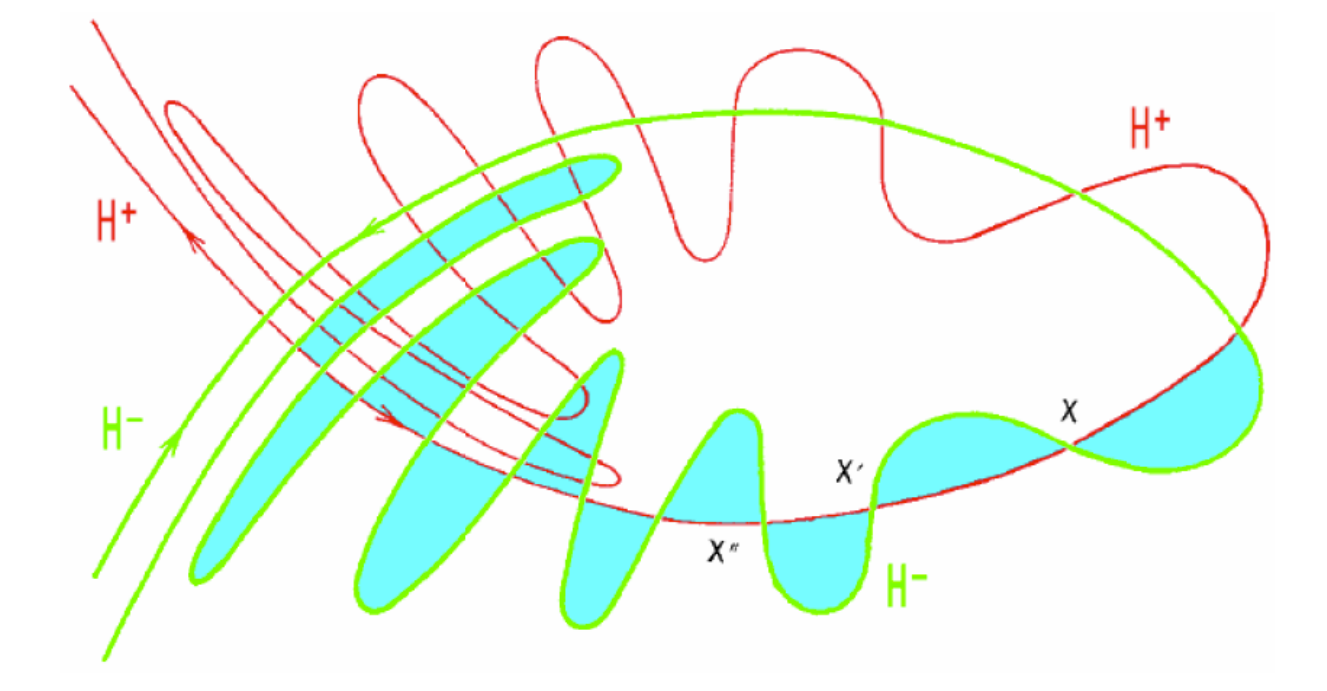}} &
\multirow{-5}{*}{\includegraphics[width=5cm]{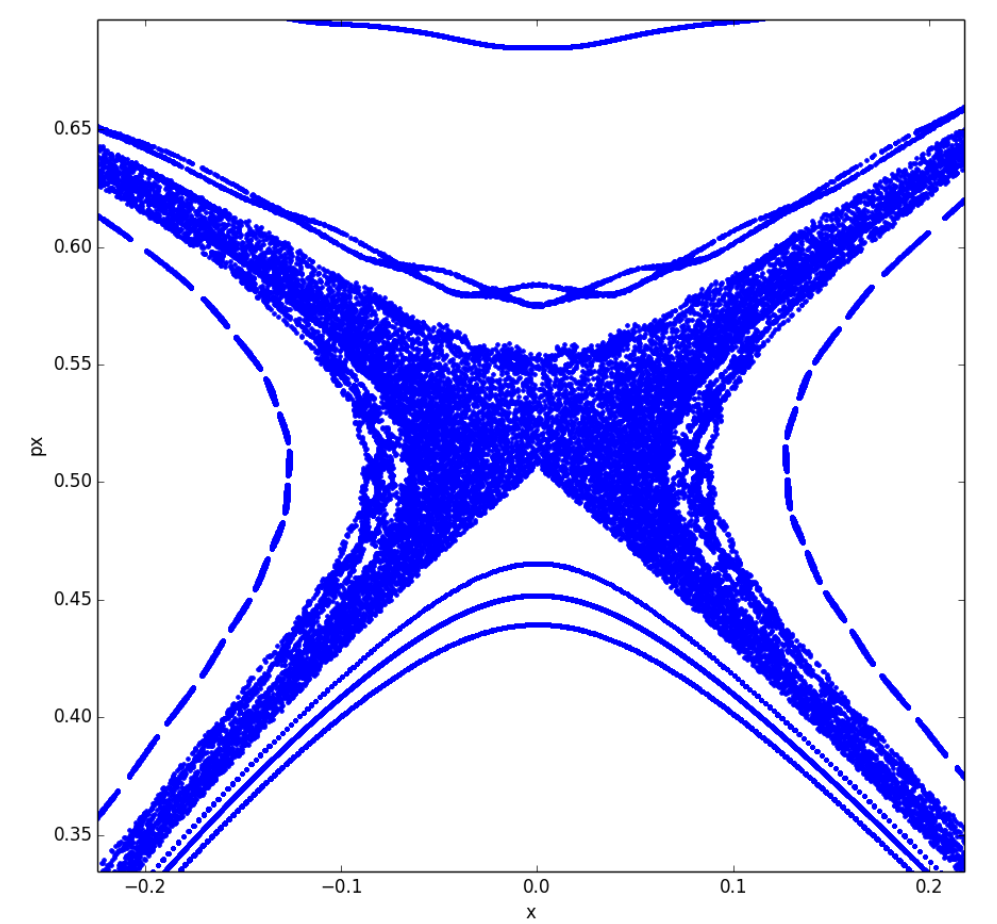}}\\
{\includegraphics[width=5cm]{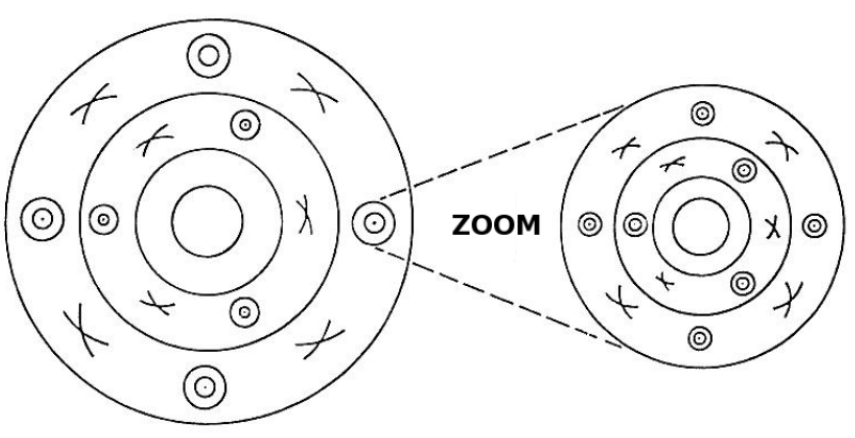}} &
\end{tabular}
\caption{Sketch of the separatrix splitting~\cite{Tabor} (left, top), the creation of self-similar fixed points~\cite{LicLie} (left, bottom) and a zoom in the separatrix of the 4th order resonance of Fig.~\ref{fig:octrot} (right).}
\label{fig:separ}
\end{center}
\end{figure}

Unstable fixed points are indeed the source of chaos when a perturbation is added. When this perturbation becomes larger, the motion around the separatrix has a very complex form. A sketch of this motion close to the separatrix (adapted by~\cite{Tabor}) is presented in Fig.~\ref{fig:separ} (left, top), where the splitting of the separatrix is shown. In fact the trajectories entering into the unstable fixed point ({\it homoclinic orbits} or stable manifolds) and the ones moving away from it ({\it heteroclinic orbits} or unstable manifolds) start forming tangles, which are very complicated structures, with the stable and unstable manifolds intersecting an infinite number of times. There is a mathematical method to prove this splitting, which implies the non-integrability of the system and the on-set of chaos, through the so called {\it Melnikov function}~\cite{Melnikov1,Melnikov2}. This function measures the distance between stable and unstable manifolds in the Poincar\'e map. When this measure is equal to zero, the manifolds cross each other transversally and the trajectories will become chaotic.

The theorem of Poincar\'e-Birkhoff ~\cite{PoincBirk} states that under perturbation of a resonance, only an even number of fixed points survives (half stable and the other half unstable). These fixed points get destroyed when the perturbation is getting increased, and so-on and so-forth. This makes the phase-space to look self-similar, as shown schematically in Fig.~\ref{fig:separ} (left, bottom). This fractal-like structure can be observed when one zooms closely to the chaotic zone of a destroyed separatrix like the one obtained by the octupole map close to the 4th order resonance, as shown in Fig.~\ref{fig:separ} (right).

\subsection{Resonance overlap criterion}
\begin{figure}[ht]
\begin{center}
{\includegraphics[width=5cm]{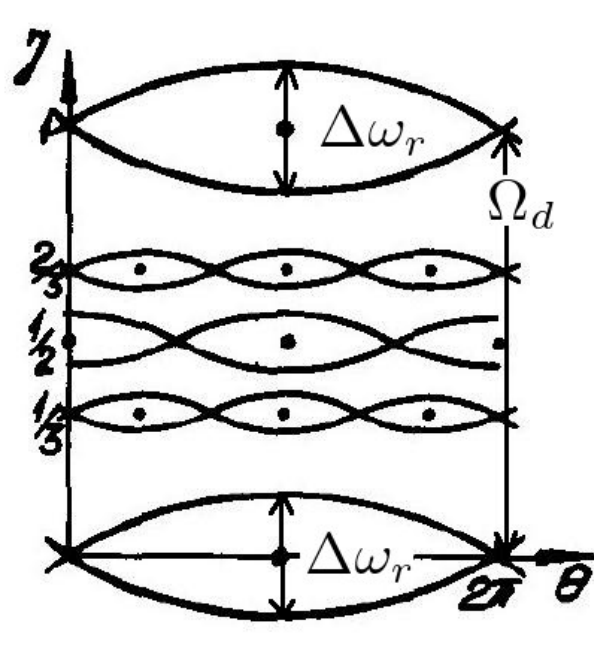}}
\caption{Sketch of the resonance overlap criterion~\cite{Chirikov1}.}
\label{fig:overlap}
\end{center}
\end{figure}

When the perturbation is increased, the resonance island width grows and resonances can overlap allowing diffusion of particles.
Chirikov~\cite{Chirikov1,Chirikov2,Chirikov3} proposed a criterion for the overlap of two neighbouring resonances and the onset of orbit diffusion.
Using the resonant Hamiltonian~\eqref{eq:resonham},
 the distance between two resonances for a two degrees of freedom system   $k_1\omega_1+k_2\omega_2=0$ and $k'_1\omega_1+k'_2\omega_2=0$ can be written as
\begin{equation}
\delta\hat{J}_{\;k,k'} = 
\frac{2\left(\frac{1}{k}-
  \frac{1}{k'}\right)}{\left|\frac{\partial^2 \bar{H}_0({\mathbf{\hat{J}}})}{\partial{\hat{J}}^{\;2}}\biggl|_{\hat{J}=\hat{J}_{0}}\right|}
	\label{eq:overlap}
\end{equation}
with $k=k_1+k_2$ and $k'=k'_1+k'_2$ the resonance orders. The simple overlap criterion is when the sum of the half widths of the two resonances $\Delta\hat{J}_{k\;max}$ and $\Delta\hat{J}_{k'\;max}$ are greater then the distance between the resonances, i.e. $\Delta\hat{J}_{k\;max}+\Delta\hat{J}_{k'\;max} \geq \delta\hat{J}_{k,k'}$. In that case, the separatrices of the two resonances can cross each other ("overlap") and then the particles can find a route to diffuse, as shown Fig.~\ref{fig:overlap}. 
Considering that the chaotic layer has a finite width and that secondary islands appear, Chirikov introduced a heuristic “two thirds” rule  $ \Delta\hat{J}_{k\;max}+\Delta\hat{J}_{k'\;max} \geq \frac{2}{3} \delta\hat{J}_{k,k'} $.
Although the criterion was widely applied in dynamical system models with 2 degrees of freedom (or 1 degree of freedom, plus time), its geometrical nature makes its extension to systems of more than 2 degrees of freedom not practical.

\subsection{Increasing dimensions}

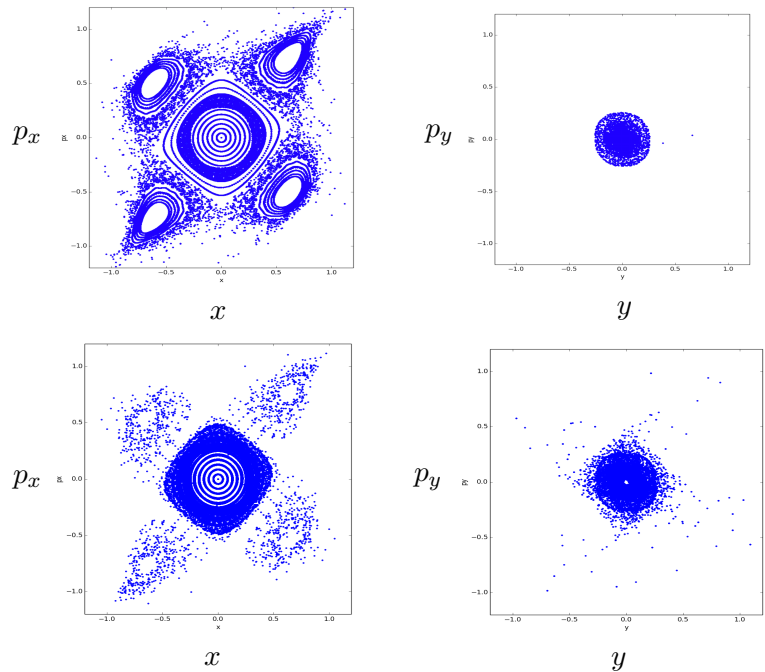
\begin{figure}[ht]
\begin{center}
\begin{tikzpicture}
 \node (img1)  {\includegraphics[width=4.15cm]{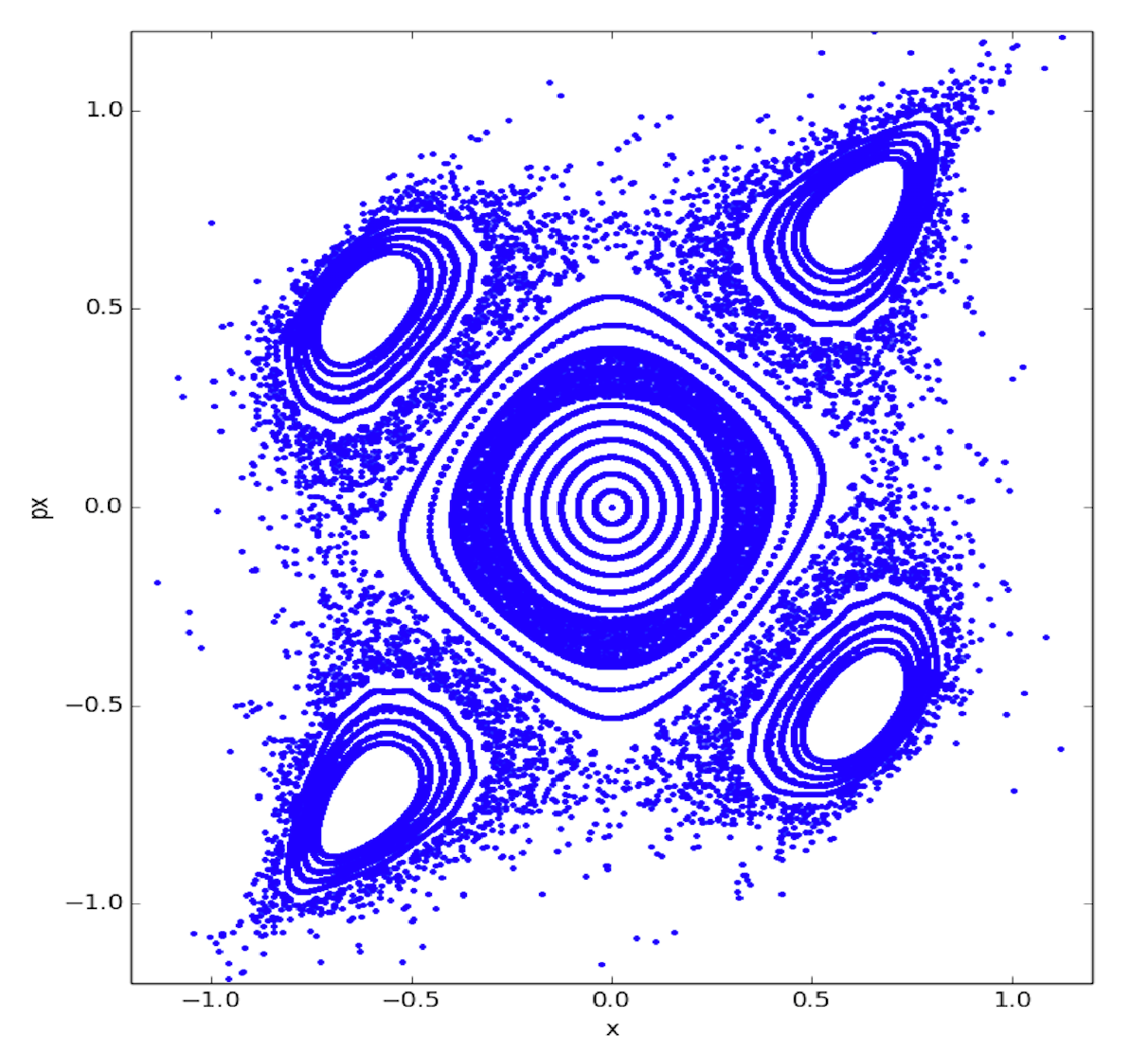}};
 \node[below=of img1, node distance=0cm, xshift=0.1cm,yshift=1cm,font=\color{black}] {$x$};
 \node[left=of img1, node distance=0cm, rotate=0, anchor=center,xshift=0.8cm,yshift=0.1cm,font=\color{black}] {$p_x$};
 \node[right=of img1,yshift=0cm] (img2) {\includegraphics[width=4cm]{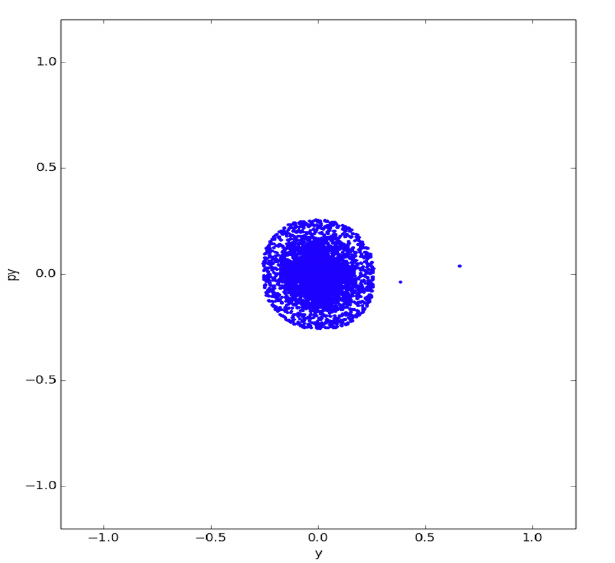}};
  \node[below=of img2, node distance=0cm, xshift=0.1cm,yshift=1cm,font=\color{black}] {$y$};
 \node[left=of img2, node distance=0cm, rotate=0, anchor=center,xshift=0.8cm,yshift=0.1cm,font=\color{black}] {$p_y$};
 \end{tikzpicture}\\
\begin{tikzpicture}
 \node (img1)  {\includegraphics[width=4cm]{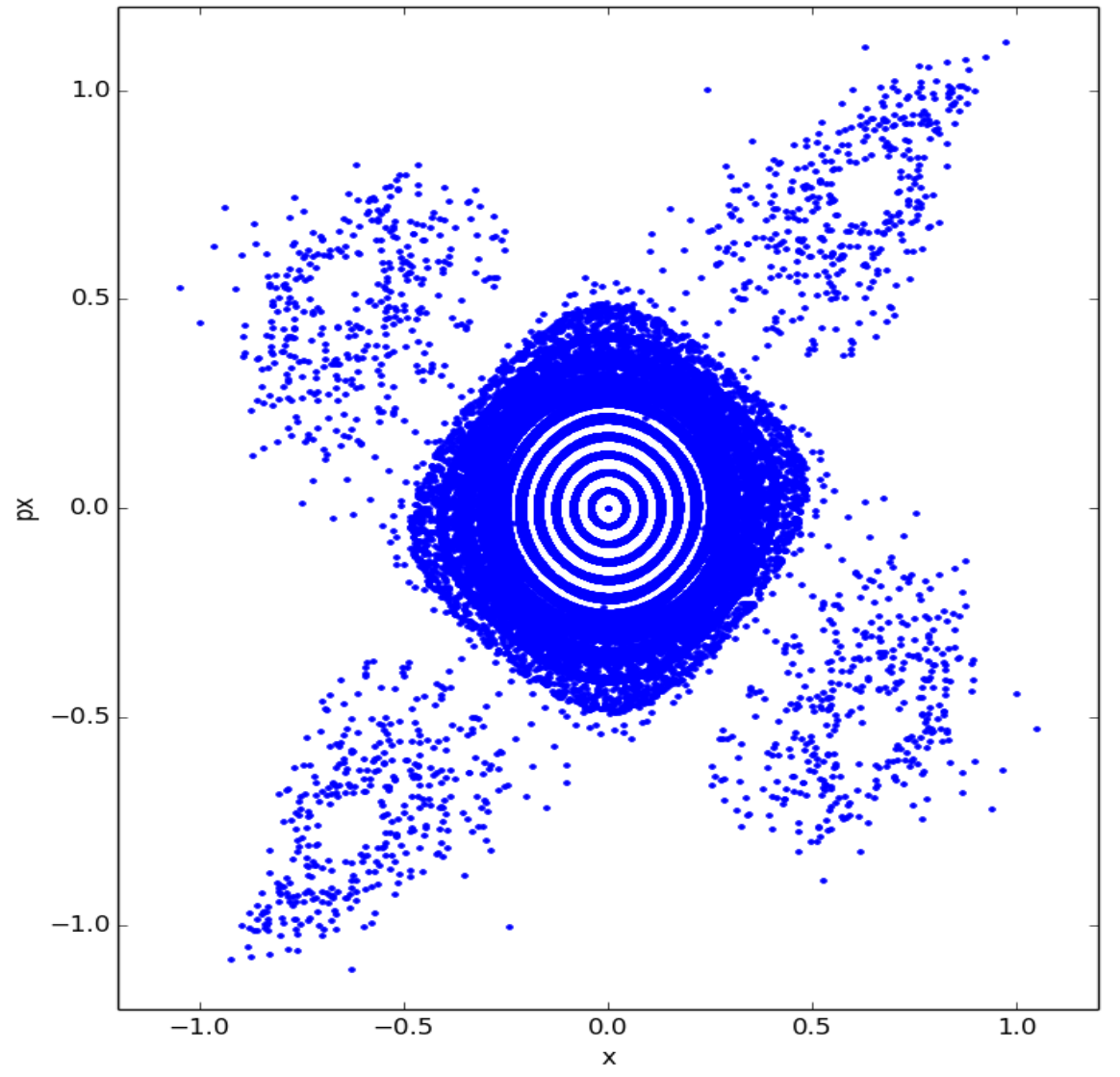}};
 \node[below=of img1, node distance=0cm, xshift=0.1cm,yshift=1cm,font=\color{black}] {$x$};
 \node[left=of img1, node distance=0cm, rotate=0, anchor=center,xshift=0.8cm,yshift=0.1cm,font=\color{black}] {$p_x$};
 \node[right=of img1,yshift=0cm] (img2) {\includegraphics[width=4.15cm]{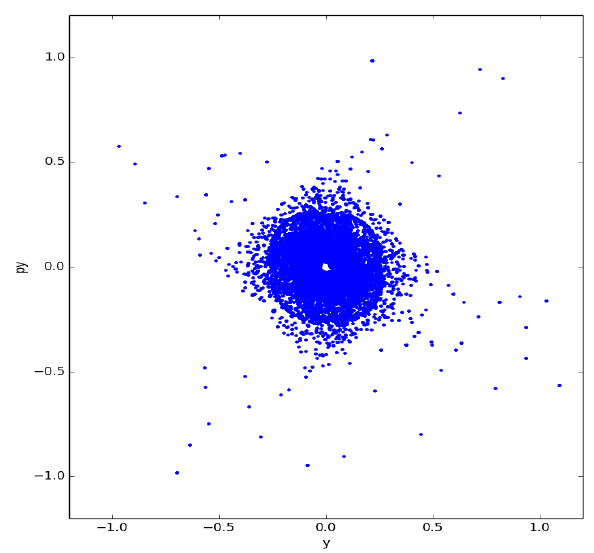}};
  \node[below=of img2, node distance=0cm, xshift=0.1cm,yshift=1cm,font=\color{black}] {$y$};
 \node[left=of img2, node distance=0cm, rotate=0, anchor=center,xshift=0.8cm,yshift=0.1cm,font=\color{black}] {$p_y$};
 \end{tikzpicture}
\caption{Poincar\'e maps $(x,p_x)$ (left) for a simple model including an octupole kick and a rotation in phase space, for two different vertical initial conditions $y_0,p_{y0}$ and phase advances of $(\mu_x,\mu_y) = (0.22,024)$.}
\label{fig:octrot_nozero_y}
\end{center}
\end{figure}

 For $(y_0,p_{y0})\ne(0,0)$, i.e. by adding another degree of freedom, the chaotic motion is enhanced, as shown in Fig.~\ref{fig:octrot_nozero_y},
 where the simple octupole map is not plotted for two non-zero vertical initial conditions. Indeed, the larger is the vertical amplitude, the more
 chaotic seems the phase space. At the same time, the analysis of the phase space trajectories on surfaces of section becomes difficult to interpret, as the trajectories are 4-dimensional objects projected on a 2-dimensional space (a plane).

\section{Chaos detection methods}

A series of methods exist for detecting chaotic orbits and applied in beam dynamics, including the estimation of the 
dynamic aperture (DA) or particle survival~\cite{Bruning,Chaoetal, Willeke95}, the computation of Lyapunov exponents~\cite{Schmidt91, Giov}
the variance of unperturbed action (\`a la Chirikov)~\cite{Chirikov3,Tennyson}, the estimation of diffusion coefficients based
or a statistical approach following the Fokker-Planck equation~\cite{Bruningphd,Irwin,Senelisson}, etc. We will briefly discuss a few
of these methods with an emphasis on Frequency Map Analysis (FMA) (for a review of the method in particle accelerators see~\cite{chaosme}).

\subsection{Dynamic aperture}

The most direct way to evaluate the non-linear dynamics performance of a ring is the computation of Dynamic Aperture (DA).
Actually, the Hamiltonian for magnetic elements, with fields defined through the usual multi-pole expansion is not bounded, 
so particles inside the chaotic regions of the system can diffuse to very large amplitudes (i.e. infinity), or in
a real accelerator will be certainly lost in the vacuum pipe. The focus of 
non-linear beam dynamics is to estimate and possibly explore mitigation measures
in order to increase the area in real space where
particles survive after some time, i.e. certain number of turns depending on the given application, 
the boundary of which defines the DA. This quantity or particle survival rates can be measured in 
a real ring, or computed through
numerical integration of the equations of motions (``particle tracking"), for several initial conditions,
with codes optimised for this task~\cite{madxptc, sixtrack}. The tracking codes need to be symplectic, 
for guaranteeing area preserving properties of the Hamiltonian flow.  One usually starts with 4D (only transverse) tracking but 
certainly needs to simulate 5D (constant energy deviation) and finally 6D (with synchrotron motion included).
As multi-pole errors may not be completely known, one has to track through several machine models built 
by random distribution of these errors. 

\begin{figure}[htp]
    \centering
    \includegraphics[keepaspectratio, width=0.75\columnwidth]{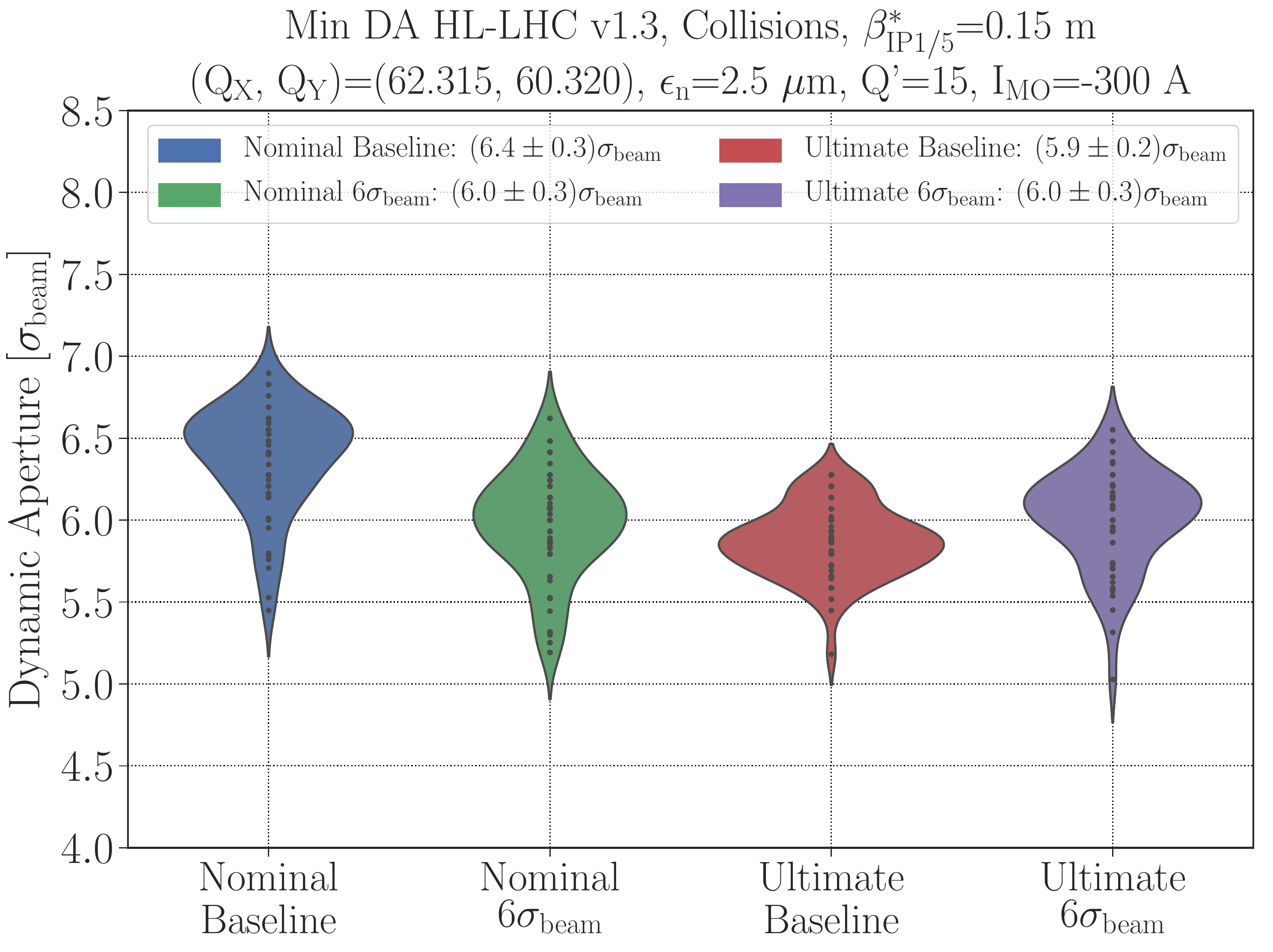}
    \caption{Distribution of the minimum DA for \num{60} different machine configurations at the end of leveling for both baseline and adaptive (i.e. lower crossing angle compared to the baseline in order to meet the DA requirement) scenarios and the two leveled luminosity targets. The worst configuration is found at less than \SI{1}{\mathrm{\sigma_{beam}}} from the statistical average, with the overall spread in the range of \num{0.2}-\SI{0.3}{\mathrm{\sigma_{beam}}}~\cite{NikosYannis}.}
    \label{fig:errorsColl}
\end{figure}

An example of such a study is plotted in Fig.~\ref{fig:errorsColl}~\cite{NikosYannis},
for the validation of the HL-LHC operational scenario at the end of luminosity levelling (i.e. with minimum $beta^\star = 15$~cm), including
the beam-beam effect and the impact of the non-linear errors. The statistical population arises from the minimum DA (over 5 initial conditions) 
of each of \num{60} different magnet error configurations (random seeds). Each violin depicts the distribution of the minimum DA for all seeds 
for a given scenario. The average values, as well as the rms spread of the minimum DA, are quoted on the legend. The spread of the minimum DA 
for any given scenario is at the level of \SI{0.3}{\mathrm{\sigma_{beam}}}. This result is within the simulation noise, therefore the impact 
of the magnetic imperfections are assumed to be in the shadow of the effect of the beam-beam interactions at any DA target.

%


\begin{figure}[htp]
    \centering
    \includegraphics[keepaspectratio, width=0.5\columnwidth]{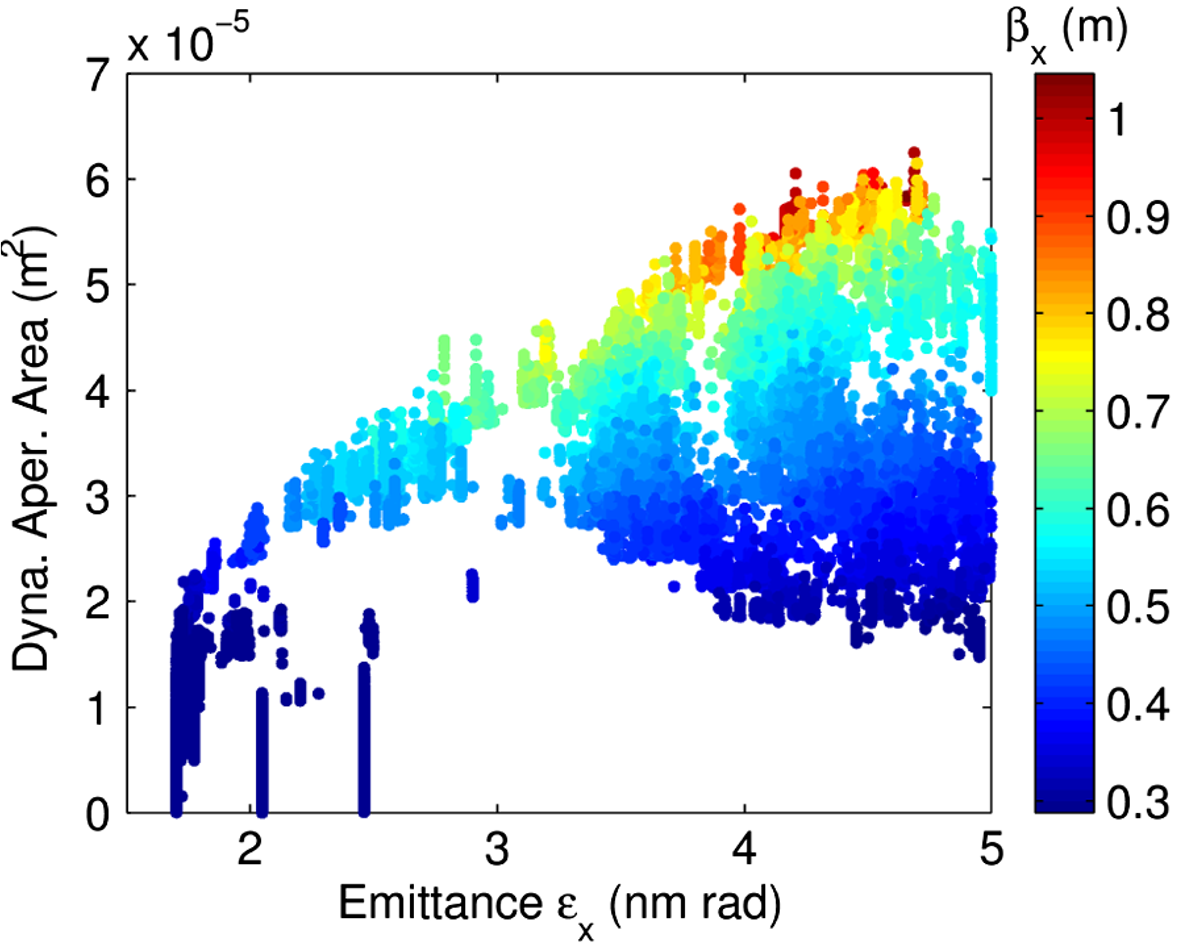}
    \caption{Optimization of the ALS lattice using MOGA with the solution front in the objectives space, composed by the horizontal emittance versus DA area, colour-coded with the horizontal beta function~\cite{ref:MOGA3}.}
    \label{fig:MOGA}
\end{figure}

In recent years, Multi Objective Genetic Algorithms (MOGA)~\cite{ref:MOGA1, ref:MOGA2, ref:MOGA3} have been popularised 
to optimise linear but also non-linear dynamics of electron low emittance storage rings. MOGA use various knobs 
such as quadrupole strengths, chromaticity sextupoles and correctors with some constraints and generate populations
which evolve in generations based on natural evolution paradigms such as inheritance, mutation, selection, and crossover. 
An example for the optimisation of the Advanced Light Source (ALS) lattice is being shown in Fig.~\ref{fig:MOGA}, where the target is 
the ultra-low horizontal emittance, but also high dynamic aperture with a constraint on optics, which in this example is
the horizontal beta function in the ID location, for this specific example. The population creates an optimal front (called the Pareto front), 
where the best solution following these constraints can be chosen.

\begin{figure}[htp]
    \centering
    \includegraphics[keepaspectratio, width=0.75\columnwidth]{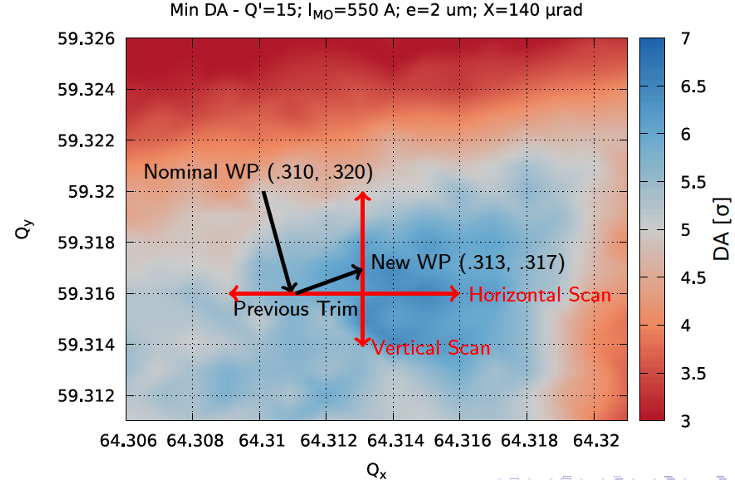}
    \caption{LHC run2 working point scan color-coded with the DA~\cite{ref:pele1}.}
    \label{fig:pele1}
\end{figure}
\begin{figure}[htp]
    \centering
    \includegraphics[keepaspectratio, width=0.75\columnwidth]{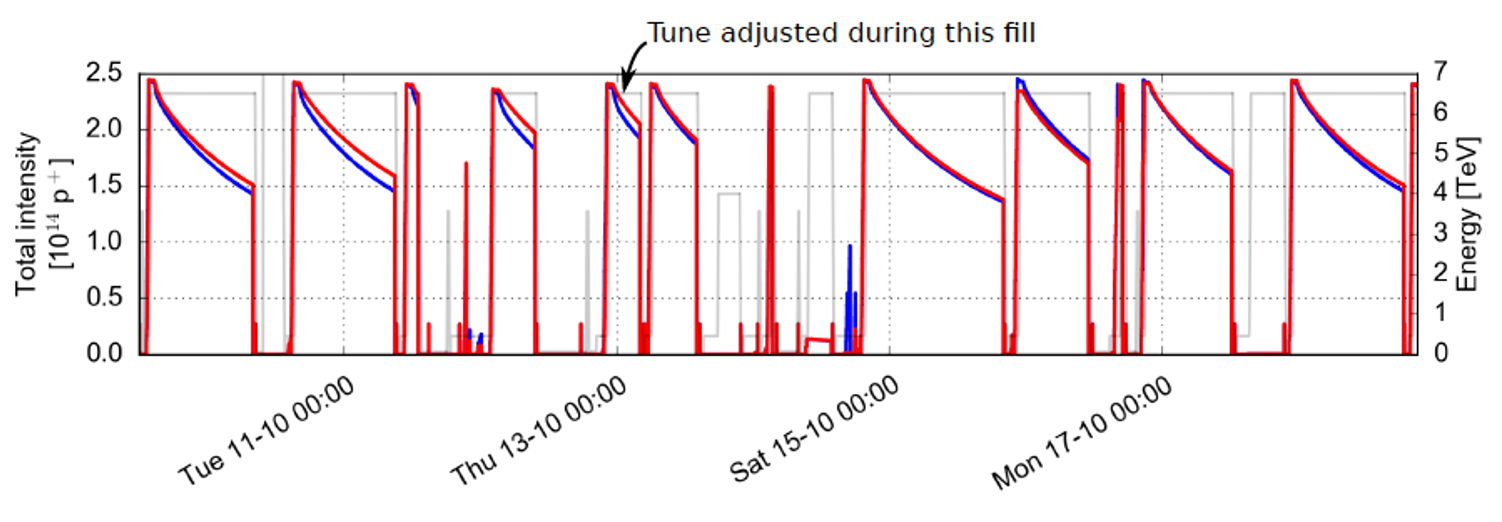}
    \caption{Intensity of Beam1 (blue) and Beam 2 (red) for consecutive fills of the LHC~\cite{ref:pele1}.}
    \label{fig:pele2}
\end{figure}

In the same line, the DA estimation can be better valued as a lifetime maximisation indicator, if used in a comparative way 
by varying certain lattice parameters. This approach was heavily used for the LHC~\cite{ref:pele1,ref:pele2} and HL-LHC~\cite{ref:pele3,NikosYannis}.
An example for the LHC during Run2 is given in Fig.~\ref{fig:pele1}, where the working point scan color-coded by the DA is produced
in order to reveal the reasons of the poor lifetime of beam 1, in particular after the reduction of the crossing angle at the end of 2016.
These DA simulations showed that the nominal fractional working point at $(0.31,0.32)$ was not optimal, in particular for the strong octupoles and 
15 units of chromaticity,
employed during LHC operation. The working point maximising the DA could be found in an area closer to the diagonal.
A subsequent working point scan was performed in operation and proved indeed beneficial for the beam lifetime. This was clearly shown
in operation, as presented in Fig.~\ref{fig:pele2}, where the intensity evolution of both beams is shown for consecutive LHC fills,
with a clear improvement of Beam 1 (blue), when the new working point was applied.

\subsection{Lyapunov exponent}
\begin{figure}[htp]
    \begin{center}
\begin{tikzpicture}
 \node (img1)  {\includegraphics[trim=15 15 0 0,clip,width=8cm]{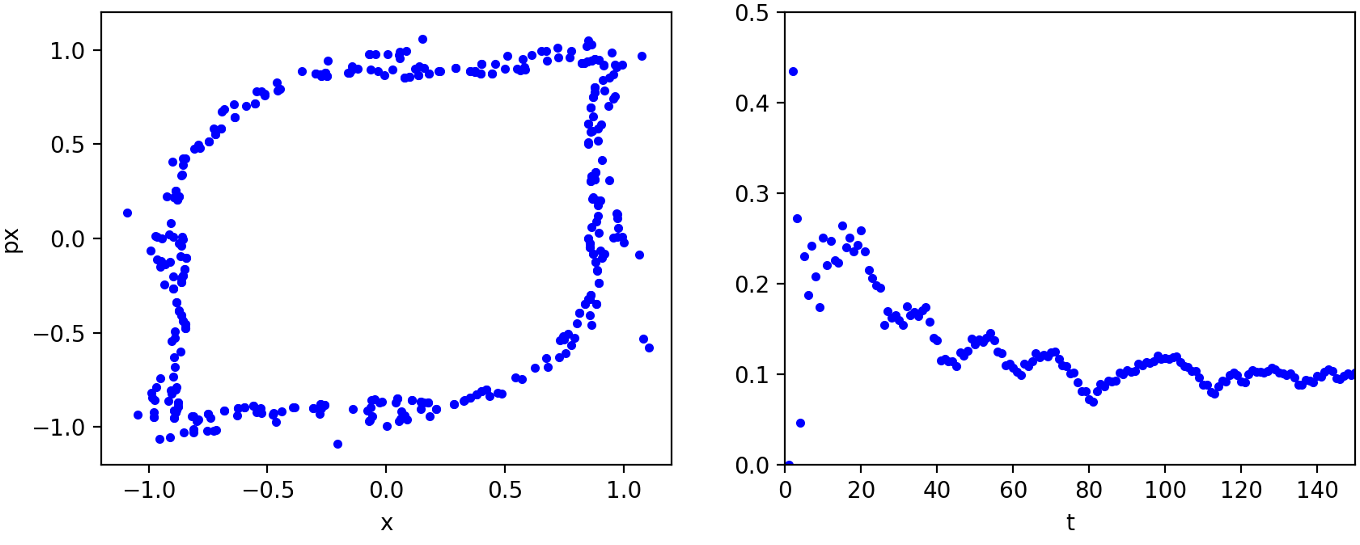}};
 \node[below=of img1, node distance=0cm, xshift=-1.9cm,yshift=1.2cm,font=\color{black}] {$x$};
 \node[below=of img1, node distance=0cm, xshift=2.2cm,yshift=1.2cm,font=\color{black}] {$t$};
 \node[left=of img1, node distance=0cm, rotate=0, anchor=center,xshift=0.8cm,yshift=0.1cm,font=\color{black}] {$p_x$};
 \node[right=of img1, node distance=0cm, rotate=0, anchor=center,xshift=-5.1cm,yshift=0.1cm,font=\color{black}] {$\lambda$};
 \end{tikzpicture}\\
\begin{tikzpicture}
 \node (img1)  {\includegraphics[trim=15 15 0 0,clip,width=8cm]{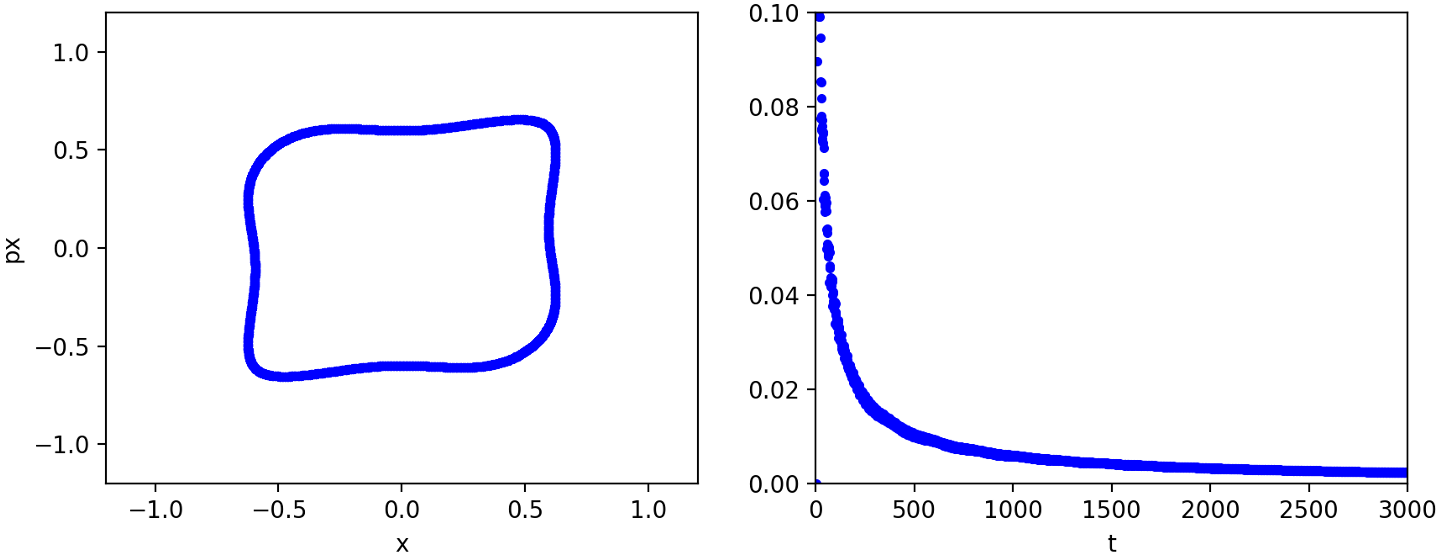}};
 \node[below=of img1, node distance=0cm, xshift=-1.9cm,yshift=1.2cm,font=\color{black}] {$x$};
 \node[below=of img1, node distance=0cm, xshift=2.2cm,yshift=1.2cm,font=\color{black}] {$t$};
 \node[left=of img1, node distance=0cm, rotate=0, anchor=center,xshift=0.8cm,yshift=0.1cm,font=\color{black}] {$p_x$}; 
 \node[right=of img1, node distance=0cm, rotate=0, anchor=center,xshift=-5.2cm,yshift=0.1cm,font=\color{black}] {$\lambda$};
 \end{tikzpicture}
\caption{Maximum Lyapunov Exponent for a chaotic (top) and regular (bottom) trajectory.}
    \label{fig:Lyap1}
\end{center}
\end{figure}
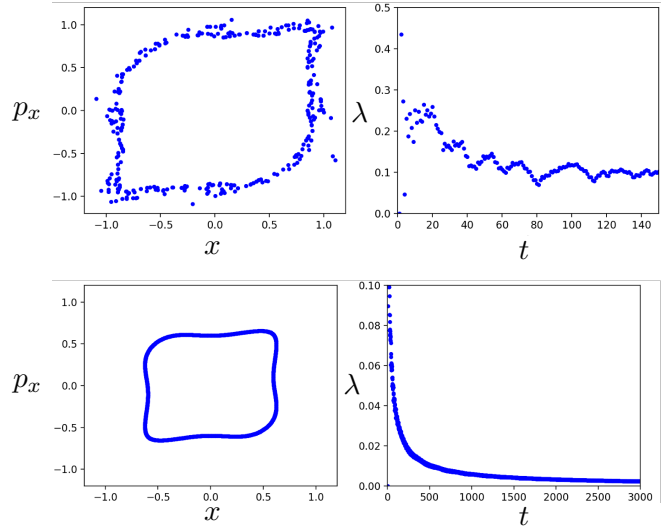

By definition, chaotic motion implies sensitivity to initial conditions of nearby orbits. In fact, two infinitesimally close chaotic trajectories 
in phase space with initial deviation vector ${\displaystyle \delta \mathbf {Z} _{0}}$  will end-up diverging with rate
${\displaystyle |\delta \mathbf {Z} (t)|\approx e^{\lambda t}|\delta \mathbf {Z} _{0}|}$, where with $\lambda$  is the maximum Lyapunov exponent~\cite{Lyap}.
As the rate can be different for different phase space variables, there are as many exponents as the phase space dimensions, forming
the so called Lyapunov spectrum. The largest one is the Maximal Lyapunov exponent (MLE) which is defined as
\begin{equation}
{\displaystyle \lambda =\lim _{t\to \infty }\lim _{\delta \mathbf {Z} _{0}\to 0}{\frac {1}{t}}\ln {\frac {|\delta \mathbf {Z} (t)|}{|\delta \mathbf {Z} _{0}|}}}
\end{equation}
The MLE converges towards a positive value for a chaotic orbit, as shown in Fig.~\ref{fig:Lyap1} (top), showing indeed exponential
divergences of nearby trajectories. In the case of a regular orbit (Fig.~\ref{fig:Lyap1}, bottom), the MLE converges to zero, as the trajectories
will never diverge.

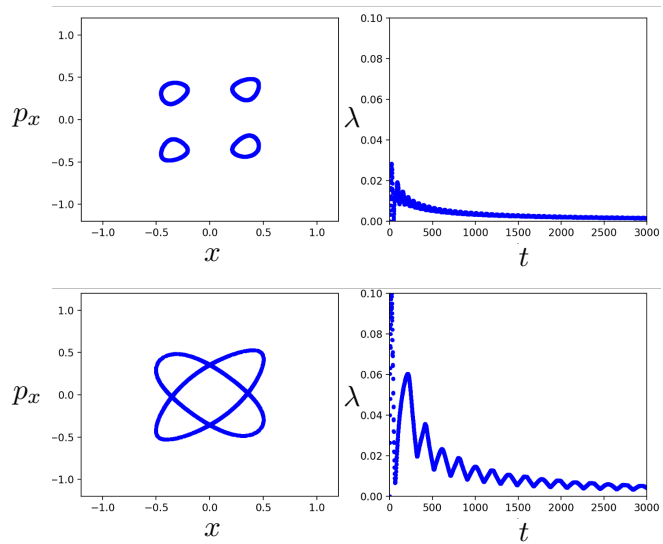
\begin{figure}[htp]
    \begin{center}
\begin{tikzpicture}
 \node (img1)  {\includegraphics[trim=5 15 0 0,clip,width=8cm]{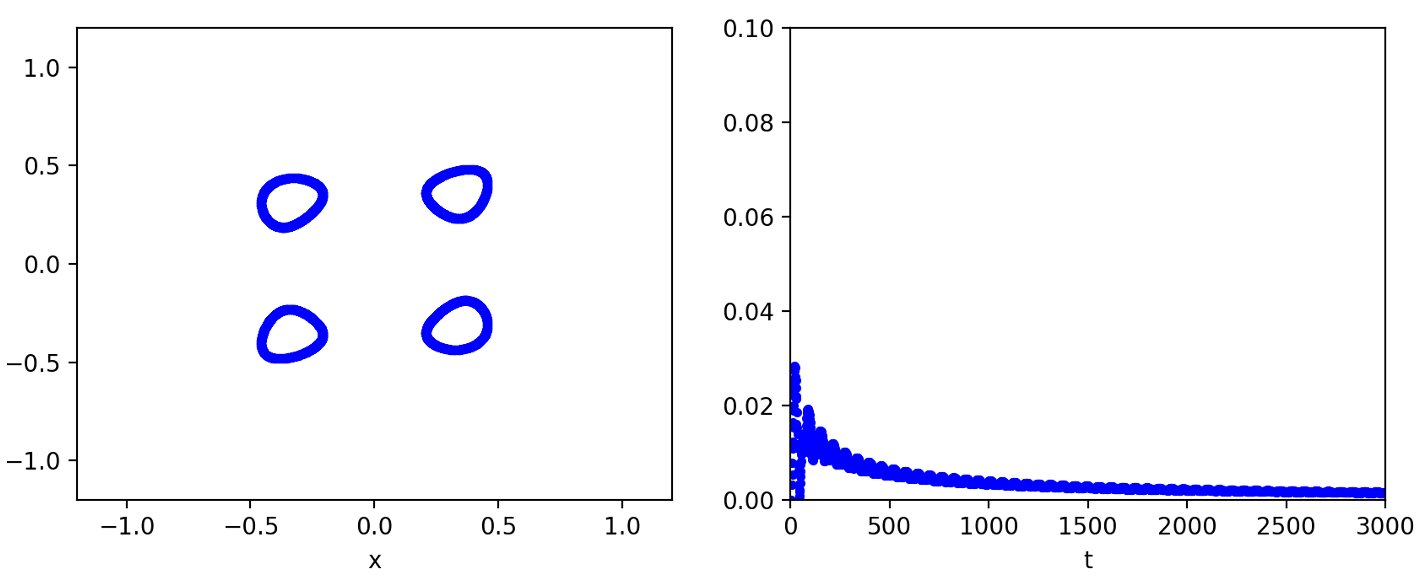}};
 \node[below=of img1, node distance=0cm, xshift=-1.9cm,yshift=1.2cm,font=\color{black}] {$x$};
 \node[below=of img1, node distance=0cm, xshift=2.2cm,yshift=1.2cm,font=\color{black}] {$t$};
 \node[left=of img1, node distance=0cm, rotate=0, anchor=center,xshift=0.8cm,yshift=0.1cm,font=\color{black}] {$p_x$};
 \node[right=of img1, node distance=0cm, rotate=0, anchor=center,xshift=-5.2cm,yshift=0.1cm,font=\color{black}] {$\lambda$};
 \end{tikzpicture}\\
\begin{tikzpicture}
 \node (img1)  {\includegraphics[trim=5 15 0 0,clip,width=8cm]{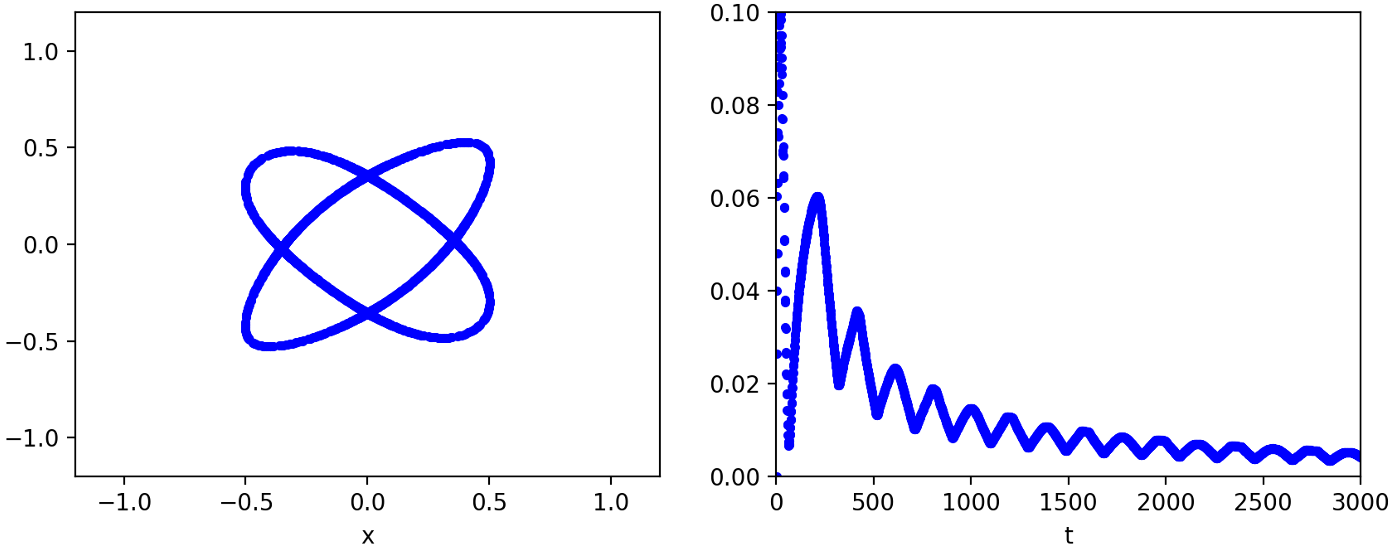}};
 \node[below=of img1, node distance=0cm, xshift=-1.9cm,yshift=1.2cm,font=\color{black}] {$x$};
 \node[below=of img1, node distance=0cm, xshift=2.2cm,yshift=1.2cm,font=\color{black}] {$t$};
 \node[left=of img1, node distance=0cm, rotate=0, anchor=center,xshift=0.8cm,yshift=0.1cm,font=\color{black}] {$p_x$}; 
 \node[right=of img1, node distance=0cm, rotate=0, anchor=center,xshift=-5.2cm,yshift=0.1cm,font=\color{black}] {$\lambda$};
 \end{tikzpicture}
\caption{Maximum Lyapunov Exponent for a trajectory close to a resonance
 (top) and one close to the separatrix (bottom).}
    \label{fig:Lyap2}
\end{center}
\end{figure}

The difficulty for using MLE in order to distinguish ordered from chaotic motion is indeed due to its definition, which
requires very large and potentially infinite integration time, in particular for trajectories that are close to a resonance Fig.~\ref{fig:Lyap2} (top) 
or close to the separatrix (Fig.~\ref{fig:Lyap2} bottom). In both of these cases, the MLE converges more slowly 
towards zero. There are indeed several numerical methods which allow the efficient calculation of the Lyapunov exponent
or its variants~\cite{Benet,Skokos}, mostly applied to dynamical system beyond beam dynamics.

\subsection{Frequency Map Analysis}

%
%
The brute-force approach of DA estimations is extensively used in beam
dynamics.  However, it is not easy to correlate the excited resonances with
the extent of the chaotic region, that limit the DA in order to provide correcting measures,
in particular by just observing trajectories in the physical or phase space. 
The Frequency Map Analysis (FMA) attempts to connect the two effects by moving to the frequency space.
This method can be used as an early chaos indicator but also for computing numerically resonance strengths. It has been
introduced by Laskar in astronomy~\cite{Laskar88,Laskar90, LaskarRobutel,PapLas96, PapLas98} and subsequently
used in several physics applications ranging from atomic physics~\cite{Uzer} to
Hamiltonian toy models~\cite{LaFrCe,Laskar93,DumasLaskar}. During the last two decades, it became 
a standard tool of beam dynamics analysis for a variety of accelerators, 
such as hadron colliders~\cite{frmap,Yannis,PapZim99, 
PapZim02,Luo2012}, synchrotron light sources~\cite{Laskarrobin,Laskarprl00,NaLa03,expfrmap03,expfrmap04}, 
high-intensity rings~\cite{YannisSNS,BartosikPS2}, B-factories~\cite{Liuzzo} and 
linear collider damping rings~\cite{Fanouria}.  The method relies on the high precision
estimation~\cite{ref:NAFF1} of 
the associated frequencies of motion (or "tunes" in the accelerator jargon), 
which are supposed to be invariant (i.e. integrals of motion) in the case of quasi-periodic motion,
as stated by the KAM theory~\cite{Kolm,Arn,Mos}. In this respect, the variation of the frequencies
over time~\cite{Laskar93,DumasLaskar,frmap} can provide an early
stability indicator that has the advantage of connecting resonant phase-space structures and their overlap
with diffusion of chaotic trajectories. 

The back-bone of the method is the Numerical Analysis of Fundamental Frequencies (NAFF) algorithm 
\cite{Laskar88,Laskar90} or each variants~\cite{SUSSIX}. The idea is to derive a
quasi-periodic approximation, in form of a truncated  to order $N$ Fourier series,
\begin{equation}
 f'_j(t) = \sum^N_{k=1} a_{j,k} e^{i\omega_{jk}t} \; ,
\end{equation}
with $f'_j(t), a_{j,k}\in \mathbb{C}$ and $j=1,\dots,n$,
of a complex phase space function $f_j(t)= q_j(t) + i p_j(t)$, formed by a pair of
conjugate variables of a general $n$-degrees 
of freedom Hamiltonian system, which are determined by usual numerical
integration or experimentally measured, for a finite time span $t=\tau$. The next step is to
keep from the quasi-periodic
approximation the fundamental frequencies of motion, corresponding most of the times to the frequency of the dominant Fourier component $\omega_{j1}$, for each degree of freedom. In this respect, the frequency vector $\frac{\boldsymbol \omega}{2\pi} = {\boldsymbol \nu} =  (\nu_1,
\nu_2, \dots, \nu_n)$ can be constructed, which, up to numerical accuracy
\cite{ref:NAFF1}, parameterises the KAM tori in the stable regions of a
 Hamiltonian system~\footnote{Strictly speaking, for the one-to-one correspondence of phase-space amplitudes, i.e. actions and their associated frequencies the system should be non-degenerate, i.e. the tune-shift with amplitude should be monotonic. This mathematical restriction does not limit though the applicability of the method.}. Then, the frequency map, an extended version of the so-called tune-footprint, can be produced~\cite{LaFrCe, Laskar93, DumasLaskar,
Laskarrobin}, by repeating the procedure for
a set of initial conditions which are transversal to the orbits of
interest. As an example, all the momenta ${\boldsymbol  p}$
 could be kept constant, and explore the positions ${\boldsymbol q}$ to
produce the map ${\mathcal F_\tau}$:
\begin{equation}
{\mathcal F_\tau}\;:
\begin{matrix} 
{\mathbb R}^{n}&\longrightarrow &{\mathbb R}^{n} \\
{\boldsymbol q}|_{{\boldsymbol p}={\boldsymbol p_0}}&\longrightarrow
&{\boldsymbol \nu} \;.
\end{matrix}
\end{equation}
The dynamics of the system is then analysed by studying the regularity
of this map on frequency or initial condition space. In addition, each initial condition 
can be associated with a
diffusion indicator, by computing the frequency vector for two equal and
successive time spans, which correspond to half of the total
integration time $\tau$. The amplitude of the diffusion vector,
\begin{equation}
{\boldsymbol D}|_{t=\tau} = {\boldsymbol
  \nu}|_{t\in(0,\tau/2]}-{\boldsymbol \nu}|_{t\in(\tau/2,\tau]} \;,
  \label{Diffvec}
\end{equation}
 can be used for characterising the extent of the chaotic behaviour and diffusion
of each orbit.
Through this representation,  the
traces of  the resonances can be viewed in the physical space, as well, and set a 
threshold for the minimum DA. Moreover,  a diffusion
quality factor defined as the average of the local diffusion
coefficient to the initial amplitude of each orbit, over a domain $R$
of the phase space:
\begin{equation}
D_{QF} = \big\langle \; \frac{{|{\boldsymbol
      D}|}}{|\boldsymbol q_0|} \;  \big\rangle_R \;.
      \label{DiffQF}
\end{equation}
This quantity can be used for the comparison of different designs  
and the optimisation of the correction schemes proposed.

\section{Application of frequency map analysis in accelerator models}
\label{accelFMA}

\subsection{Frequency maps for the LHC}

The long term stability of the beam is the major concern for the
design of a hadron collider, as the LHC or its high-luminosity upgrade HL-LHC not only during
the long injection period of more than $10^7$ turns needed to fill the LHC
with more than 2800 bunches per beam, in its nominal configuration, but also during collisions
which take place for ~10 hours. The particle trajectories are thereby perturbed
strongly by non-linear magnet fields, mainly attributed to the
multipole errors of the super-conducting magnets. In addition, collective stability requirements necessitate  
the use of high strength in the arc octupoles and sextupoles, providing the necessary tune-shift with amplitude and
chromaticity to guarantee Landau damping, even in the presence and the continuous action of the transverse damper. 
In addition, at collision beam-beam effects but also e-cloud which are intrinsically non-linear can enhance
diffusion of large amplitude particles and reduce beam lifetime, ideally dominated by the particle burn-off,
as they collide in the detectors.
\begin{figure*}[ht]
\begin{center}
    \includegraphics*[height=6.4cm,width=7.1cm]{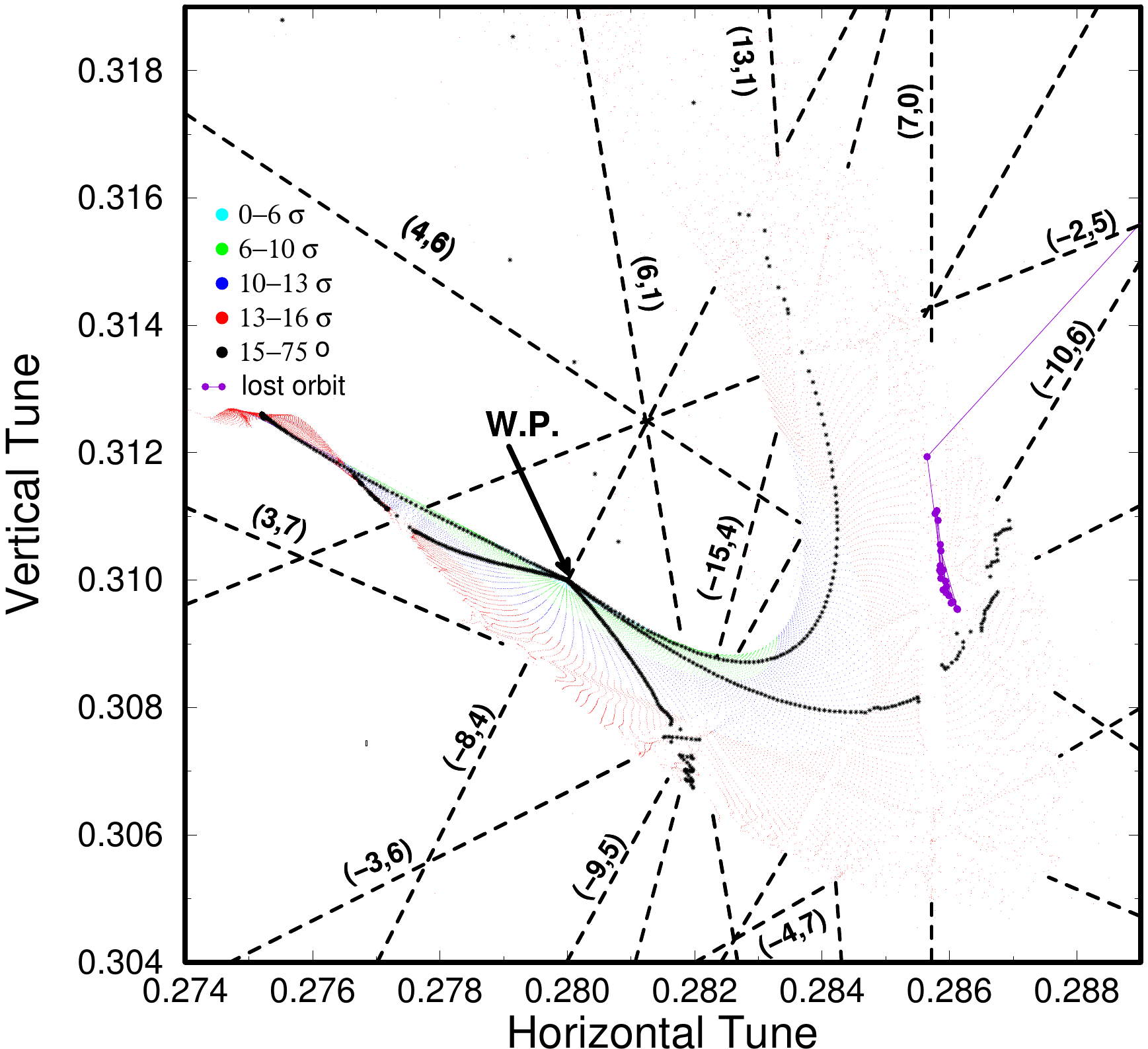}\hskip 1.cm
    \includegraphics*[height=6.4cm,width=7.1cm]{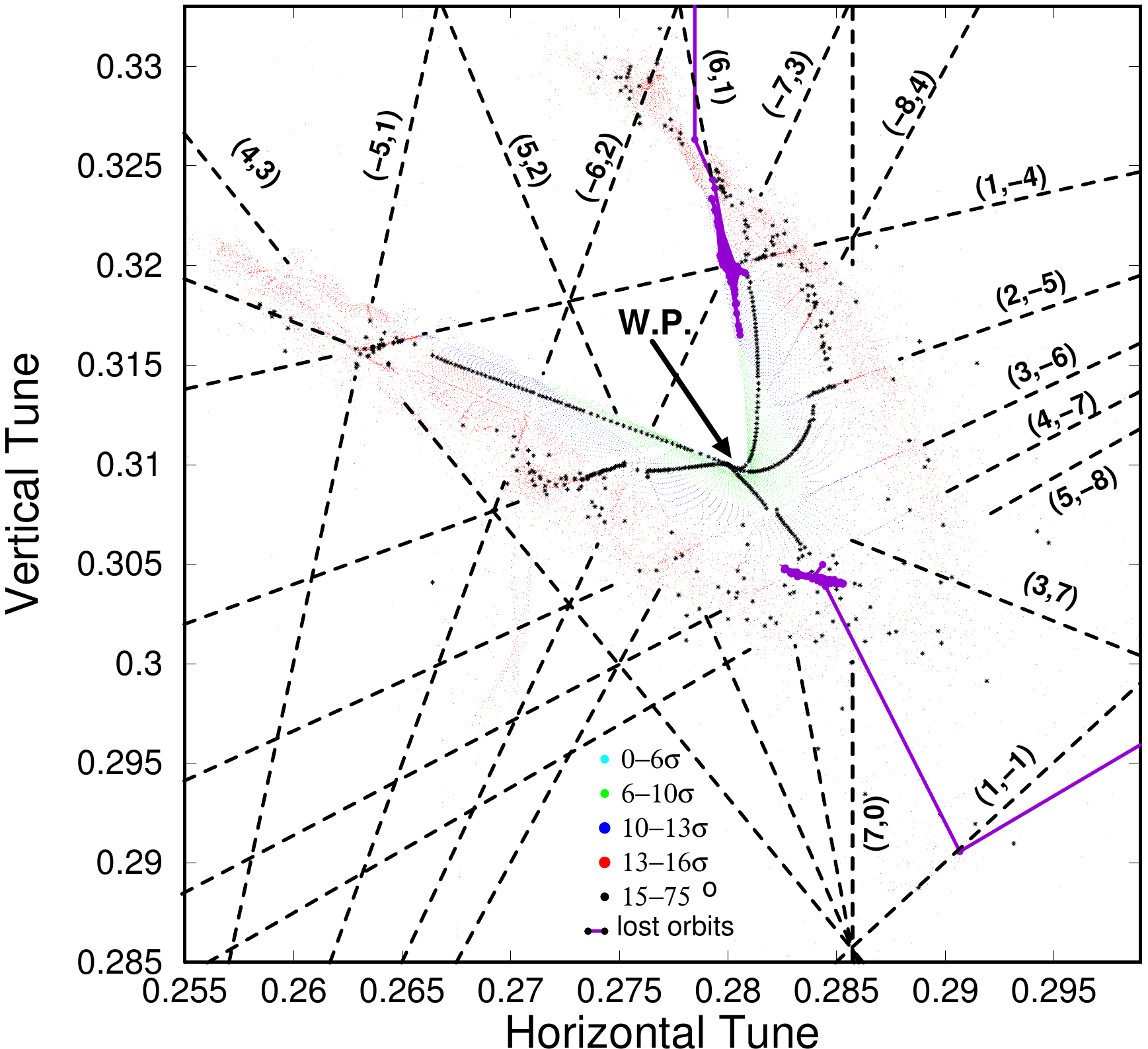}  \ \vskip 0.3cm
    \includegraphics*[height=6.4cm,width=7.1cm]{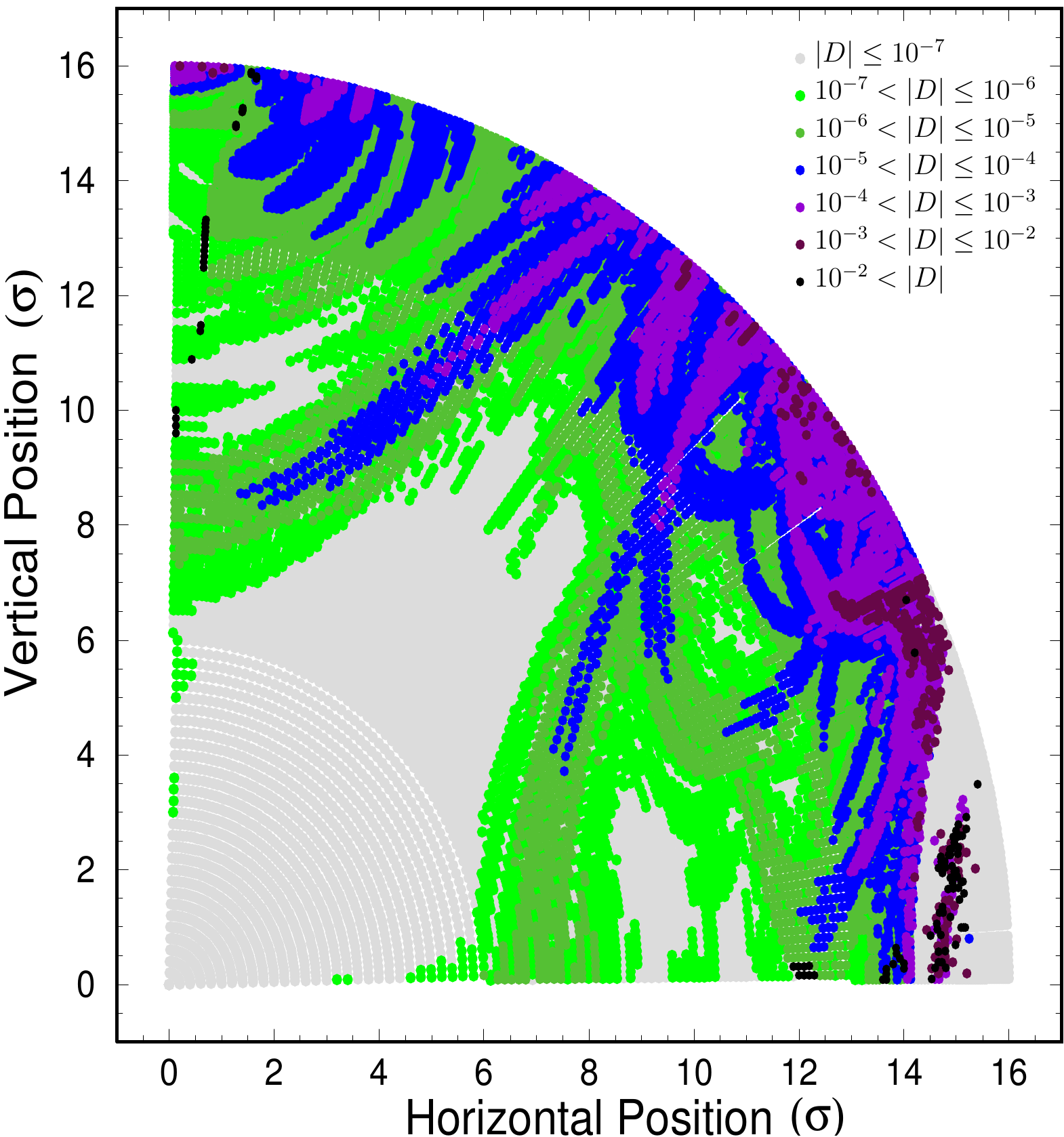}\hskip 1cm
    \includegraphics*[height=6.4cm,width=7.1cm]{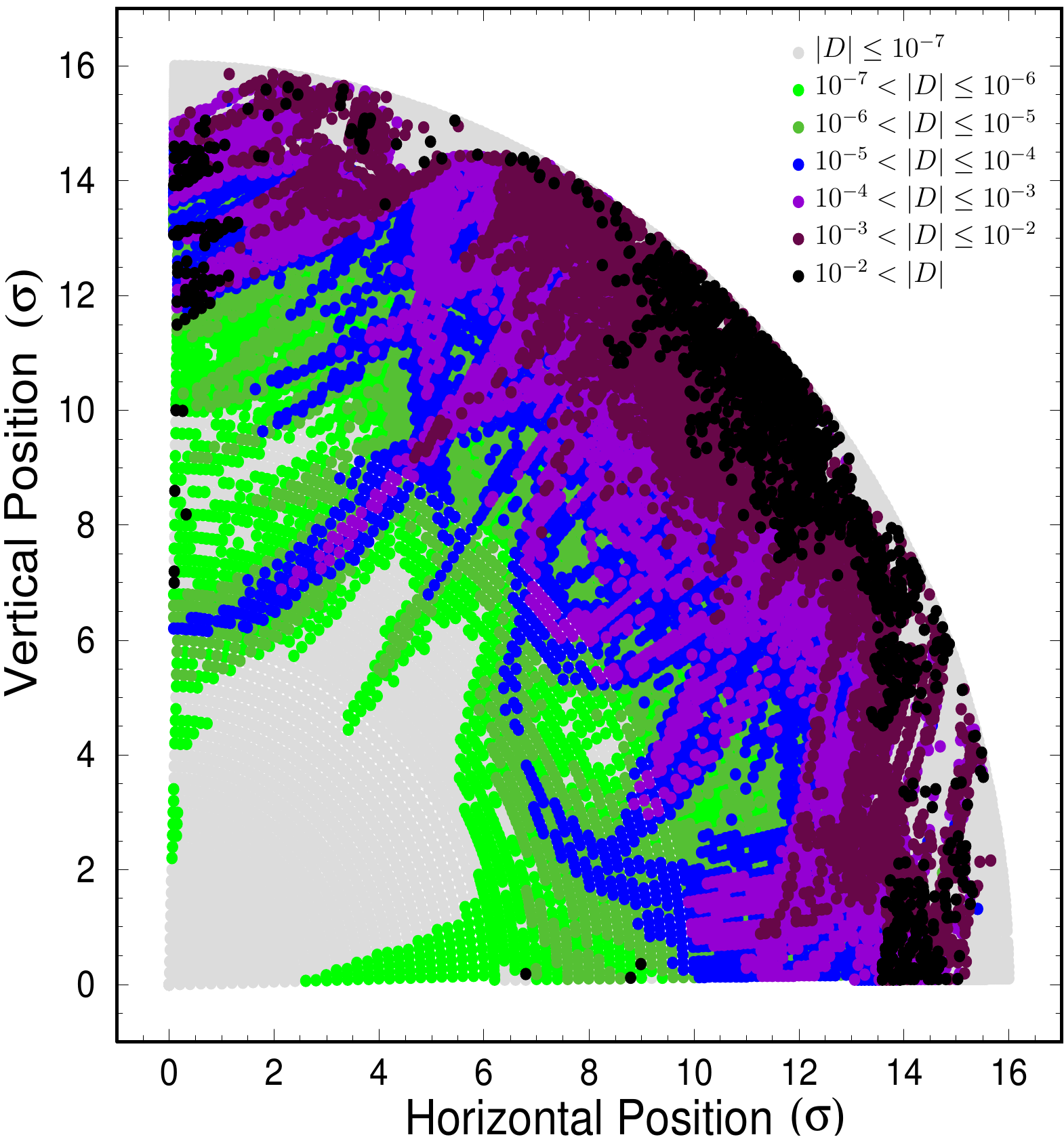} 
\end{center}
\caption{Frequency (top) and diffusion (bottom) maps for the LHC  target
  error table without (left) and with (right) the high
  $a_4$ value on the dipoles~\cite{frmap}.}
\label{frmap5}
\end{figure*} 
In order to reach the target DA of 12 rms beam sizes $\sigma_x=\sigma_y=\sigma$,  for the LHC injection
optics, giving a necessary safety factor of 2 with respect to the
physical aperture restriction of the beam at $6~\sigma$
 (location of collimators
protecting the super-conducting magnets), 
a target magnetic multi-pole error table was proposed~\cite{Koutch}. A frequency map 
for this error table and for the nominal tunes ($Q_x=64.28$, $Q_y=59.31$) is shown
in the top left part of  Fig.~\ref{frmap5}~\cite{frmap}.  This specific machine gives an average
DA of around $13~\sigma$ and a minimum of $12~\sigma$, values which are close to the average 
and minimum DA over all the 60 random realisation of the magnet errors. Each point in the 
frequency space corresponds to a particular on-momentum orbit tracked over 1000 turns. 
The different colours in the map correspond to orbits with
different initial position amplitudes (from $0-16\sigma$)
and the black dots label initial conditions with   
different amplitude ratios (from $15^\circ$ to $75^\circ$). The orderly spaced
points correspond to regular orbits whereas the dispersed points to
chaotic ones. This plot is a numerical construction of the so called Arnold web~\cite{Arn}, the
 network of resonances $a\nu_x+b\nu_y+c=0$, which appear as
distortions of the map (empty and filled lines) and can be easily
identified. For example, the importance of three 7th order
resonances ($(a,b) = (7,0), (6,-1)\,\text{and}\,(-2,5)$) is put in evidence. 
Especially the crossings of the resonant lines are the phase space areas, through which
particles can easily diffuse: as an example, the evolution of
the frequency vector of an orbit starting close to the crossing of the $(7,0)$
with the $(-3,6)$ and $(4,6)$ resonances is shown (purple dots), as estimated over consecutive
time-spans.  The orbit diffuses along
the unstable manifold of the 7th order resonance and
is lost after a few thousand turns. This is a clear demonstration of
the destructive nature of this resonance with respect to the DA of this model
and came to some as a surprise that such a high-order resonance could
limit beam dynamics.

One of the main issues in the specification of the LHC injection
optics, was the correction of the systematic part of the lowest order
multipole errors of the super-conducting dipoles, which limit the
DA~\cite{Koutch}, by using magnetic coils (``spool
pieces") placed at the ends of the dipoles. In the case of  more realistic  error 
table with increased normal and skew octupoles, there was an important loss of the 
DA~\cite{skew} with respect to the target error table. A
frequency map for the same random ``seed" as for the previous case with the
increased skew octupole error in the dipoles is shown in the top right plot of
Fig.~\ref{frmap5}. The frequency maps now looks much more distorted. The
most remarkable feature is the huge increase of the tune variation with amplitude, to the
point that particles are diffusing towards the 
$(1,-1)$-resonance (which can be also the $(2,-2)$, $3,-3$, etc.) in the right bottom corner of the map. 
On the other hand, particles close to horizontal motion at the top of the map are approaching the
$(0,3)$ resonance and the ones close to vertical motion the $(4,0)$. 
This finding has been confirmed with Normal Form analysis~\cite{skew}. 
The DA could be recovered by tuning the skew octupole spool pieces
such as to cancel the $(1,-1)$ resonance~\cite{skew}.

The global dynamics of these two cases can be also explored in the physical
space of the system by mapping each initial condition with the diffusion vector~\eqref{Diffvec}, 
the amplitude of which can be used for characterising the diffusion rate
of each orbit. In Figs.~\ref{frmap5} (bottom),  the points in the
configuration space are plotted using a different colour coding corresponding to
different diffusion indicators in logarithmic scale: from grey for
stable ($|{\boldsymbol D}|\le 10^{-7}$) to black for strongly 
chaotic particles ($|{\boldsymbol D}|>10^{-2}$), that actually are lost
within that short integration time. Through this
representation, the traces of 
the resonances in the physical space are clearly visible, and thereby a 
threshold for the minimum DA can be set.

\begin{figure}[t]
\begin{center}
\rotatebox{0}{\scalebox{0.44}{{\includegraphics*{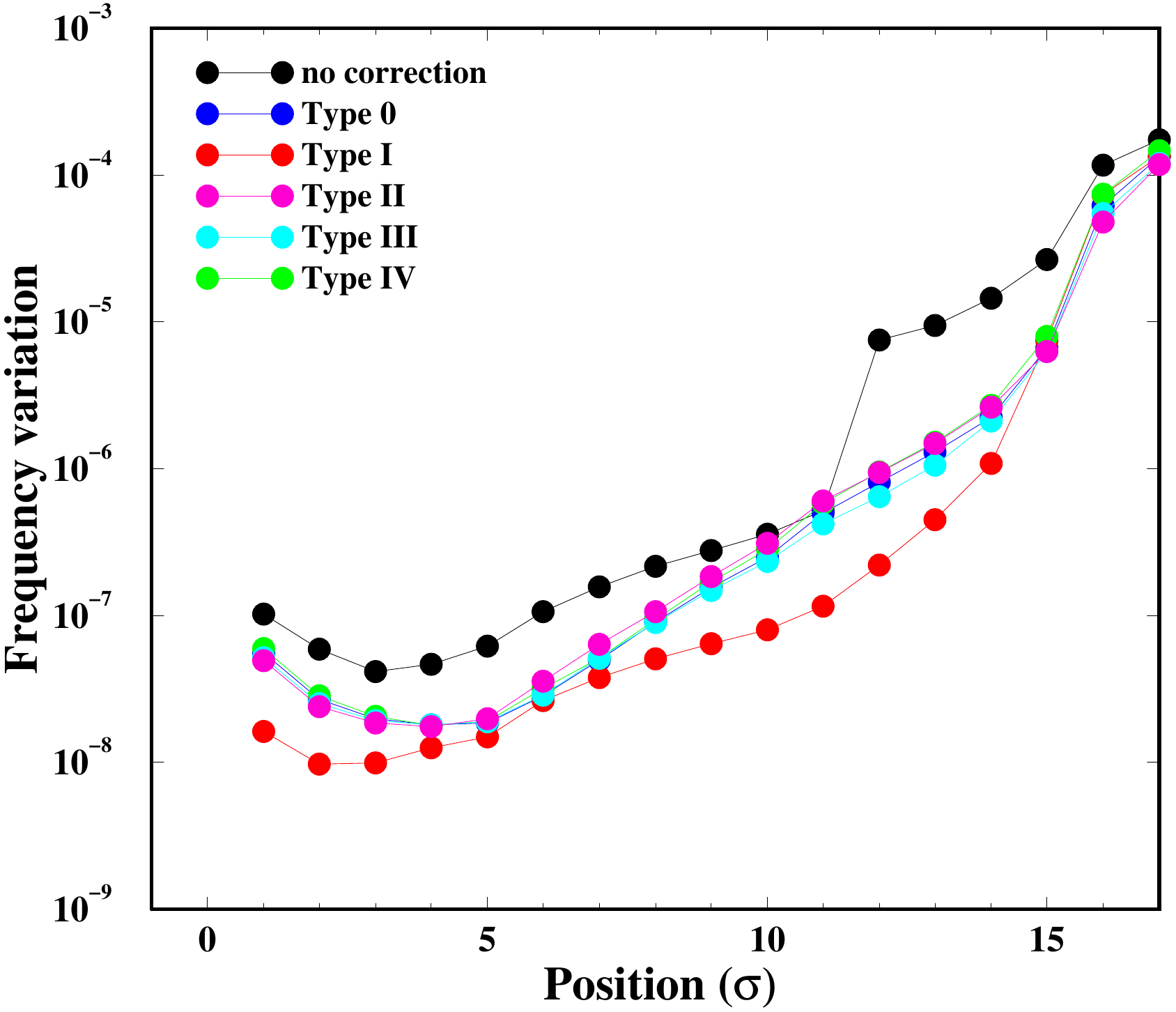}}}}\hskip .5cm
\rotatebox{0}{\scalebox{0.44}{{\includegraphics*{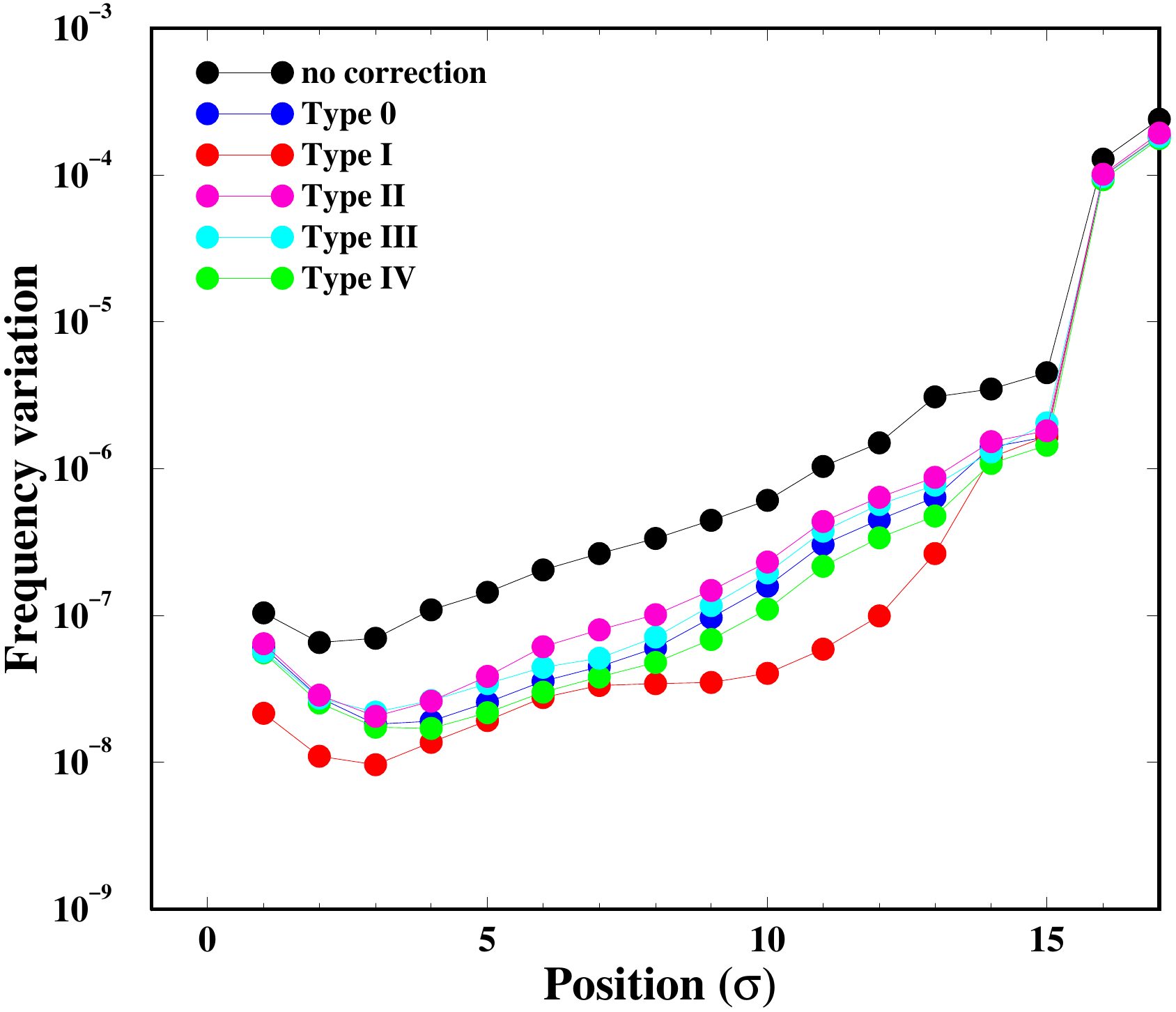}}}}
 \caption{Evolution of the frequency variation averaged over all
directions, with the particles' amplitude (in $\sigma$) for $\delta
p/p=7\times10^{-4}$,  and for two different tunes: $(Q_x,Qy)=(0.28,0.31)$  (left) and $(Q_x,Qy)=(0.21,0.24)$ (right)~\cite{Yannis}.}
\label{diffu}
\end{center}
\end{figure}

The diffusion
quality factor~\eqref{DiffQF}  is a very efficient global chaos indicator for comparing different designs  
and optimising the correction schemes proposed~\cite{Yannis,BartosikPS2,ref:MOGA3}. For
example, for the normal octupole $b_4$ and decapole $b_5$ correctors in the LHC, 
five schemes where proposed, regarding the positioning and the amount of
the correctors~\cite{Yannis}.
Frequency maps were produced for all the correction
cases and two working points.
In Fig.~\ref{diffu}, the diffusion quality factor averaged over the angles 
is plotted in logarithmic scale  versus the amplitude, for both working points,
 for all correction schemes and for the non-zero momentum deviation. These plots confirmed
 that all the correction schemes are quite similar and indeed necessary, following the comparison
 with the diffusion quality factor with no correction
 (black dots). They also indicated that the nominal (but most expensive) solution of including
 correctors in all the super-conducting dipole (blue dots)
was not performing better than the one with correctors in every second dipole (red dots), which
actually presented a slightly better diffusion quality factor. Based on this study, the baseline
correction for the LHC was decided tp include correctors in every second dipole, a rather
cost effective solution. Finally, this study demonstrated that the diffusion quality factor  
is correlated with other global chaos indicators, such as the resonances driving terms norm 
and the dynamic aperture~\cite{Yannis}.

\subsection{Working point choice through frequency maps}
\begin{figure}[htb]
\centering
\includegraphics*[trim= 10 230 10 260, clip,width=150mm]{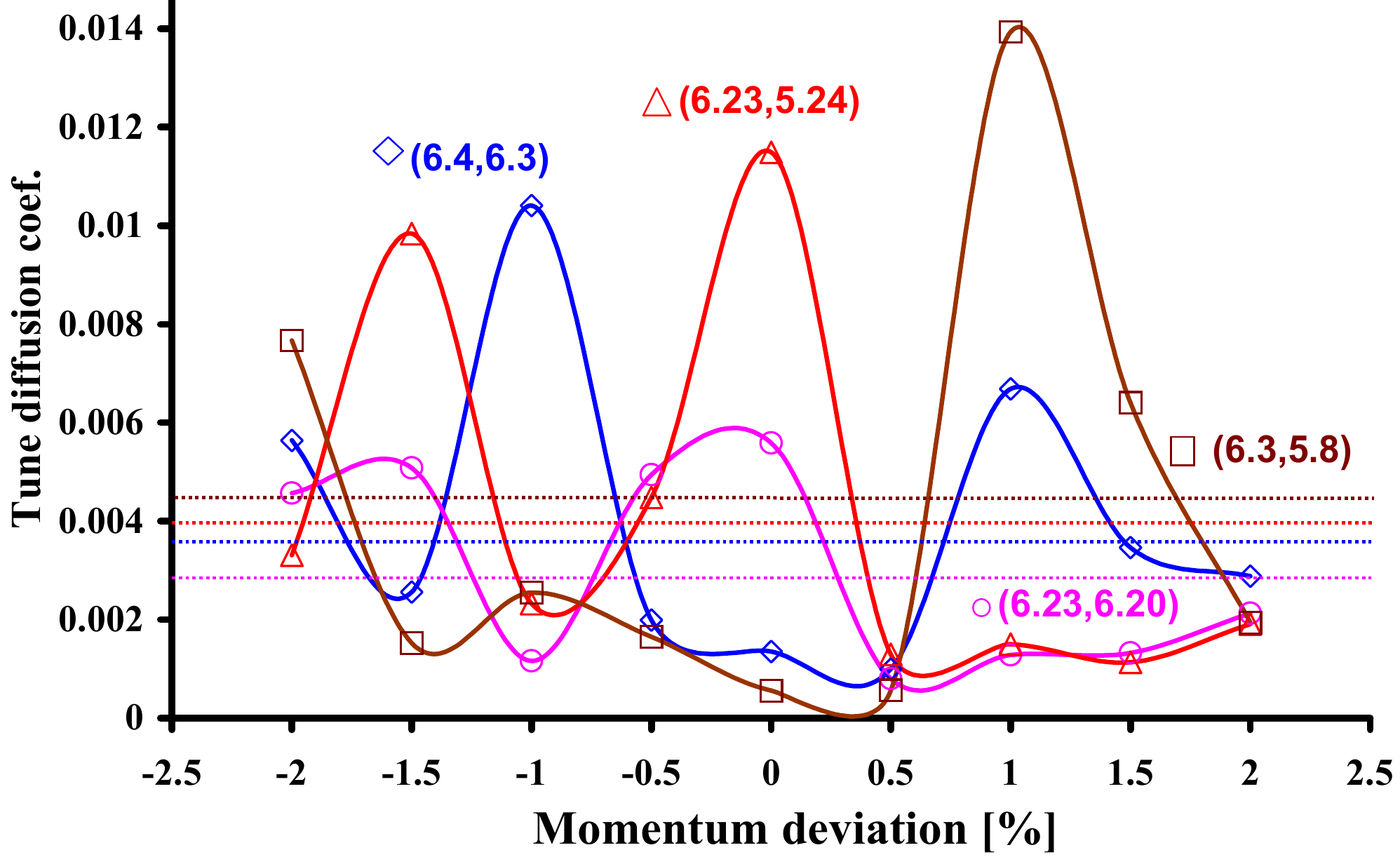}
\caption{Tune diffusion coefficients for four SNS working points versus the
different momentum deviations~\cite{YannisSNS}.}

\label{fig:tunecomp}
\end{figure}
Frequency map analysis is naturally very powerful for choosing the
best working point of an accelerator. An example is given in 
Fig.~\ref{fig:tunecomp}, where the value of the tune diffusion
coefficient  is plotted versus the momentum deviation
$\delta p/p$, for all working points of the Spallation Neutron Source accumulator
ring~\cite{YannisSNS},  currently in operation in Oak-Ridge
National Laboratory and holding the beam power world record.  The single-particle dynamics of this ring is dominated by edge effects
in the magnets, and especially the quadrupole
fringe fields, which are octupole-like~\cite{Forest1,ref:ForestFringeFields,ref:PTW}, at leading order, and given by the hard-edge
Hamiltonian for a quadrupole with strength $Q$ :
\begin{equation}
  H_{f} = \frac{\pm Q}{12B\!\rho(1\!+\!\frac{\delta p}{p})}
          (y^3 p_y - x^3 p_x + 3 x^2 y p_y - 3 y^2 x p_x)\;\;,
\end{equation}
which does not only depend on positions $x,y$ but also on momenta $p_x,p_y$. The associated octupole-like tune-shift at leading order can be computed analytically~\cite{ref:YPA}
\begin{equation} 
\begin{pmatrix}\delta\nu_x \\ \delta\nu_y \end{pmatrix} = 
 \begin{pmatrix} a_{hh} & a_{hv} \\ a_{hv} & a_{vv} \end{pmatrix}
 \begin{pmatrix} 2 J_x \\  2 J_y \end{pmatrix}\;\;,
\end{equation}
with the ``anharmonicity” coefficients (or ``torsion" in the non-linear dynamics wording)\begin{equation} 
\begin{aligned}
a_{hh} &= \frac{-1}{16\pi B\!\rho}\sum_i \pm Q_i\beta_{xi}\alpha_{xi} \\
a_{hv} &= \frac{ 1}{16\pi B\!\rho}\sum_i \pm Q_i
                        (\beta_{xi}\alpha_{yi}-\beta_{yi}\alpha_{xi}) \\
a_{vv} &= \frac{ 1}{16\pi B\!\rho}\sum_i \pm Q_i\beta_{yi}\alpha_{yi}
\end{aligned}
\;\;,
\end{equation}
depend on the optics functions $\beta_{x,y},\alpha_{x,y}$ at the location $i$ of the fringe field.

In order to study their impact to SNS dynamics, four working points were selected and compared
corresponding to the different curves of Fig.~\ref{fig:tunecomp}, using the diffusion indicator
issued by FMAs.
 The peak values on the diffusion indicators,
for all working points correspond to areas of the phase space that are
perturbed due to 4th order resonances, showing  the
destructive effect of quadrupole fringe fields. The dotted lines on the
plots represent the average values of the diffusion indicators for all
tracked momentum deviations. It is clear that $(6.23,6.20)$ is the best
choice, followed by $(6.40,6.30)$. Their performance can be further
improved by using the available multi-pole correctors~\cite{YannisSNS},
for correcting the normal and skew 3rd order resonances, in the case
of $(6.40,6.30)$, and the 4th order normal resonances in the case of
$(6.23,6.20)$.  The other two  working points have the disadvantage of crossing
major resonances, which are very difficult to correct. Based on this
study, the nominal working point of the SNS ring was chosen to be
$(6.23,6.20)$ and was successfully used in commissioning and operation until today.

Similar working point optimisation was employed for different type of rings such as SuperB, a lepton
collider designed at INFN-LNF~\cite{Liuzzo}, or the PS2 ring~\cite{BartosikPS2}, an upgrade study 
of the CERN Proton Synchrotron.

\subsection{Chaotic behaviour due to the long range beam-beam interaction}

In a colliding-beam storage ring, one of the largest perturbations affecting the motion of
beam particles is the collision with the opposing beam. This interaction occurs, unavoidably,
in the form of head-on collisions between bunches of the two beams at designated interaction
points. Hadron colliders employ long trains of closely spaced bunches, and individual bunches
encounter many others of the opposing beam at various long-range collision points, where the
beams are not fully separated into two separate
vacuum pipes. In general, the effect of the long-range collisions depends on the
ratio of the beam separation to the local rms beam size, and on the total number of long-range
collision points. For example, on either
side of the two LHC main collision points, a beam encounters about 15 long-range collisions
with an approximate average separation between the closed orbits of the two beams of 9.5 rms
beam sizes.

\begin{figure}[htb]
\begin{center}
{\includegraphics[height=8cm,width=8cm]{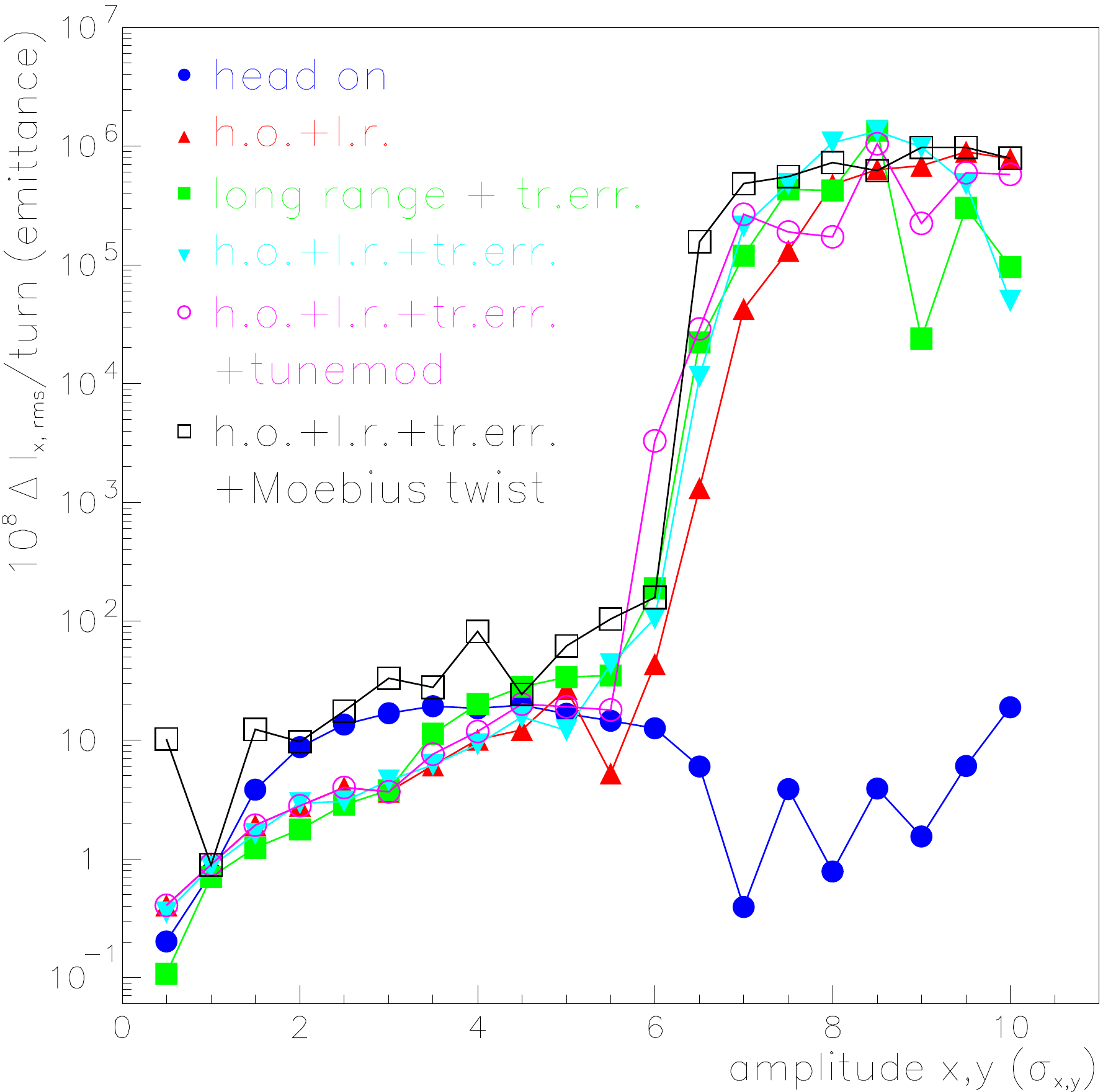}}
\end{center}
\caption{The change of action variance per turn
as a function of the starting amplitude. Compared are
the cases: head-on collisions only (dark blue), 
head-on and long-range collisions (red),
long-range collisions plus triplet field errors (green)
both types of collisions plus  triplet
field errors (light blue), the additional effect of a tune
modulation at the synchrotron frequency (22~Hz)
of amplitude $10^{-4}$ (pink), the additional effect
of a M\"{o}bius twist (black)~\cite{PapZim99}.}

\label{diffam}
\end{figure}
\begin{figure}[htb]
\begin{center}
{\includegraphics*[trim= 30 50 90 320, clip,height=6.8cm,width=7.5cm]{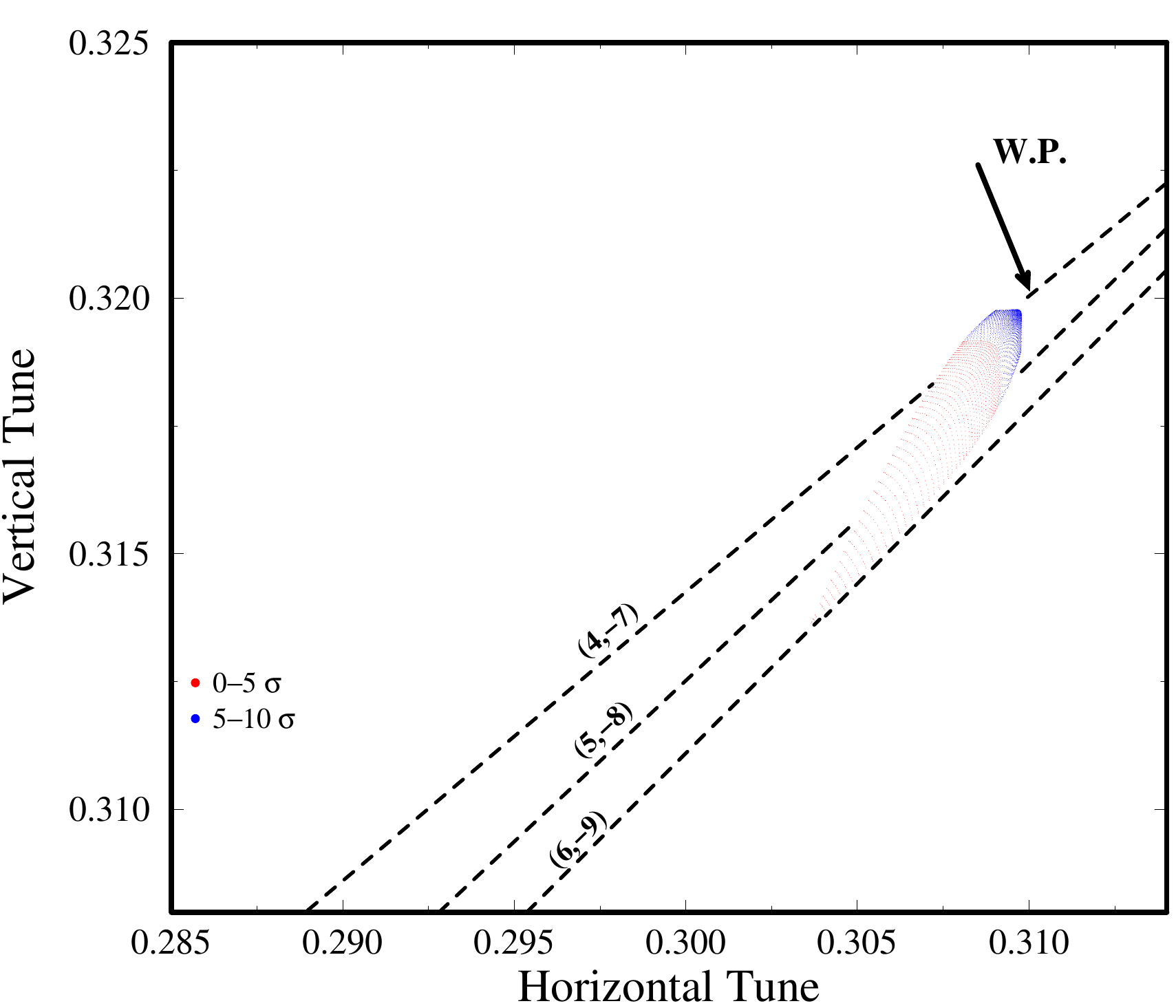}}
{\includegraphics*[trim= 30 50 90 320, clip,height=6.8cm,width=7.5cm]{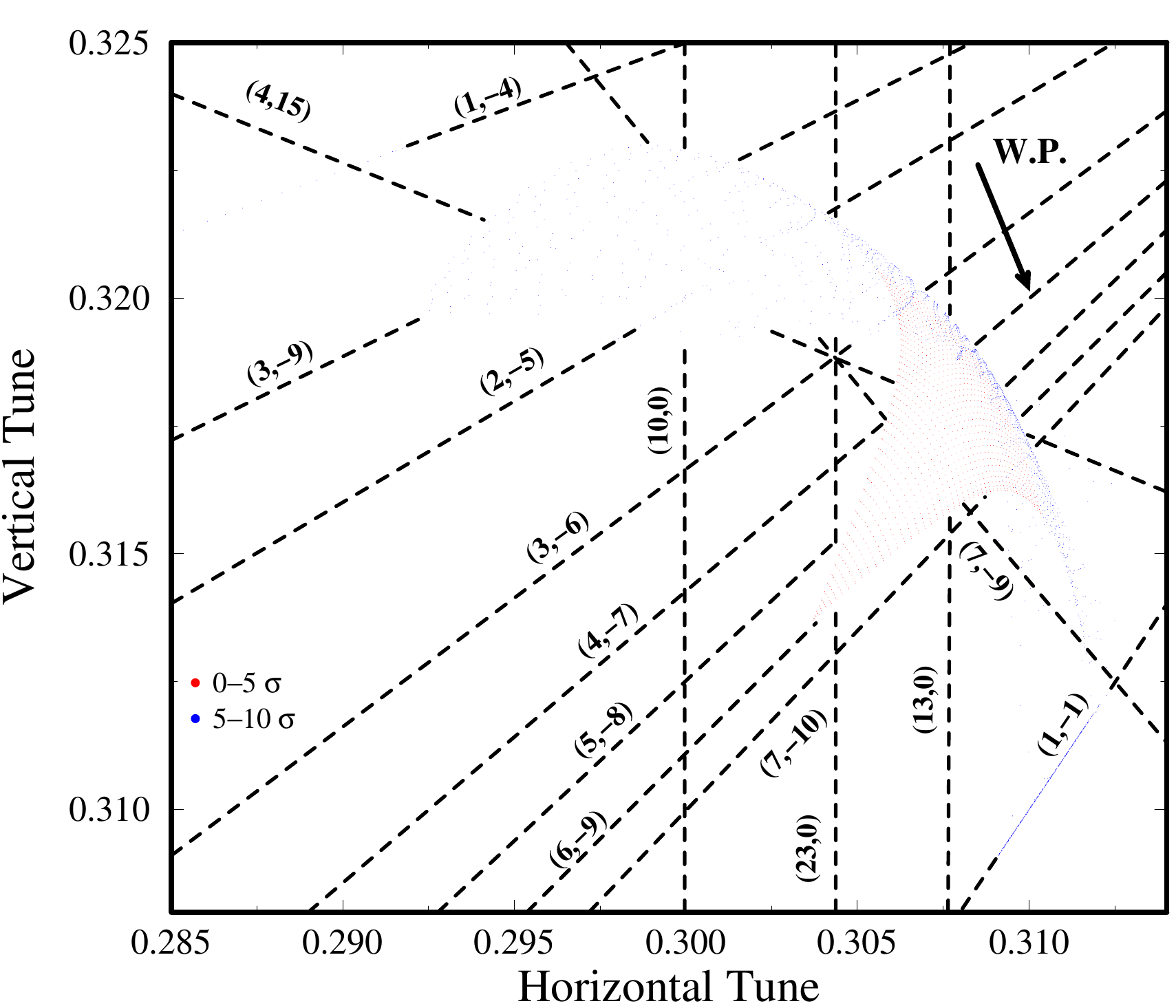}}\\
{\includegraphics*[height=6.8cm,width=7.5cm]{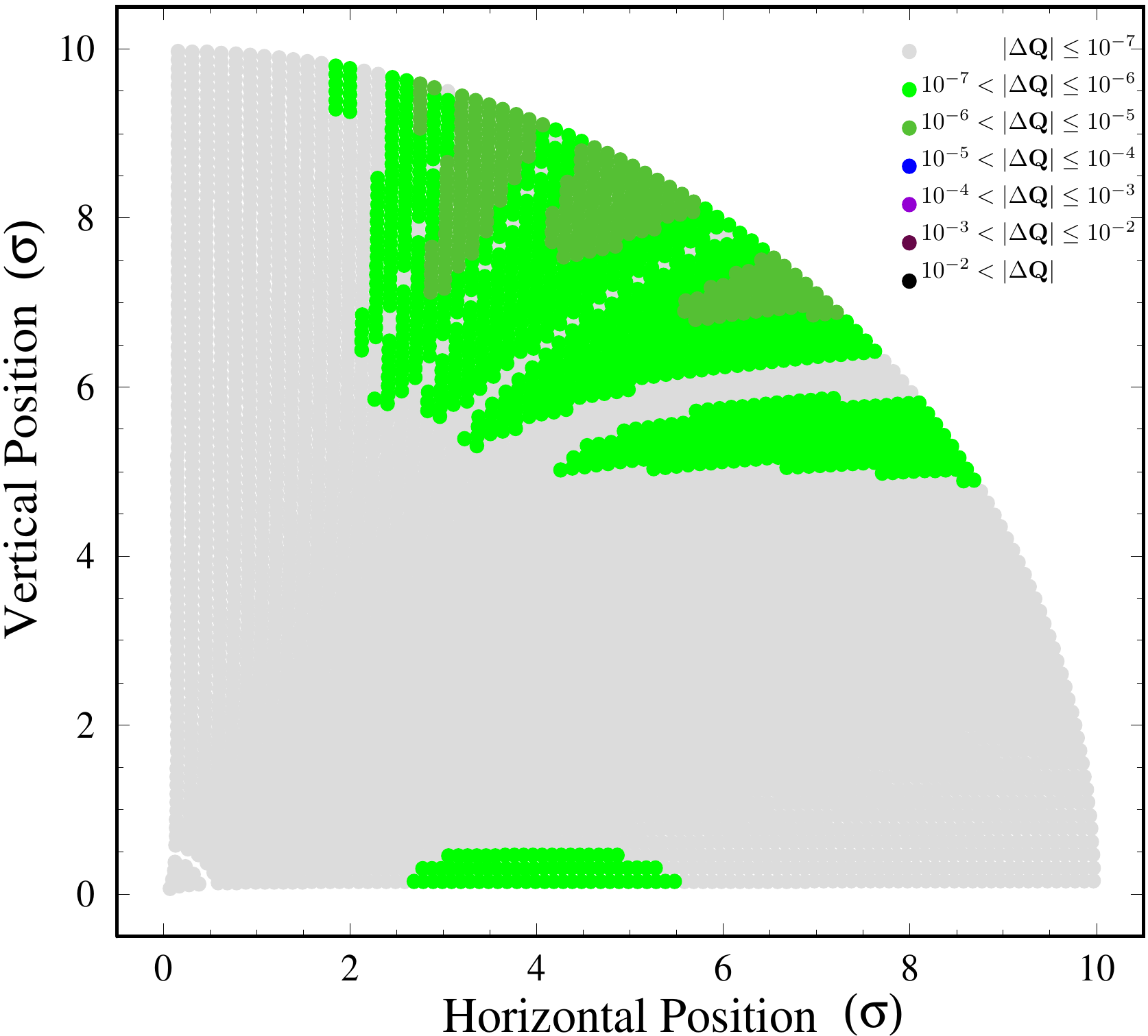}}
{\includegraphics*[height=6.8cm,width=7.5cm]{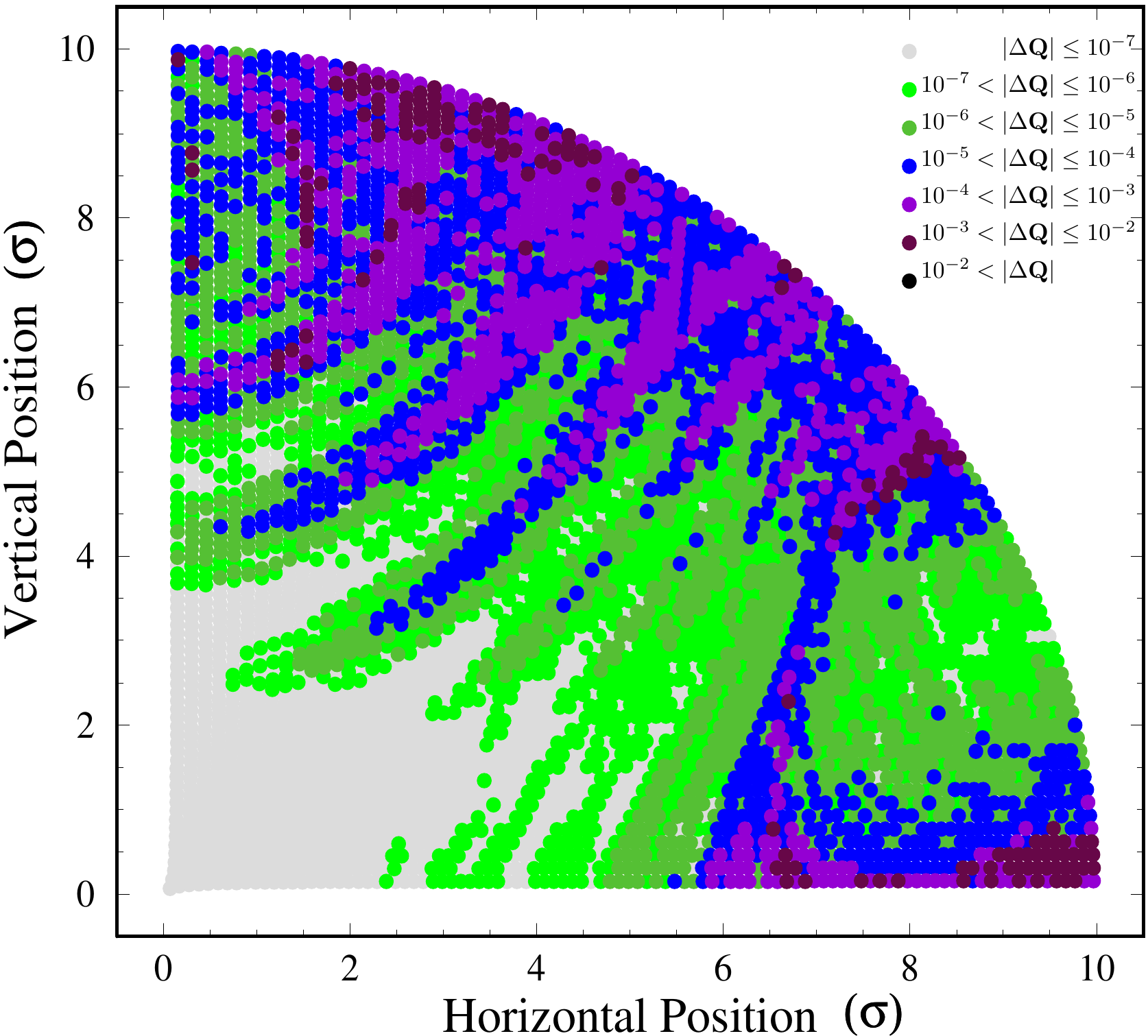}}
\end{center}
\caption{Frequency (top) and diffusion (bottom) maps for head-on
collisions only (left)  and including long-range collisions, as well (right)~\cite{PapZim99}.}

\label{footprint}
\end{figure}

Simulations predicted that the long-range collisions in hadron colliders give rise to a well
defined border of stability~\cite{Irwin,PapZim99}. This was also recently verified by measurements
in the LHC. Following standard ideas popularised by Chirikov 
and his collaborators \cite{Chirikov2,Chirikov3}, a diffusion
coefficient can be estimated in the chaotic region of phase space
by calculating the variance of the
unperturbed actions for a large number of turns and 
averaging over the quasi-randomly varying betatron phase variable as
\begin{equation}
D(J) = \frac{1}{2 \pi} \int_{0}^{2 \pi} d\phi
           \ [\Delta J (\phi) ]^{2}\nonumber \;\;.
\end{equation}
Fig.~\ref{diffam} displays the change 
of the action variance, in terms of beam sizes, as computed 
by beam-beam simulations which consider
the particle motion in a 4-dimensional transverse phase space for a model with 2 interaction
points  and parameters similar to those of the LHC~\cite{PapZim99}. The stability border is insensitive
to the presence of the head-on collision (filled circles with dark blue curve), and only marginally
affected by the nonlinear field errors in the final-triplet quadrupoles (squares with green curve)
or by a small additional tune ripple (empty circles with pink curve). This ``diffusive aperture'' with
long-range collisions is equally insensitive to transverse closed-orbit offsets between the two
beams at the head-on collision points~\cite{PapZim99}.

The top part of figure~\ref{footprint} presents frequency maps obtained by tracking single particles
over 1000 turns under the influence of beam-beam effects. Red dots
represent particles with initial transverse amplitudes up to 
5~$\sigma_{x,y}$, whereas blue dots show results for 
initial amplitudes up to 10~$\sigma_{x,y}$.
The dramatic effect of the long-range collisions is revealed through
the comparison of the maps obtained with (right) and without (left) the long-range effects. 
Up to initial particle amplitudes of around 6~$\sigma_{x,y}$, the effect of
the head-on collisions dominates. Then, the long-range effect
takes over and the frequency map flips, as the tune shift with
amplitude changes direction. This non-monotonic dependence of the tune
with respect to the amplitude is potentially dangerous for the
stability of particles beyond this limit~\cite{Laskar93,ref:NAFF2}.
The conclusions of the previous paragraph regarding the dominant
destabilising role of the long-range collisions are also confirmed
in the diffusion maps at the bottom plots of Fig.~\ref{footprint}.

\begin{figure}
\begin{center}
{\includegraphics*[height=7.5cm,width=7.5cm]{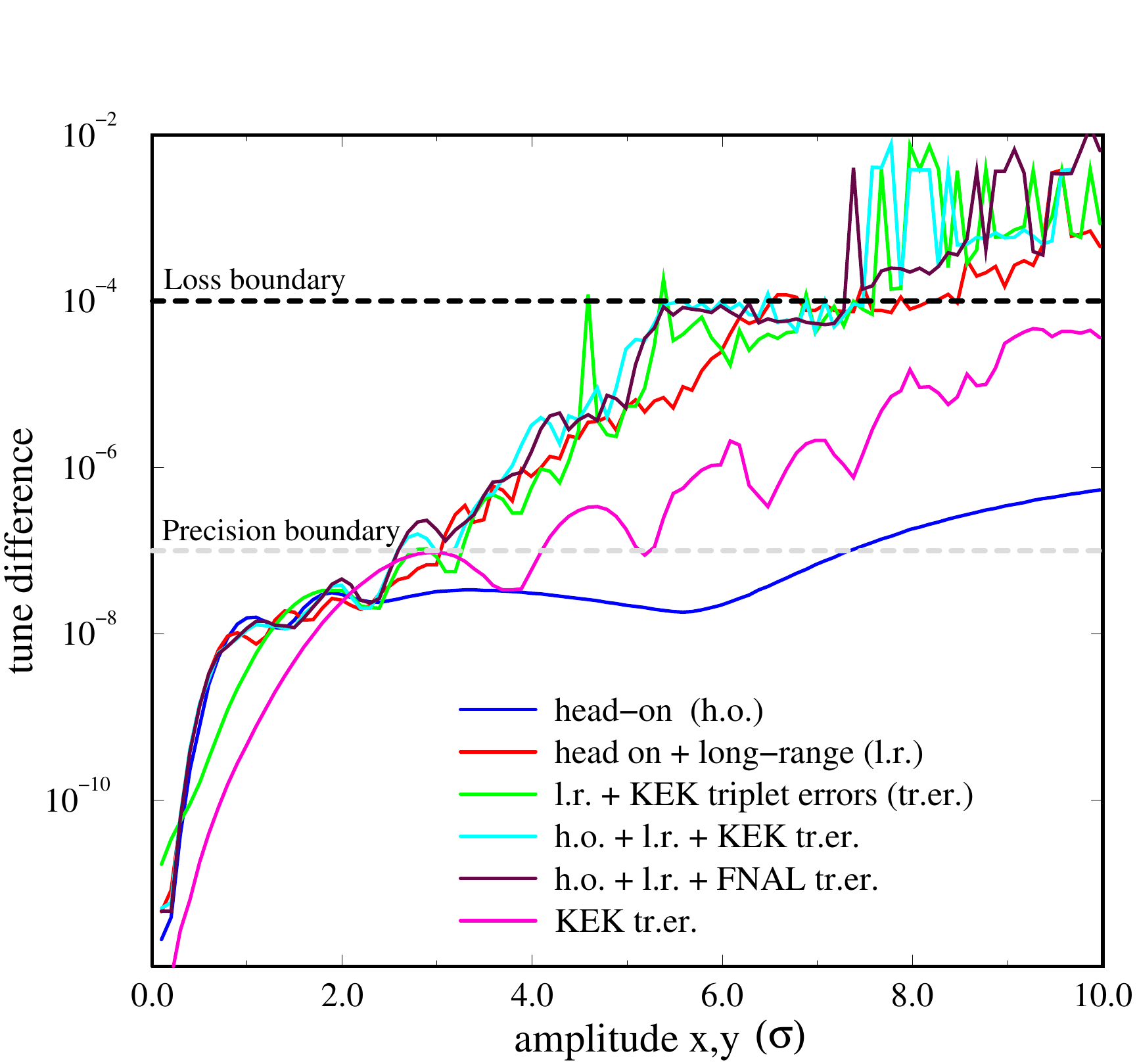}} \hskip 1cm
\end{center}
\caption{The change of frequency per 500 turns
averaged over all initial amplitude ratios $x-y$ as a function of the starting
amplitude. Compared are the cases: head-on collisions only; 
head-on and long-range collisions;
long-range collisions plus  triplet field errors;
both types of collisions plus  triplet
field errors; both types of collisions plus  triplet
field errors; only  triplet errors~\cite{PapZim99}.
}
\label{df}
\end{figure}

In Fig.~\ref{df}, the diffusion quality factor versus the amplitude,
averaged over all initial amplitude ratios is plotted, 
for different combinations of beam-beam and  triplet nonlinearities (quadrupole magnets focusing
the beam at the collision point). 
There are two thresholds marking the precision boundary and a  particle loss boundary for tune changes bigger than
$10^{-4}$. Considering linear frequency diffusion over time, this corresponds to one unit in frequency within $10^7$ turns, which
certainly induces particle loss. For all the cases where long-range collisions and triplet
field errors are included, the loss boundary is located at the same
point, around $5.5 \sigma_{x,y}$. For the case where the triplet field
errors are not added to the beam-beam effect, the threshold is reached a
little further, around $6\sigma_{x,y}$. The case with only 
triplet errors is clearly more stable, but indeed there is still
a visible effect
for larger initial amplitudes. For the case with only the head-on effect included,
there is no apparent chaotic diffusion present, as the tune
variation is very close to the precision limit of the method.

For the mitigation of the long range beam-beam effect , the use of electron lenses or DC wires was proposed and their effectiveness was tested  experimentally~\cite{wire1, wire2}. 
The positive experimental results of wire demonstrators at the LHC (IP1 and IP5). Using these compensators in different experimental studies (that are accompanied by numerical ones),
it was shown a significant improvement in beam lifetime~\cite{wire2}. These results encouraged using the wire demonstrators in the upcoming Run 3 of the LHC and are also under study
for their eventual implementation in the HL-LHC~\cite{wire3}.

\subsection{Dynamics of the CLIC Pre-damping rings}

The main limitation of the DA in the low emittance lattices comes from the non-linear effects induced 
by the strong sextupole magnet strengths, which are introduced for the  correction of the tune change with momentum (chromaticity). 
Following first order perturbation theory~\cite{ref:nonlinearsls}, the strength of a resonance $n_x Q_x+n_y Q_y=p$ of order $n$, with $|n_x|+|n_y|=n$ 
the order of the resonance and $p$ any integer, vanishes within an ensemble of $N_c$ cells, if the resonance amplification factor is~\cite{ref:ResFreeLat}
\begin{equation}
 \label{eq:resonancecancellation}
  \left| \sum_{\mathrm{p=0}}^{\mathrm{N_c-1}} e^{\mathrm{ip(n_x\mu_{x,c}+n_y\mu_{y,c})}} \right|=
  \sqrt{\frac{1-\cos[N_c(n_x\mu_{x\mathrm{,c}}+n_y\mu_{y\mathrm{,c}})]}{1-\cos (n_x \mu_{x\mathrm{,c}}+n_y \mu_{y\mathrm{,c}})}}=0\;\;,
\end{equation}
with $\mu_{x,y}$ the horizontal and vertical phase advance of the cell. Note that the tune is just the total phase advance for one turn.
The previous condition is achieved when $N_c(n_x\mu_{x\mathrm{,c}}+n_y\mu_{y\mathrm{,c}})=2k\pi$, provided the denominator of 
Eq.~\eqref{eq:resonancecancellation} is non zero, i.e.: $n_x\mu_{x\mathrm{,c}}+n_y\mu_{y\mathrm{,c}}\neq 2k'\pi$, with $k$ and $k'$ 
any integers. 
From this, a part of a circular accelerator will not contribute to the excitation of any non-linear resonances, except
of those defined by $\nu_x \mu_x +\nu_y \mu_y = 2 k_3 \pi$, if the phase
advances per cell satisfy the conditions: $N_c \mu_x = 2k_1\pi$ and $N_c \mu_y = 2k_2\pi$, where $k_1$, $k_2$ and $k_3$ are 
any integers. Prime numbers for $N_c$ are interesting, as there are less resonances satisfying both diophantine conditions simultaneously.


\begin{figure}[pht]
\centering\includegraphics[width=.45\linewidth,height=.32\linewidth]{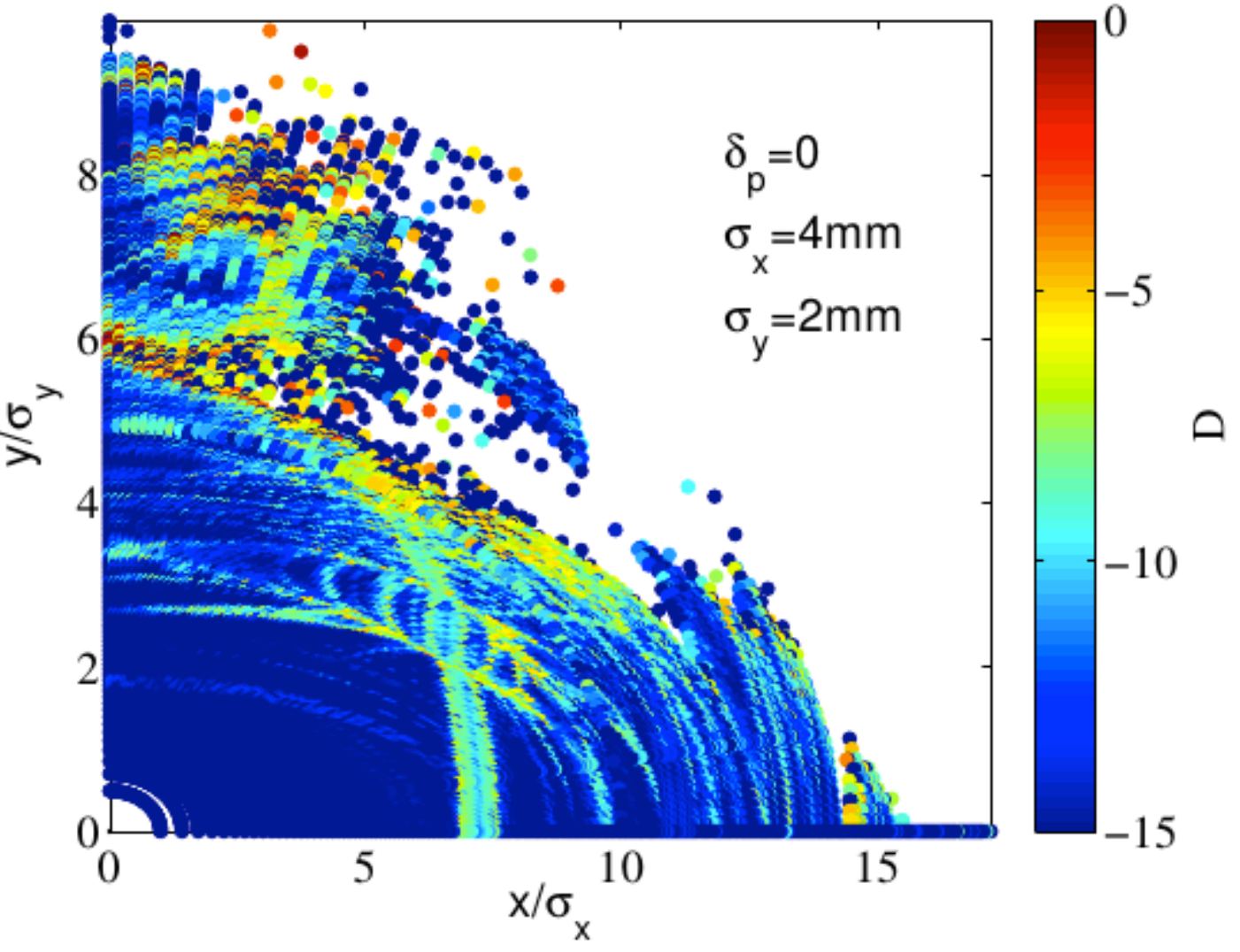}
\centering\includegraphics[width=.45\linewidth,height=.32\linewidth]{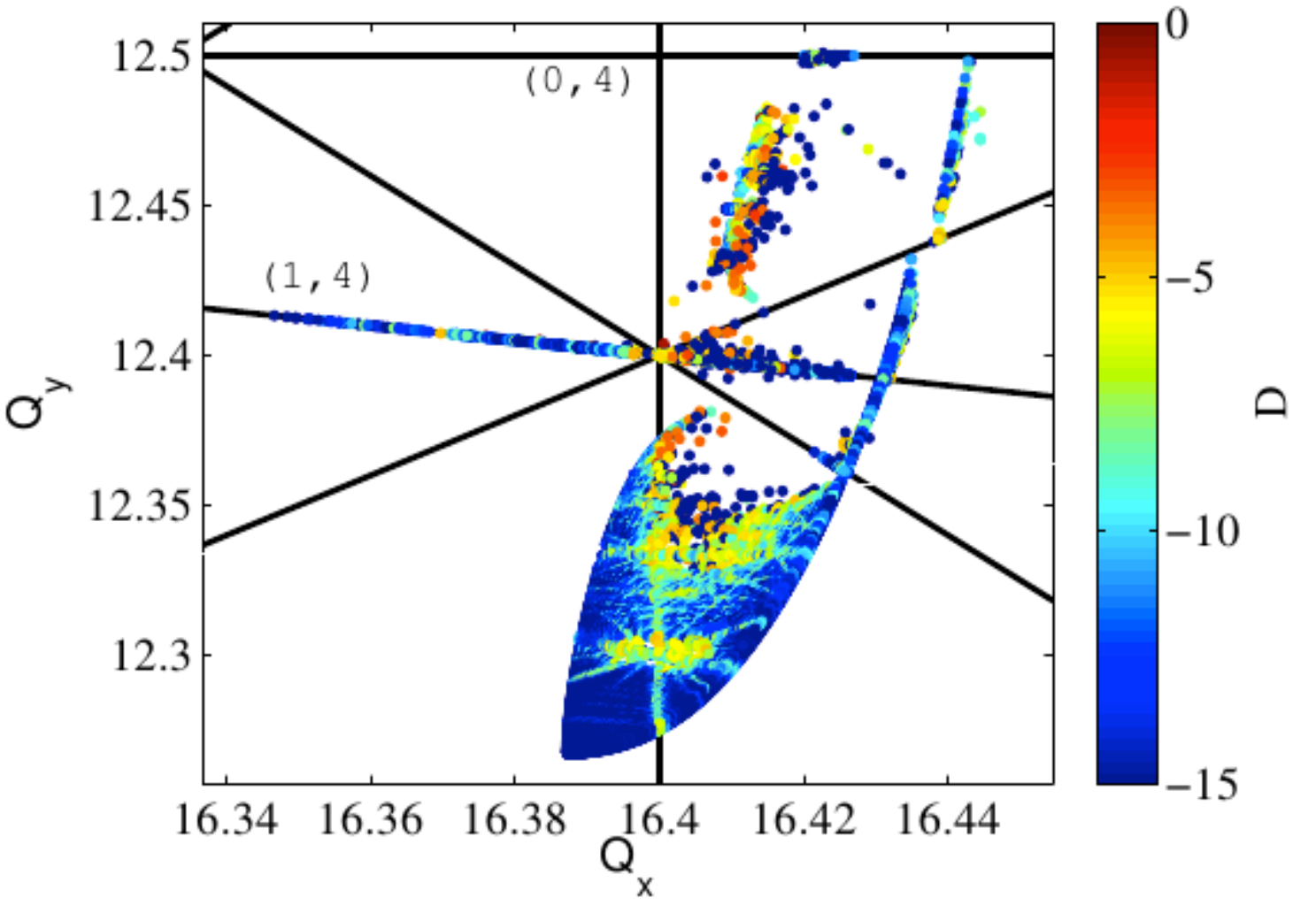} \\
\caption{Frequency maps (right) and diffusion maps (left) for on momentum particles, in the CLIC PDRs, for the working point (16.39,12.26)~\cite{Fanouria}.
}
\label{fig:FM-PDR}
\end{figure}

The nonlinear optimisation of the pre-damping rings (PDR), 
in the injector complex of the future Compact Linear Collider (CLIC),
was based on this ``resonance free lattice concept"~\cite{Fanouria}.
Tracking of particles with different initial conditions for 1024 turns, was performed with MADX-PTC~\cite{madxptc},
for a model of the lattice including sextupoles and fringe fields. 
Fig.~\ref{fig:FM-PDR} presents  frequency and diffusion maps for trajectories that survived over 1024 turns, color-coded with the 
diffusion coefficient of Eq.~\eqref{Diffvec}, for on-momentum particles. 
From the frequency maps it is observed that the tune is crossing the $(1,4)$ resonance, which is not eliminated 
by the resonance free lattice and the phase advances chosen ($\mu_x=5/17,~\mu_y=3/17$). 
This seems to be the main limitation of the DA. The shape of the frequency maps, especially at high amplitudes, 
does not have the triangular shape expected by the linear dependence of the tune shift to the action,  
and foldings appear. This occurs when terms of higher order in the
Hamiltonian become dominant over the quadratic terms, as the amplitude increases. 
This behaviour is expected due to the suppression of the lower order resonances, following the resonance free lattice 
concept.

\section*{Acknowledgements}
I would like to thank F.~Antoniou, F.~Asvesta, H.~Bartosik, W.~Herr, J.~Laskar, N.~Karastathis, S.~Kostoglou, S.~Liuzzo, L.~Nadolski, D.~Pellegrini, D.~Robin, C.~Skokos, C.~Steier, F.~Schmidt, G.~Sterbini, A.~Wolski, F.~Zimmermann for their collaboration and their various contributions to the material provided for this lecture. I would like to thank finally the CAS director H.~Schmickler for his patience regarding the very late delivery of this report.


%
%
%
%
%
%

\end{document}